 \documentclass[final,3p,times]{elsarticle}

\usepackage{amssymb,bm,amsmath,amsfonts,color}
\usepackage[version=3]{mhchem}
\newcommand{\whirl}{\kern0.5ex\vcenter{\hbox{$\scriptstyle\circlearrowright$}}\kern0.5ex}
\newcommand{\eff[1]}{({#1_{\textit{eff}}})}
\newcommand{\barred[1]}{\overline{#1}}
\usepackage{multirow}

\begin{document}

\begin{frontmatter}



\title{Multiple-scale stochastic processes: decimation, averaging and beyond}


\author{Stefano Bo}

\address{Nordita\\
KTH Royal Institute of Technology and Stockholm University,
Roslagstullsbacken 23, SE-106 91 Stockholm, Sweden }

     \author{Antonio Celani}
           
\address{Quantitative Life Sciences, The Abdus Salam International Centre for Theoretical Physics (ICTP), Strada Costiera 11, I-34151 - Trieste, Italy }
\begin{abstract}
The recent experimental progresses in handling microscopic systems {   have} allowed to  probe them at levels where 
fluctuations are prominent, calling for stochastic modeling in a large number of physical, chemical and biological phenomena.
 This has provided fruitful applications for established stochastic methods and motivated further developments.  
These systems 
often involve processes taking place on widely separated time scales.
For an efficient modeling one usually focuses on
the slower degrees 
of freedom and it is of great importance to accurately eliminate the fast variables in a controlled fashion, carefully accounting
for their net effect on the slower dynamics. This procedure in general requires to perform two different operations: decimation and coarse-graining.  We introduce the asymptotic methods that form the basis of this procedure and discuss their application to a series of physical, biological and chemical examples.
We then turn our attention to functionals 
of the stochastic trajectories such as residence times, counting statistics, fluxes, entropy production, etc. which 
have  been increasingly studied in recent years. 
For such functionals, the elimination of the fast degrees of freedom can present
additional difficulties and naive procedures can lead to blatantly inconsistent results.
Homogenization techniques for functionals are less covered in the literature
and we will pedagogically present them here, as natural extensions of the ones employed for the trajectories. We will also discuss recent applications of these techniques to the thermodynamics of small systems and their interpretation in terms of {   information-theoretic}  concepts.
\end{abstract}


\end{frontmatter}



\tableofcontents

 \clearpage

 \section{Motivations and examples: Stochastic processes with multiple time scales.}
 
 When modeling a process one of the most challenging issues to address is to determine 
which variables are the fundamental ones that characterize the system
and which ones are just details that can be safely disregarded. Once these relevant variables are identified, one would like to have a dynamic description of their evolution.

Neglecting the inessential processes is crucial for providing
a neat, useful description of the considered phenomena.
Several methods are available to reduce the complexity of involved systems especially when the scales on which the various interacting processes take
place are different. 
The use of stochastic formalisms in modeling physical, chemical or biological processes is 
on this line of reasoning. Rather than following the microscopic erratic evolution of
some processes  their net contribution
is accounted for by a random term that preserves the relevant statistical properties
that affect the overall behavior of the system.

Think for instance of Brownian motion: the erratic motion of a mesoscopic particle 
immersed in a fluid. 
A complete description would require the solution of the equation of motion of the Brownian particle
and of those of the many fluid particles with which it collides. Modeling the collisions 
as a white noise term with an amplitude that depends on the physical 
properties of the fluid and the particle greatly simplifies the formalism and yet provides
an excellent description of the statistics of the phenomenon.

Similarly, when considering chemical reactions \cite{gillespie2000chemical}, a probabilistic treatment in terms of a chemical master equation
provides an efficient formalism which is immensely simpler than the
the solution of the complete molecular dynamics of the reagents.

Once stochastic systems are taken as the starting models their complexity can also be reduced.
Various methods exist to extract the relevant features form large stochastic systems involving
several components. 
Here, we focus on the case in which different time scales are involved.
This provides a 'natural' distinction between the different processes.

A paradigmatic example is again given by Brownian motion, and namely by the strong friction (overdamped) limit of
Langevin-Kramers dynamics, in which the velocity
degrees of freedom very rapidly converge to equilibrium. 
As discussed above, the collisions
with the fluid particles can be described as a noisy force acting on the particle so that
the system obeys the following set of equations 
\begin{equation}\label{eq:LK}
\begin{array}{lll}
&\dot{{\bm X}}_t&= {\bm V}_t  \\
m&\dot{{\bm V}}_t&= -\gamma {\bm V}_t + \sqrt{2k_B T \gamma}\, {\bm \xi}_t
\end{array}
\end{equation}
where ${\bm \xi}_t$ is a Gaussian, zero-mean white noise, i.e. $\langle \xi^i_t\xi^j_{t'} \rangle =\delta^{ij}\delta(t-t')$,
$T$ is the fluid temperature, $k_B$ the Boltzmann constant and $\gamma$ the viscous friction coefficient. According
to Stokes law we have that $\gamma=6\pi\eta a$ where $a$ is the radius of the Brownian particle and ${\cal\eta}$ the fluid dynamic viscosity.
At the microscale friction dominates the motion. Indeed, a spherical particle of radius $a=1\,\mu m$ traveling at a speed 
of $50\,\mu m\,s^{-1}$ in water experiences such a strong damping that
if no external force is applied it would stop in less than a microsecond ($\sim m/\gamma$) and travel only $0.1 \,\AA$ \cite{purcell1977life}. 
This shows how inertia plays a negligible role in the dynamics and how relaxation takes place on a fast time-scale.
One is therefore tempted to discard it altogether 
and to revert to an Aristotelian description of the motion.
We would then have a description involving only the positional degrees of freedom 
and neglecting the velocity ones.
This is indeed possible and goes under the name 
of overdamped approximation
\begin{equation}\label{eq:OD}
\begin{array}{lll}
\dot{\bm X}_t & = & \sqrt{2 k_B T / \gamma } \; {\bm \xi}_t
\end{array}
\end{equation}
and provides an accurate description of the process on time scales 
longer than the relaxation ones ($t\gg m/\gamma$).

In biology there are usually many more time scales on which intertwined processes take place.
Just to mention an example consider the cell response to a change in the environment.
Such response involves several subsequent reactions starting 
from the detection of the signal by membrane receptors, which takes a fraction of a second, then triggering
 the signaling cascades, on the minutes time-scale, finally resulting in the
organization of the suitable transcriptional response, which takes place in several hours \cite{phillips2009physical}.

Also chemistry is rich in multiple time-scale reaction networks
where the characteristic times of the single reactions
may span several orders of magnitudes. 
Such networks are hard to describe in detail and the presence of processes
which are much faster than others makes
simulations via the Gillespie algorithm \cite{gillespie1976general}
computationally very demanding. Indeed, 
one has to simulate a large amount of fast events before
the slow ones (which usually are the relevant ones) start taking place.
When fast and slow reaction
are treated on the same footing the procedure cannot be efficient (see e.g. \cite{rao2003stochastic}).

One would rather aim at describing the dynamics on the slower time scales not by fully following
the details of the fastest processes but just accounting for their overall contributions
to the slower processes. Can this be done in a systematic way without making recourse to uncontrolled approximations ?

Asymptotic methods provide indeed the appropriate tool to derive systematically the slow dynamics by
consistently eliminating/grouping the fast degrees of freedom in order to obtain an effective Markov process that describes the transitions between suitably defined ``slow states" \cite{pavliotis2008multiscale}. 
For instance, the overdamped approximation of the Langevin-Kramers dynamics is consistent with the adiabatic elimination of the 
velocity degrees of freedom.

We will provide a detailed, systematic description of asymptotic methods for both discrete and diffusive Markov processes 
in the first part of this report. Examples from simple biochemical networks and processes on more complex graphs will be presented to show how to apply these techniques. For diffusive processes, we will discuss applications to population genetics and to Brownian motion in inhomogeneous fluid environments.

A much less explored question is what can possibly be said about observables that depend on the trajectories of the system. 
These include but are not limited to: 
residence and first-passage times, counting statistics, fluxes, entropy production, etc.
Is it possible to obtain a correct description of the slow evolution of these quantities ? What level of knowledge about the details of the fast dynamics is required ? 
In the second part of this report we will address these questions by extending the asymptotic methods presented in the first part.

A specific instance of this problem that deserves special attention is entropy production. Systems that are amenable to a stochastic modeling admit indeed a thermodynamic description where fluctuations  play a prominent role (see \cite{esposito2010three,lebowitz1999gallavotti,seifert2012review,sekimoto2010stochastic,van2013stochastic,kurchan1998fluctuation,ritort2008nonequilibrium} for the definitions).
Such stochastic thermodynamic framework has proven to be effective in the description 
of small systems involving few components studying the energy exchanges accompanying their evolution. 

The problem of deriving an effective thermodynamic description for slow dynamics has been addressed in several works (see e.g. \cite{esposito2012stochastic,kawaguchi2013fluctuation,puglisi2010entropy,rahav2007fluctuation,santillan2011irreversible,Nakayama2015,Ford2015,Celani2012,Bo2014,Lan2015,Marino2016}).
In section~\ref{sec:entro_block} and ~\ref{sec:diff_ent} and we will address this problem and show by the application of asymptotic techniques that under quite general non-equilibrium conditions it is not possible to give  a correct description of thermodynamics just in terms of the slow variables.  
Indeed, neglecting altogether the fast processes, while it leads to consistent equations for the effective dynamics,
can nevertheless overlook the irreversibility of some processes compromising the thermodynamic interpretation of the effective system.
(In order to reassure the reader, we anticipate that detailed balance will be shown to be sufficient to warrant consistency between 
the microscopic and the macroscopic thermodynamics.).

\begin{figure}[!h]
  \includegraphics[width=0.9\textwidth]{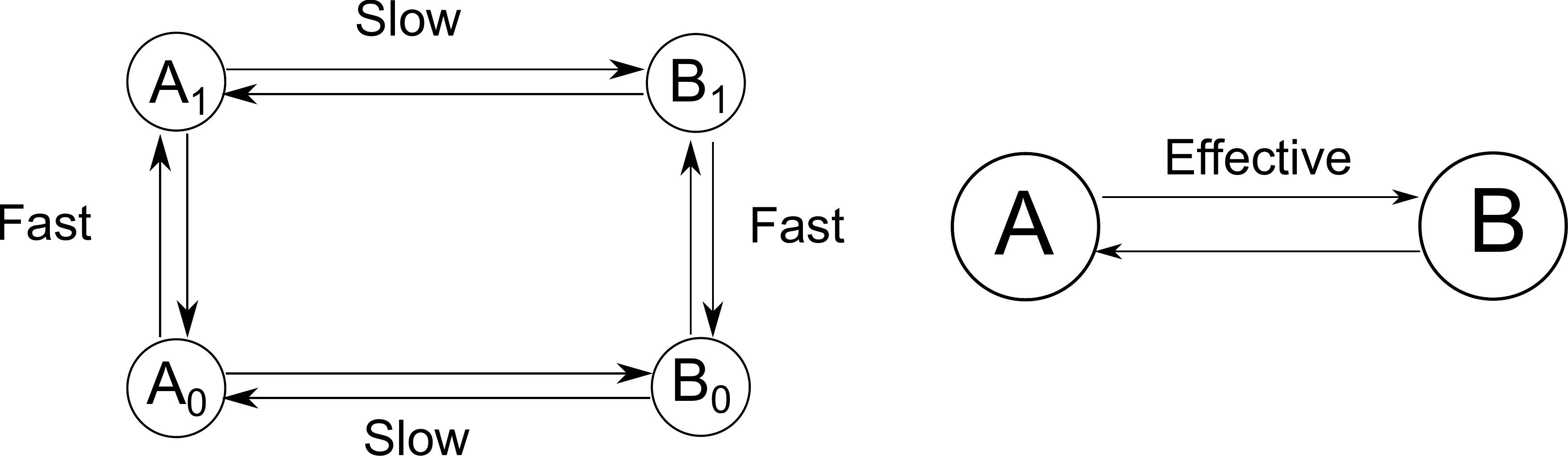}
\caption{\label{fig:sketch} 
Left: Full dynamics involving two time scales. Right: Effective dynamics}
\end{figure}
One can already get an intuition of this issue by considering the toy system sketched in figure \ref{fig:sketch} where we have four
states and four transition channels. State $A$ rapidly switches between two possible conformations which can slowly convert to a state $B$.
Also state $B$ oscillates between two states. Given the difference in the transition rates, by looking at the system on slower
scales (or with low temporal resolution), one would not distinguish the substates and basically observe the effective dynamics involving  $A$ and $B$.
This description provides a good approximation of the dynamics.
However, thermodynamically the full description and the effective one are crucially different.
The complete one can feature cycles at the steady state whereas the reduced one, displaying only two states, cannot.
This implies that the full system may be in a non-equilibrium steady state and that, on the contrary, for the effective system stationarity must correspond to equilibrium.

As embarassingly simple as this example may seem, it nonetheless has the merit of highlighting the risks in defining thermodynamics
starting from the effective dynamics without being aware of what the 
eliminated details are. At the same time, by showing what are the details of the fast processes that must not be neglected we 
can provide a tool for consistently deriving the effective description of entropy production.

We close this introduction with a disclaimer. This report makes no attempt at providing an exhaustive review of previous results in the subject. Rather, it aims at providing a personal viewpoint with the hope that  the reader will find in it a systematic way to approach these problems and gain some working knowledge about the relevant techniques.

 \clearpage
 \part{Dynamics}
 \section{Discrete Markov chains}\label{sec:dyn_discrete}
{   \paragraph{The question}
Is it possible to exploit time-scale separation to obtain an effective Markovian dynamics on a reduced state space?
Under what conditions? What are the relevant properties of the initial system that have to be kept into account? 
}\\
Several processes in physics \cite{van1992stochastic}, biology \cite{Elowitz2002,Paulsson2005} and chemistry  \cite{Goutsias2013} are best modeled by
continuous-time Markov chains.
 Studying the complete dynamics of complex Markovian networks becomes easily unfeasible as the number of
states increases. It is then of great importance to attempt to reduce such complexity by exploiting the structure of the network. This can be achieved by identifying groups of states that share
some features and deriving an effective dynamics between such groups or by directly eliminating some transient states based on their low probability of occupation.
However, in general, the resulting dynamics between blocks will not be Markovian and the waiting times for jumps will follow a phase type distribution reflecting
the Poissonian transitions between the internal states. 
Systems featuring separated time scales are of particular interest.
On the one hand, they pose serious issues for  simulation purposes (e.g. via the Gillespie algorithm \cite{gillespie1976general}) as one has to generate a large number of the fast events before observing the slow ones. Such problem is often referred to as stiffness and has attracted considerable interest~\cite{Haseltine2002,rao2003stochastic,Cao2005,Peles2006,Gillespie2009,Sanft2011,Gopich2006,Mastny2007,Shahrezaei2008,pigolotti2008coarse,sinitsyn2009adiabatic,esposito2012stochastic,Popovic2015,Bo2014,Jia2016}).
On the other hand, they suggest a natural structure for model reduction.
The reduction may be achieved via decimation: the elimination of transient states that rapidly relax to vanishing probabilities (see e.g.\cite{pigolotti2008coarse,Mastny2007,Shahrezaei2008,Popovic2015}).
The time separation may also allow
to define blocks of rapidly equilibrating states which then  undergo a Markovian effective dynamics on slower time scales,
a procedure which we shall refer to as averaging~\cite{Cao2005,esposito2012stochastic,Bo2014,pavliotis2008multiscale}.
In the present section we shall detail how a multiple-scale approach can be applied to systematically
 address systems presenting two widely separated time scales. This approach extends the one given in Refs.~\cite{pavliotis2008multiscale,Bo2014}.
 As examples we will provide applications of the general procedures to the problem of studying the fluctuations
 of protein numbers in cells, sensing and chemical kinetics.
\subsection*{Preliminaries}
 Consider a continuous-time Markov chain
 displaying transitions between discrete states with given rates.
In general, the evolution of such systems is described by a master equation:
\begin{equation}\label{eq:discr_me}
\frac{dp^{i}_{m}}{dt} = \sum_{j}\left( K^{i}_{j} p^{j}_{m} -
K^{j}_{i}  p^{i}_{m}\right)  
\end{equation}
where $p^{i}_{m}(t,t')$ denotes the probability of being in state $i$ at time $t$ having started from state $m$ at time $t'$ and 
$K^{i}_{j}$ the transition rate from state $j$ to state $i$. \footnote{Note that  in this section, whenever dealing 
with discrete processes, we assume no implicit summation for the repeated indices.}
In compact notation  \begin{equation}\label{eq:me}
\frac{d \mathsf{p}}{dt} = \left( \mathsf{K} - \mathrm{diag}(\mathsf{1^\dagger K}) \right) \mathsf{p} 
= \mathsf{L p}
\end{equation}
where $\mathsf{1^\dagger}$ {   is}  a row vector with all entries equal to $1$ and, consequently, 
$\mathrm{diag}(\mathsf{1^\dagger K})$ is a diagonal matrix with the elements $i, i$ given by the exit rate from 
$i$: $e_i=\sum_j K^j_i$. 
 \\
\subsection*{{   Time-scale} separation}
 Let us now turn our attention to systems with transition rates of different magnitude.
 For the sake of clarity we shall discuss the case in which the rates belong to two different
 classes, a slow one and a fast one. To make such separation explicit we will denote the ratio between slow and fast rates as $\epsilon$. The corresponding master equation then takes the following form: 
  \begin{equation}\label{eq:me_ts}
\frac{d \mathsf{p}}{dt} = \epsilon^{-1}\mathsf{M p} + \mathsf{L^{(0)} p}
\end{equation}
where $\mathsf{M}$ generates the fast process and $\mathsf{L^{(0)}}$
the slow one\footnote{Note that the time-scale separation is accounted for by the parameter $\epsilon$ and that  $\mathsf{M}$ and $\mathsf{L^{(0)}}$ are of the same order of magnitude}.
Both $\mathsf{M}$ and $\mathsf{L^{(0)}}$ conserve probability:  
$\mathsf{1^\dagger M}=\mathsf{1^\dagger L^{(0)}}=0$.
Given a pair of states, a transition in one direction can be much faster than its reversed one. 
This means that for such states we will have 
$M^j_i\neq 0$ but $M^i_j=0$. 
As a consequence, the directed graphs associated to the fast and slow transitions may not be 
strongly connected.
Recall that a directed graph is said to be strongly connected if, for any given pair of its vertices (e.g. $i$ and $j$),  there is a path connecting them in each direction (e.g. $i\to\ldots\to j$, $j\to\ldots \to i$, it need not be the same path).\\
 As anticipated in the introduction, the existence of different time scales can justify certain approximations (which become exact in the limit
 of infinite time-scale separation, {\it i.e.}, $\epsilon\to 0$) that simplify considerably the dynamics.
 The kind of simplification and its extent depends mostly on the structure  of the fast connections and can be determined by inspection of the network composed solely of the fast transitions, {\it i.e.}, by the algebraic properties of $\mathsf{M}$. If the fast network is strongly connected, it is not possible to 
 exploit the time-scale separation to reduce the complexity of the complete system by defining some suitable effective Markovian system. 
 On the contrary, if the graph is not strongly connected,  $\mathsf{M}$ is a reducible matrix and it is possible to 
 express it in its normal form, {\it i.e.},  as a lower block triangular matrix by permutations of its rows and columns. Each  {   block} along the diagonal represents a strongly connected component.
 The network composed  of the strongly connected components and the  transitions across them is called a condensation of the
 graph and it is a directed acyclic graph.
 Inspection of the graph condensation allows to identify the blocks that will take part in the effective dynamics.
 They are the ones with no
 outgoing transitions: the sinks of the directed acyclic graph.
 Indeed, the blocks with outgoing connections do not conserve probability but "leak" it on the fast scale
leading to a vanishing probability of occupation and will not be present in the effective description.
 A transient block is characterized by a negative sum of its rows, indicating that it is not conserving probability.
 The elimination of transient blocks (or states) is often referred to as decimation.
Summarizing, we have the following recipe for simplifying the dynamics in presence of a 
time-scale separation when the graph composed of the fast transitions is not strongly connected:
\begin{enumerate}
\item Inspect if the fast transitions represent a {   strongly} connected graph.
\item Identify the blocks, {\it i.e.}, the strongly connected components of the graph associated to
$\mathsf{M}$ (the fast transitions generator) 
\item Eliminate the blocks that do not conserve probability, {\it i. e.}, keep only sinks of the graph condensation
\item Derive the effective dynamics between the remaining blocks on the slower time-scale
\end{enumerate}
 \begin{figure}[!h]
  \includegraphics[width=0.9\textwidth]{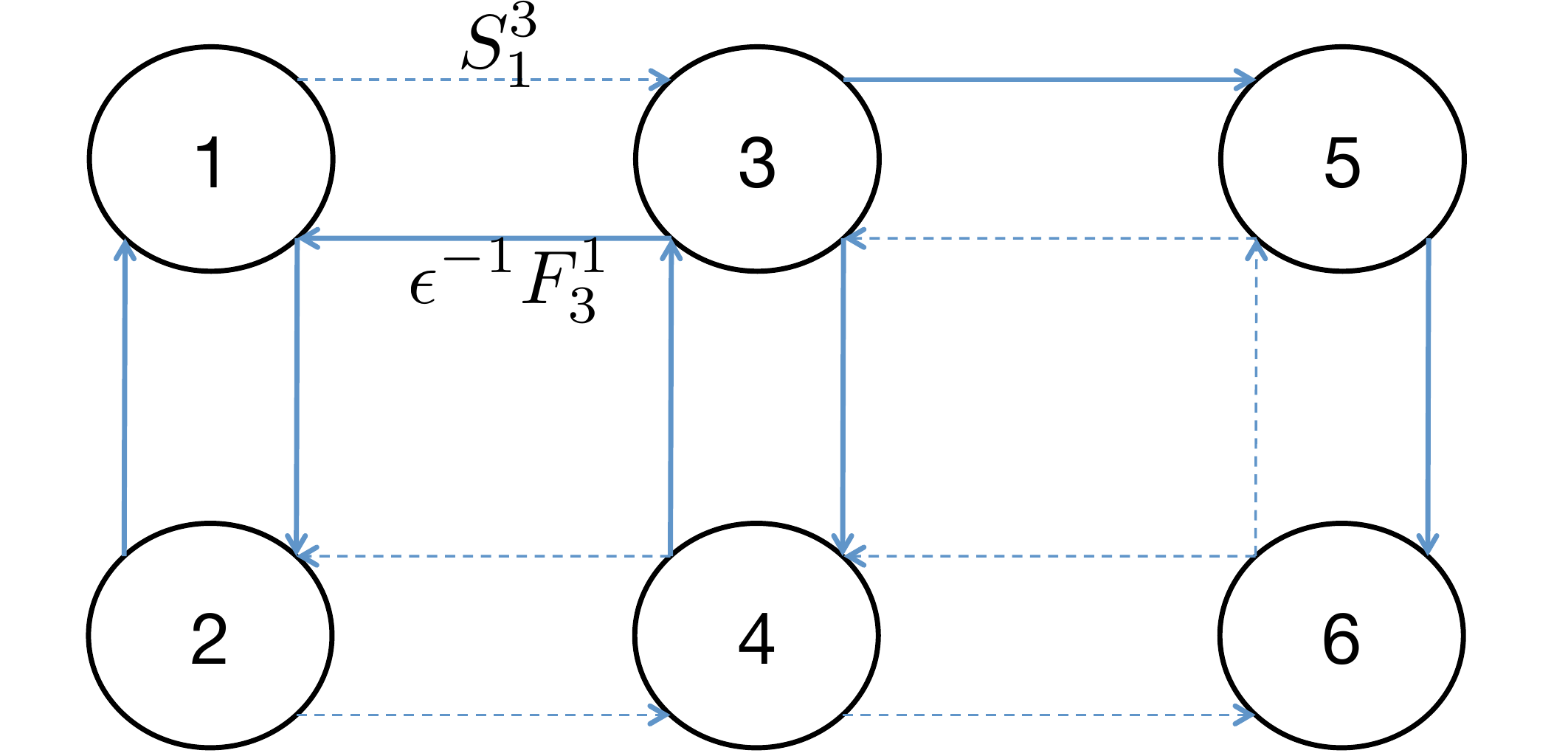}
\caption{\label{fig:dec_gen} {  Continuous}-time Markov chain involving 6 states. The dashed lines represent transitions occurring at low rates whereas the solid ones fast transitions.}
\end{figure}
Notice that for the first three steps, only knowledge about the fast transitions is required.
The asymptotic approach we present allows to eliminate the rapidly emptying blocks and to derive
an effective dynamics between the remaining ones.
We shall present this general procedure, via a general example
where, for the sake of simplicity we do not consider the dependence on the initial state so
that the probability is simply a vector {   $p^i$}.
\subsection{General example}
Consider the Markov chain described in figure~\ref{fig:dec_gen}
 where the solid lines represent fast rates and the dashed ones
slow rates.
 The generators read respectively:
 \begin{equation}\label{eq:M}
  M=
 \left(
 \begin{array}{c c c c c c}
 -F^2_1	& F^1_2 	 & F^1_3		&0	&0	& 0 \\
  F^2_1	& -F^1_2 	 & 0		&0	&0	& 0 \\
  0	&0	&-\left(F^1_3+F^4_3+F^5_3\right)&F^3_4	&0	&0\\
  0	&0	&F^4_3	&-F^3_4	&0	&0\\ 
  0	&0	&F^5_3	&0	&-F^6_5	&0\\
  0	&0	&0	&0	&F^6_5	&0
 \end{array}
 \right)
 \end{equation}
 and
  \begin{equation}
  L^{(0)}=
 \left(
 \begin{array}{c c c c c c}
 -S^3_1	& 0 	 & 0		&0	&0	& 0 \\
  0	& -S^4_2 	 & 0		&S^2_4	&0	& 0 \\
  S^3_1	&0	&0	&0	&S^3_5	&0\\
  0	&S^4_2	&0	&-\left(S^2_4+S^6_4\right)	&0	&S^4_6\\ 
  0	&0	&0	&0	&-S^3_5	&S^5_6\\
  0	&0	&0	&S^6_4	&0	&-\left(S^4_6+S^5_6\right)
 \end{array}
 \right)
 \end{equation}
 where the $F$ and $S$ are of the same order (the bookkeeping parameter $\epsilon$ takes care of their
 magnitude difference) and used to label respectively fast and slow transition rates.
 \paragraph{Step 1}
 By inspection of the network involving fast transitions only (solid lines in figure 2
 )
 we see that it is not strongly connected. 
  \paragraph{Step 2}
 Indeed, the matrix $M$ of eq. \ref{eq:M} is reducible and,
 by permutation of its states (to $(3,4,5,1,2,6)$), it can be written as a block triangular matrix 
 \begin{equation}\label{eq:Mmod}
  \tilde{M}=
 \left(
 \begin{array}{c c  c  c  c  | c}
 -\left(F^1_3+F^4_3+F^5_3\right)&F^3_4	&\multicolumn{4}{|c }{}	\\
 F^4_3	&-F^3_4	&\multicolumn{4}{|c}{  
 \text{\huge0}
   }	\\ 
 \cline{1-3}
  F^5_3&0  &\multicolumn{1}{|c |}{-F^6_5}	&\multicolumn{3}{|c }{}\\
  \cline{3-5}
 F^1_3	&0& 0& \multicolumn{1}{|c }{-F^2_1}	& F^1_2 	 & 		 \\
  0	&0  &0 &\multicolumn{1}{|c }{F^2_1}	& -F^1_2 	 & 			 \\
  \cline{4-6}
  0	&0	&F^6_5	&0 & 0 &0\\
 \end{array}
 \right)
 \end{equation}
 showing the underlying block structure.
 Four blocks can be identified: the one including states  $3,4$ which we shall name $b$, the one with $1, 2$
 called $a$ and two blocks involving single states, $5$ and $6$ that we will name respectively $c$ and $d$. They represent the strongly connected components of the graph
 describing the fast transitions. 
  \begin{figure}[!h]
  \includegraphics[width=0.9\textwidth]{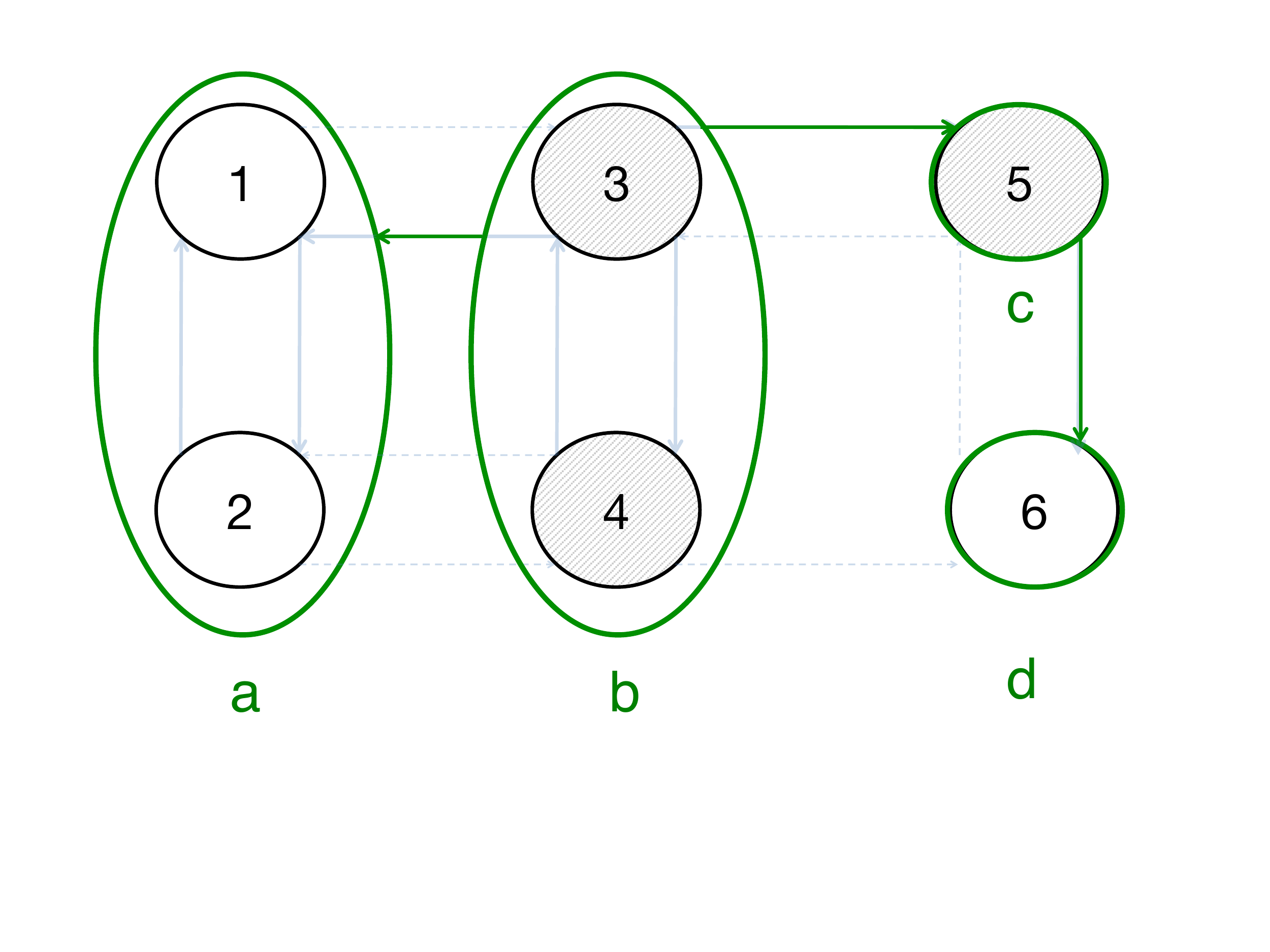}
\caption{\label{fig:dec_gen_cond} Condensation of the fast transitions graph of figure \ref{fig:dec_gen}.
{   Four} blocks (strongly connected components) are identified and shown in green.
The fast transitions across blocks are highligthed with green arrows. Only blocks $a$ and $d$ are sinks: they have have no fast outgoing transitions.}
\end{figure}
  \paragraph{Step 3}
  The condensation of the fast connections graph is shown in figure \ref{fig:dec_gen_cond}.
  Blocks $b$ and $c$ have outgoing fast transitions whereas blocks $a$ and $d$ are sinks of the condensation.
  Indeed, summing the first two lines of the fast transition matrix (\ref{eq:Mmod}), one sees that block $b$ does not conserve probability
 and will therefore relax to a vanishing probability of occupation on fast time scales. Also block $c$ (state $5$)  "leaks" probability. Hence, the blocks $b$ and $c$ can be eliminated from the effective description. On the contrary, blocks $a$ and $d$ conserve
 probability and will be part of the effective dynamics on the slow scales. 
  \paragraph{Step 4}
  In order to determine the effective dynamics between the surviving blocks we will systematically exploit the time-scale
  separation employing a multiple scale technique (see e.g. \cite{pavliotis2008multiscale}). 
  The first key idea is to formally introduce the fast time-scale $\theta=\epsilon^{-1} t$  on which the rapid processes are expected to equilibrate. The second  one is to develop the probability as $p=p^{(0)} + \epsilon p^{(1)} + \ldots$.
 Applying such expansions to eq.~(\ref{eq:me_ts}), one gets a hierarchy of equations.
 At order $\epsilon^{-1}$ the system 
 reads
 \begin{equation}
\frac{dp^{(0)}}{d\theta} = \mathsf{M} p^{(0)}\;.
\end{equation}
After the initial rapid relaxation (requiring a time $O(\epsilon)$) we obtain a stationary solution
\begin{equation}
\mathsf{M} p^{(0)}=0
\end{equation} 
where $p^{(0)}$ is a  positive column vector belonging to the right nullspace of the
matrix $\mathsf{M}$. The presence of two blocks conserving probability in $\mathsf{M}$ implies
that the nullspace has multiplicity $2$. It is spanned by
  \begin{equation}\label{eq:w}
w^a=
 \left(
 \begin{array}{c }
 w^{a_1}	\\
 w^{a_2}	\\ 
 0	 \\
  0	 \\
  0\\
  0
 \end{array}
 \right)
\qquad\qquad w^d=
 \left(
 \begin{array}{c }
 0	 \\
  0	 \\
  0\\
  0\\
  0	\\
 w^{d_6}
 \end{array}
 \right)
 \end{equation}
 where $w^{i}$ is a normalized equilibrium solution of the block $\mathsf{M}_i$:
 \begin{eqnarray}\label{eq:eqw}
 w^{a_1}=\frac{F^1_2}{F^1_2+F^2_1}\qquad w^{a_2}=\frac{F^2_1}{F^1_2+F^2_1}
 \qquad w^{d_6}=1\;.
 \end{eqnarray}
 At first order then, one finds the probability to be
 \begin{equation}\label{eq:dis_p0}
 p^{(0)}=\phi_a w^a +\phi_d w^d
 \end{equation}
 where $\phi_i$ is a generic function obeying the normalization $\sum_i\phi_i=1$. We have then split the first order probability into two factors:
 the probability of being in a  block $\phi_i$ and the equilibrium (with respect to the fast dynamics)
  probability of being in a state conditional on a given block.
The equation at order $\epsilon^{0}$  reads
\begin{equation}
\frac{dp^{(1)}}{d\theta} + \frac{dp^{(0)}}{dt}
= \mathsf{M} p^{(1)} + \mathsf{L^{(0)}} p^{(0)} 
\end{equation}
and, after the {   fast} relaxation,
\begin{equation}\label{eq:me_ex_0}
 \mathsf{M} p^{(1)} = \frac{dp^{(0)}}{dt}- \mathsf{L^{(0)}} p^{(0)}\;.
\end{equation}
According to the Fredholm alternative, eq.~(\ref{eq:me_ex_0}) is solvable only if  its RHS is orthogonal to the left nullspace of $\mathsf{M}$. 
Such nullspace has the same dimension of the right one (2) and, for this example, it is spanned by:
\begin{equation}\label{eq:v}
v^a= \left(1,\,1,\,\frac{F^1_3}{F^1_3+F^5_3},\,\frac{F^1_3}{F^1_3+F^5_3},\,0,\,0\right)
\qquad\qquad v^d=
\left(0,\,0,\,\frac{F^5_3}{F^1_3+F^5_3},\,\frac{F^5_3}{F^1_3+F^5_3},\,1,\,1\right)
 \end{equation}
 where the vectors are chosen to be biorthonormal to the ones spanning the right nullspace.
 The entries of such vectors have a clear interpretation in terms of probability of being absorbed by a given block. The entries $v^a_i$ are the probability of being absorbed in block $a$ having started in state $i$. 
So, for example, the first two entries of $v^a$ indicate the probability of being absorbed in block $a$ having started from $a$, namely
from 1 and 2, which is equal to one. The  third and fourth entry are the probability of ending in $a$ having started in $b$, from 3 and 4, and are given by the probability of choosing a transition form state $3$ to state $1$ rather than
 one to state $5$: {  $\frac{F^1_3}{F^1_3+F^5_3}$}.
  
 A generic vector in the left nullspace can be written as 
  \begin{equation}
 \hat{p}^{(0)}=\hat{\phi}_a v^a +\hat{\phi}_d v^d\
 \end{equation}
 for generic $\hat{\phi}_i$.
 The solvability condition then requires that:
 \begin{equation}\label{eq:gen_eff}
\hat{p}^{(0)} \frac{dp^{(0)}}{dt}- \hat{p}^{(0)}\mathsf{L^{(0)}} p^{(0)}=0
 \end{equation}
 which, owing to the biorthonormality, reduces to the set of two equations:
{  
 \begin{eqnarray}
  \hat{\phi}_a\left\{\frac{d \phi_a}{dt}-v^a\mathsf{L^{(0)}}\left(\phi_aw^{a}+\phi_dw^{d}
\right)\right\}=
 \hat{\phi}_a\left\{\frac{d \phi_a}{dt}-\frac{1}{F^1_3+F^5_3}\left[ F^1_3S^4_6\phi_d-F^5_3(w^{a_1}S^3_1+w^{a_2}S^4_2)\phi_a
\right] \right\}&=&0\\\nonumber
  \hat{\phi}_d\left\{\frac{d \phi_d}{dt}-v^d\mathsf{L^{(0)}}\left(\phi_aw^{a}+\phi_dw^{d}
\right)\right\}=\hat{\phi}_d\left\{\frac{d \phi_d}{dt}+\frac{1}{F^1_3+F^5_3}\left[ F^1_3S^4_6\phi_d-F^5_3(w^{a_1}S^3_1+w^{a_2}S^4_2)\phi_a
\right] \right\}&=&0\;.
 \end{eqnarray}
 }
  \begin{figure}[!h]
  \includegraphics[width=0.9\textwidth]{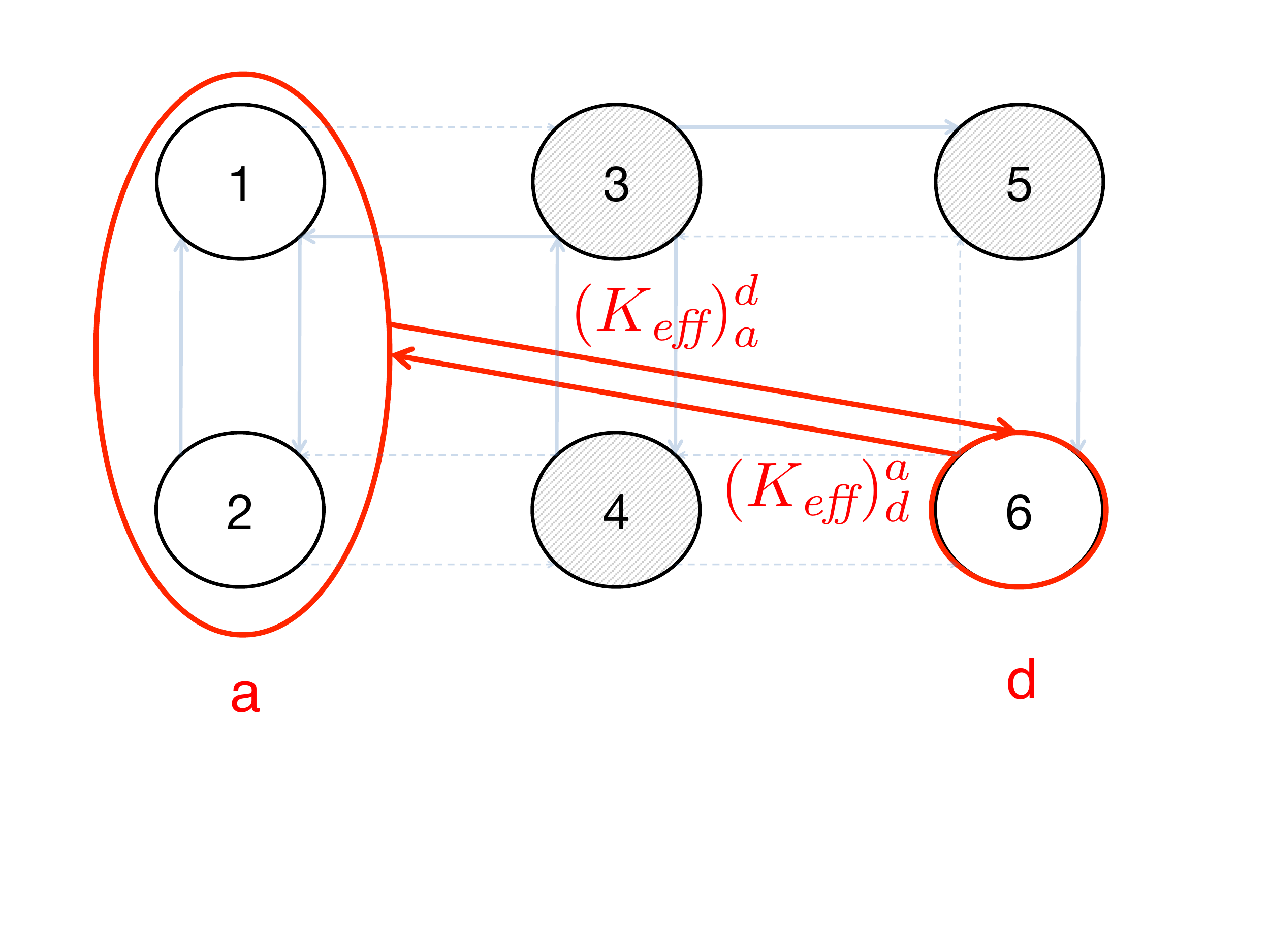}
\caption{\label{fig:dec_gen_eff}    Effective {   process} corresponding to the Markov chain described in fig.~\ref{fig:dec_gen}.
The dynamics takes place between block $a$ (consisting of $2$ states) and block $d$ (a single state) with rates
$ \eff[K]^a_d$ and $ \eff[K]^d_a$.
 State $3,\,4,\,5$ have a vanishing probability at first order and are therefore decimated from the system.
 Within block $a$ the probability of being in state $1$ and $2$ are given by the steady state 
 solution $ w^{a_1}$ and $ w^{a_2}$ as in eq.~(\ref{eq:eqw}).
}  
\end{figure}
 Since such condition has to be satisfied for any choice of $ \hat{\phi}_i$ the terms inside braces must equal zero
 giving what is the effective dynamics between the surviving blocks:   \begin{eqnarray}
  \frac{d \phi_a}{dt}= \eff[K]^a_d\phi_d- \eff[K]^d_a\phi_a\\\nonumber
   \frac{d \phi_d}{dt}= \eff[K]^d_a\phi_a- \eff[K]^a_d\phi_d
   \end{eqnarray}
   with the effective rates
    {  
     \begin{eqnarray}\label{eq:rate_eff_nb}
 \eff[K]^a_d&=&v^a\mathsf{L^{(0)}}w^{d}= -v^d\mathsf{L^{(0)}}w^{d}=\frac{F^1_3}{F^1_3+F^5_3}S^4_6\\ \nonumber
 \eff[K]^d_a&=&v^d\mathsf{L^{(0)}}w^{a}=-v^a\mathsf{L^{(0)}}w^{a}=\frac{F^5_3}{F^1_3+F^5_3}(w^{a_1}S^3_1+w^{a_2}S^4_2)\;.
   \end{eqnarray}
  The fact that $ v^a\mathsf{L^{(0)}}w^{d}= -v^d\mathsf{L^{(0)}}w^{d}$ is ensured by $v_a+v_d=(1,\ldots ,1)$  and  $\sum_i L^{(0)}_{ij}=0$.}
   Equation~(\ref{eq:gen_eff}) then identifies the effective dynamics which can  be expressed as (see fig.~\ref{fig:dec_gen_eff})
    \begin{equation}\label{eq:eff_comp}
 \frac{d\phi}{dt}= \eff[\mathsf{L^{(0)}}] \phi
  \end{equation}
  where the dimension of $\phi$ is given by the number of blocks that conserve probability
  ($2$ in this example).
  The shown procedure has allowed us to reduce the initial network involving all states ($6$)
  to a much smaller one {   in two steps}: decimating the blocks not conserving probability (state $ 3,\, 4,\, 5$)
  and averaging by grouping the states that rapidly reached  equilibrium (state $1$ and $2$) in a block.
    Before proceeding further, let us present an alternative, equivalent derivation of the effective dynamics which highlights the connection with the decimation procedure discussed, for example, in \cite{pigolotti2008coarse,Jia2016}.  Let us rearrange our states by ordering first the blocks with vanishing probabilities (transient)
  and then the ones conserving probability (recurrent).
  Our probability vectors then read: $p=(p_T, p_R)$ where $p_T$ denotes the probability of the
   transient states and $p_R$ the ones of the recurrent states. The  order of the example $(3,4,5,1,2,6)$, is already in this form. However, in general, listing first the vanishing states is more restrictive than simply ordering the matrix in block triangular order.
 By definition $p_T^{(0)}=(0\ldots 0)$ and for the chosen example $p_R^{(0)}=(w^{a_1}\phi_a, w^{a_2}\phi_a,\phi_d)$. The fast transition matrix can be written as in eq.~\ref{eq:Mmod} now highlighting
 \begin{equation}
 \tilde{\mathsf{M}}=
 \left(
 \begin{array}{c c}
\mathsf{M}_{T\to T} & 0\\
\mathsf{M}_{T\to R}& \mathsf{M}_{R\to R}
  \end{array}
 \right)
 \end{equation}
where $\mathsf{M}_{T\to T}$ contains the transition within and between transient blocks, $\mathsf{M}_{T\to R}$  the ones from the transient ones to the recurrent ones and $\mathsf{M}_{R\to R}$ within the recurrent blocks, being block diagonal. 
Then the equation at order $\epsilon^0$ (\ref{eq:me_ex_0}) can be written separately for the 
transient and recurrent components:
\begin{eqnarray}
 \mathsf{M}_{T\to T} p^{(1)}_T + \mathsf{L^{(0)}}_{R\to T} p^{(0)}_R=0\\
 \mathsf{L^{(0)}}_{R\to R} p^{(0)}_R+\mathsf{M}_{T\to R} p^{(1)}_T +\mathsf{M}_{R\to R} p^{(1)}_R=\frac{d}{dt}p^{(0)}_R 
 \end{eqnarray}
and, upon solution of the first one,
\begin{eqnarray}
\left(\mathsf{L^{(0)}}_{R\to R} -\mathsf{M}_{T\to R}({\mathsf{M}_{T\to T}})^{-1}\mathsf{L^{(0)}}_{R\to T} \right)p^{(0)}_R +\mathsf{M}_{R\to R} p^{(1)}_R=\frac{d}{dt}p^{(0)}_R\,.
 \end{eqnarray}
 Multiplying by the left eigenvectors spanning the null space of $\mathsf{M}_{R\to R}$ we obtain the effective 
 equation.
 Since each block on the diagonal of $\mathsf{M}_{R\to R}$ is a stochastic matrix the eigenvalues are composed of blocks of $1$ and   the multiplication corresponds to summing over the states within the recurrent blocks.
 The effective equation then reads
  \begin{eqnarray}
\overline{\left(\mathsf{L^{(0)}}_{R\to R} -\mathsf{M}_{T\to R}({\mathsf{M}_{T\to T}})^{-1}\mathsf{L^{(0)}}_{R\to T} \right)}\phi =\frac{d}{dt}\phi\
 \end{eqnarray}
{   where the overline is} a weighted average over the states belonging to  the given recurrent strongly connected component. 
 The connection with the previous procedure involving the multiplication by vector of the left eigenspace can be seen by considering that
$(v_T, v_R)\tilde{M}=0$ corresponds to requiring that $v_R$ is in the left eigenspace of $\mathsf{M}_{R\to R}$
which is composed of block of unit vectors and that 
$v_T=-v_R\mathsf{M}_{T\to R}({\mathsf{M}_{T\to T}})^{-1}$.
{  
\paragraph{Summary} We presented a systematic method  to derive an effective Markovian evolution on the slow scales involving fewer states.
The extent of the simplification depends
on the structure of the graph of fast transitions. If this graph is reducible it is possible to decrease the complexity of the initial problem.
The key ingredients are the strongly connected components of the fast graph that identify the building blocks of the network. Only the components
conserving probability on the fast scales (sinks of the graph condensation) will appear in the effective dynamics. 
Grouping the states into the strongly connected components is called averaging, whereas the elimination of the blocks that are not sinks is called decimation.
}
   \subsection{Block-diagonal fast dynamics}\label{sec:dyn_block}
   \begin{figure}[!h]
  \includegraphics[width=0.9\textwidth]{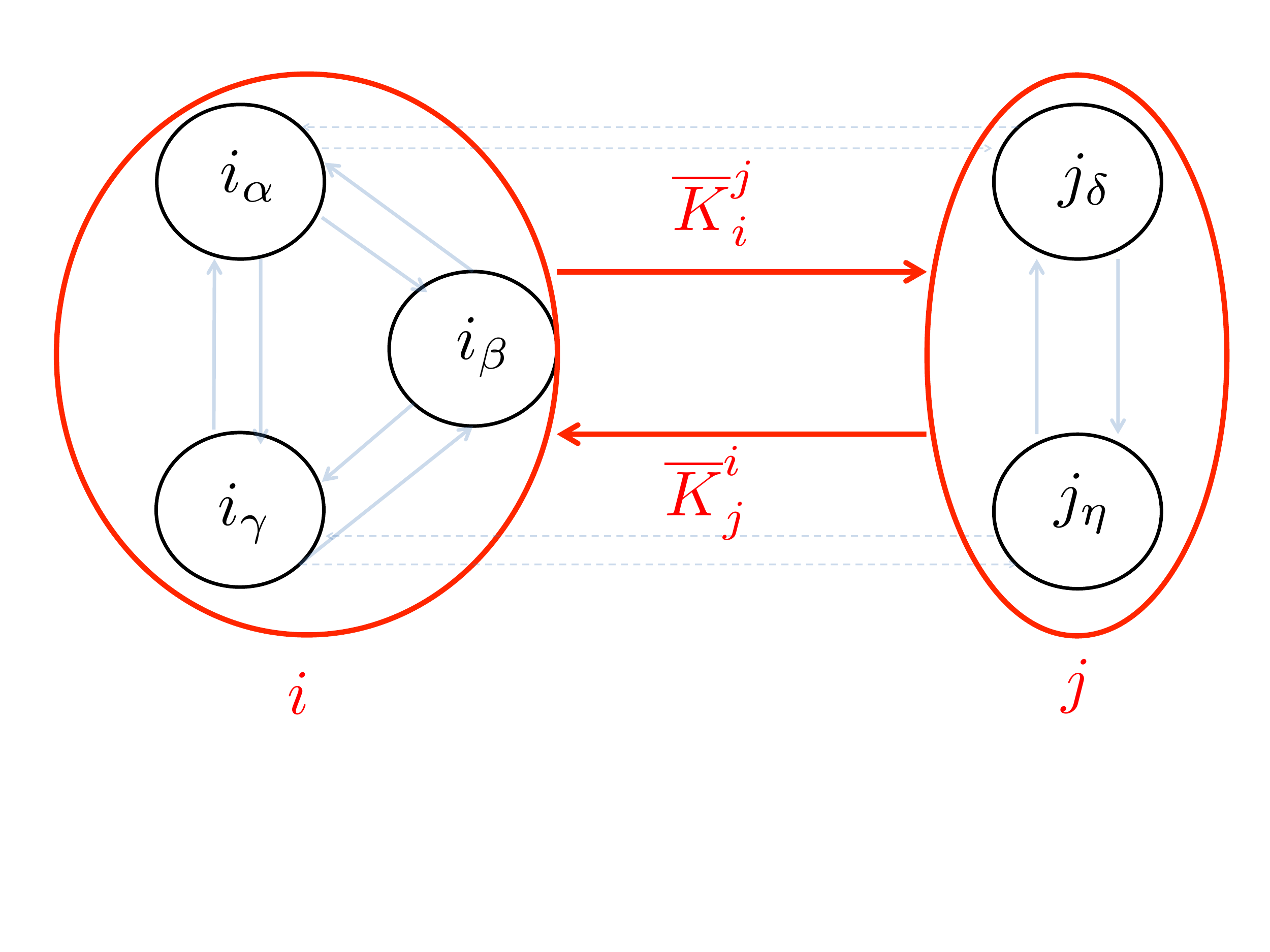}
\caption{\label{fig:blocks} Example of a system with block-diagonal fast transitions. It is possible to
identify blocks of states that are connected via fast transitions among themselves (solid lines) and
by slow ones (dashed lines) to other blocks. The blocks are highlighted in red and the red arrows
represent the effective transition across blocks as derived in eq.~(\ref{eq:eff_Rate}).}
\end{figure}
{  
\paragraph{Question} How does the effective process look like when fast transitions have a block-diagonal structure?\\
}
   When the fast transitions generator $\mathsf{M}$ can be cast in a block-diagonal
   form (see figure \ref{fig:blocks}):
   $$\newcommand*{\temp}{\multicolumn{1}{r|}{}}
K=\left[\begin{array}{cccccccccccccccc}
& &F^{i_\alpha}_{i_\beta} & & &\temp & & S^{i_\alpha}_{j_\beta}& &\temp &   &\ldots &\temp &   &\ldots &\\ \cline{1-16}
&  &S^{j_\beta}_{i_\alpha} & & &\temp & &F^{j_\alpha}_{j_\beta} & &\temp &  &\ldots&\temp &   &\ldots &\\
 \cline{1-16}
&  &\vdots& & &\temp & &\vdots& &\temp &  &\ddots &\temp &   &\vdots &\\
 \cline{1-16}
&  &\ldots& & &\temp & &\dots& &\temp &  &\ldots &\temp &   &F^{k_\alpha}_{k_\beta} &\\
\end{array}\right]
\;.
$$
    all the identified blocks will conserve probability within them and will be part
   of the effective dynamics. In other words, no state will be eliminated.
   Such case has been studied in detail in a number of publications (see e.g. \cite{esposito2012stochastic,Bo2014,Wang2015,Wang2016,rahav2007fluctuation,Macnamara2008,pavliotis2008multiscale,Peles2006}). 
 It is convenient to label the states of the system highlighting the block-structure.
 We chose to identify a state as $i_\alpha$ where the Latin letter $i=1\ldots n$ refers
to the block and the Greek subscript $\alpha=1\ldots n_i$ denotes the specific
state within the block. 
The asymptotic technique presented in the previous section will now provide the effective dynamics that connects the  $n$ slow blocks labeled by the Latin indices. 
   A major simplification introduced by the block-diagonal structure is that, since each block
   conserves probability, the left nullspace has a simple structure. The number of blocks $n$
   sets its dimension and the space is spanned by $n$ row vectors composed of ones in the states belonging to a given block and zeros elsewhere. Denoting the dimension of block $i$ as $n_i$, the left nullspace is generated by
   \begin{equation}\label{eq:block_leftev}
   \hat{\phi}_{\mathit block}=\hat{\phi}_{1}(\underbrace{1,\ldots,1}_{n_1},\underbrace{ 0,\ldots,0}_{\sum_{i=2}^{n}n_i})+
   \hat{\phi}_{2}(\underbrace{0,\ldots,0}_{n_1}\underbrace{1,\ldots,1}_{n_2}, \underbrace{ 0,\ldots,0}_{\sum_{i=3}^{n}n_i})+\ldots+
   \hat{\phi}_{n}(\underbrace{ 0,\ldots,0}_{\sum_{i=1}^{n-1}n_i}\underbrace{1,\ldots,1}_{n_{n}})\,.
   \end{equation}
    The most relevant consequence is that the product $(\hat{p}^{(0)}\mathsf{L^{(0)}}p^{(0)})_{\mathit block}$ in eq.~(\ref{eq:gen_eff}) now results in    the generator of the effective  dynamics between blocks $\eff[\mathsf{L^{(0)}}]_{\mathit block}$ being a weighted average
    of the transitions between blocks (as derived by e.g.~\cite{Cao2005,esposito2012stochastic,Bo2014}). More precisely, the effective transition rates are given by
\begin{equation}\label{eq:eff_Rate}
\eff[K]^i_j =\overline{K}^i_j = \sum_{\alpha,\beta} S^{i_\alpha}_{j_\beta} w^{j_\beta}
\end{equation}
where we recall that $w^{j_\beta}$ is the equilibrium probability of state 
$j_\beta$  under the fast dynamics of $\mathsf{M}$ conditional on being in block $j$ 
 and $S^{i_\alpha}_{j_\beta}$ is a slow transition from state ${j_\beta}$ in block $j$ to state $i_\alpha$ in block $i$.
 {  
 \paragraph{Summary}
 A block-diagonal structure of the fast transition allows for an immediate identification of the relevant states for the slow dynamics after averaging --- these are the blocks themselves. None of the blocks will be decimated.
 }
 \subsection{Examples of effective dynamics for discrete Markov chains}
    
    \subsubsection{Decimation: bursty protein production}
\begin{figure}[!h]
  \includegraphics[width=0.9\textwidth]{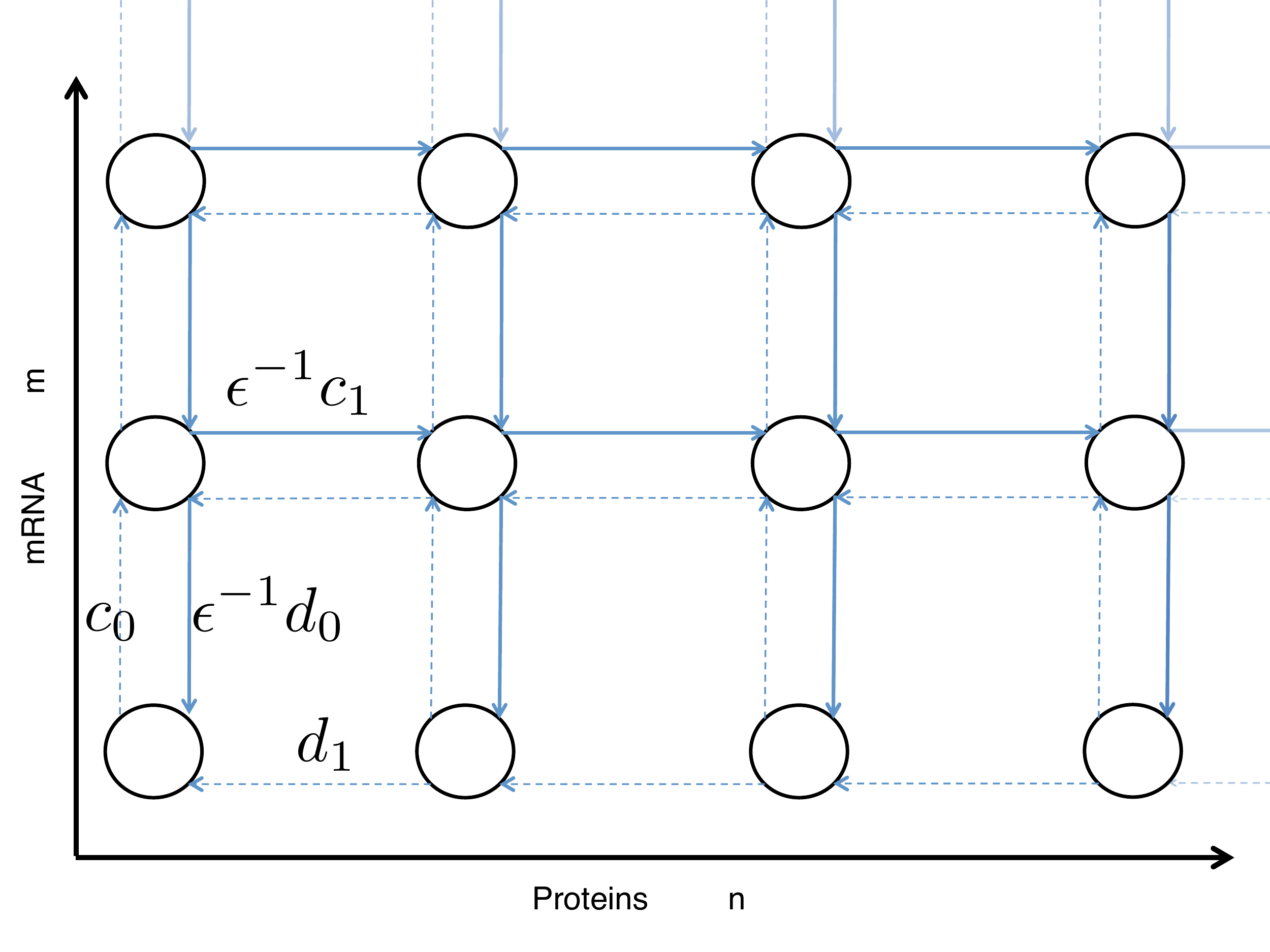}
\caption{\label{fig:prot}  Schematic view of the Markov chain representing protein production and mRNA synthesis according to eq.~\ref{eq:prot}.}    
\end{figure}

Consider a toy model of protein production from a constitutively active gene.
Protein synthesis involves two steps: transcription of the gene into a messenger RNA (mRNA) and
translation of the mRNA into the protein. 
When the gene and the mRNA molecules are present only in a few copies, a stochastic 
treatment of the process is mandatory. 
Determining the dynamics and the fluctuations of the proteins is key to investigate 
cell to cell variability and the reliability of cellular processes~\cite{Elowitz2002,Ozbudak2002,Paulsson2005}. 
When mRNA translation (protein production) and mRNA decay are faster 
than protein degradation and gene transcription it is possible to eliminate
the mRNA states from the dynamics and obtain an effective equation
for the protein dynamics which accounts for the fluctuations of the transient mRNA state  \cite{Shahrezaei2008,Popovic2015}.
The general method presented in this section can be adapted to 
such system and gives the same elimination.
The system states (identified by the number of proteins and mRNA molecules) and transitions are illustrated in fig.~\ref{fig:prot}
and obey the following master equation:
\begin{eqnarray}\label{eq:prot}
\frac{d\,p(n,m)}{dt}&=&c_0p(n,m-1)+\epsilon^{-1}c_1mp(m,n-1)+\epsilon^{-1}d_0(m+1)p(n,m+1)+d_1(n+1)p(n+1,m)\\\nonumber
&-&
\left(c_0+c_1m+d_0m+d_1n\right)p(n,m)
\end{eqnarray}
where $m$ and $n$ refer respectively to the number of mRNA molecules and proteins,
$c_0$ and $\epsilon^{-1}d_0$ are the rate of mRNA synthesis and degradation and,
$\epsilon^{-1}c_1$ and $d_1$ the ones of protein production and decay.
As the number of states is unlimited it can become rather cumbersome to  handle the probability vectors
and transition matrix directly. However, by a suitable rearrangement of terms it is possible to identify blocks and reduce the system
complexity. 
One needs to order  states by grouping them according to the number of mRNA molecules 
$\ldots$, $\underbrace{(n=0,m=M),(n=1,m=M),\ldots }_{m=M}$ $\underbrace{(n=0,m=1),(n=1,m=1),\ldots}_{m=1} $, $\underbrace{(n=0,m=0),(n=1,m=0),\ldots}_{m=0}$.
The corresponding matrix of fast transitions is then already in the block triangular form with blocks of dimension $1$ (made of single states). The only blocks conserving probability are the ones not having fast transitions outwards:   {\it i.e.}, the ones having 
 no mRNA molecules. All the states with $m>0$ will not conserve probability on the fast scale and will be decimated from the effective dynamics. 
For the sake of clarity we shall consider the case with at most one mRNA molecule (the general case of infinitely many
mRNA yields the same effective dynamics\footnote{This follows from the fact that 
the only states which will not be decimated (the non vanishing entries of $\phi^{(0)}$) are all in the block $m=0$ and that the only slow transitions from states with $m=0$ go to states with $m=1$ implying that $\mathsf{L^{(0)}} p^{(0)}$ from eq.~(\ref{eq:gen_eff})  will not have any contributions from the states with $m>1$.}). 
The matrix of fast transition reads:
\begin{equation}
\mathsf{M}=
\left(
\begin{array}{c|c}
\mathsf{M}_a & 0 \\
\hline
\mathsf{M}_b & 0
\end{array}
\right)
\end{equation}
where $0$ represents a matrix with all entries equal to zero.
$\mathsf{M}_a$ refers to fast transitions between states with one mRNA molecule ($m=1$) occurring with rate $c_1$ and the exit rate from states with $m=1$: $(d_0+c_1)$. It  is a band diagonal
matrix  with
elements\footnote{Since the lowest number of proteins is $0$, for ease of notation, in this example we
let the indices referring to proteins start at $0$. 
} $M^{ij}_a=-(d_0+c_1)\delta^{i,j}+c_1\delta^{i,j+1}$.
The fast transitions from states with $m=1$ to ones with $m=0$ are included in $\mathsf{M}_b$ which is a diagonal matrix
with entries equal to $d_0$. 
The steady state probability for these states will be given by
$ p^{(0)}=\sum_{i=0}^\infty
\phi_iw^i $ where 
  \begin{equation}\label{eq:p0_prot}
 w^i =\left(
 \begin{array}{c }
w^{i_1}	\\
 w^{i_2}
 \end{array}
 \right)
 \end{equation}
with $w^{i_1}$ being a column vector with entries equal to zero for every $i$ and 
$w^{i_2}_k=\delta_{k,i}$ a column vector with one non-zero entry in position $i$:
 \begin{equation}
 p^{(0)}=\phi_0\left(
 \begin{array}{c }
 0	 \\
 \vdots\\
  0	 \\
  \hline
  1\\
0\\
0\\
0\\
\vdots\\
0
 \end{array}
 \right)+
 \phi_1\left(
 \begin{array}{c }
 0	 \\
 \vdots\\
  0	 \\
  \hline
  0\\
1\\
0\\
0\\
\vdots\\
0
 \end{array}
 \right)+
 \phi_2\left(
 \begin{array}{c }
 0	 \\
 \vdots\\
  0	 \\
  \hline
  0\\
0\\
1\\
0\\
\vdots\\
0
 \end{array}
 \right)+\ldots
  \end{equation}
where the upper block refers to states with $m=1$
and the lower one to those with $m=0$ and $\phi_i$ denotes the probability of having $i$ proteins.
\begin{figure}[th]
  \includegraphics[width=0.9\textwidth]{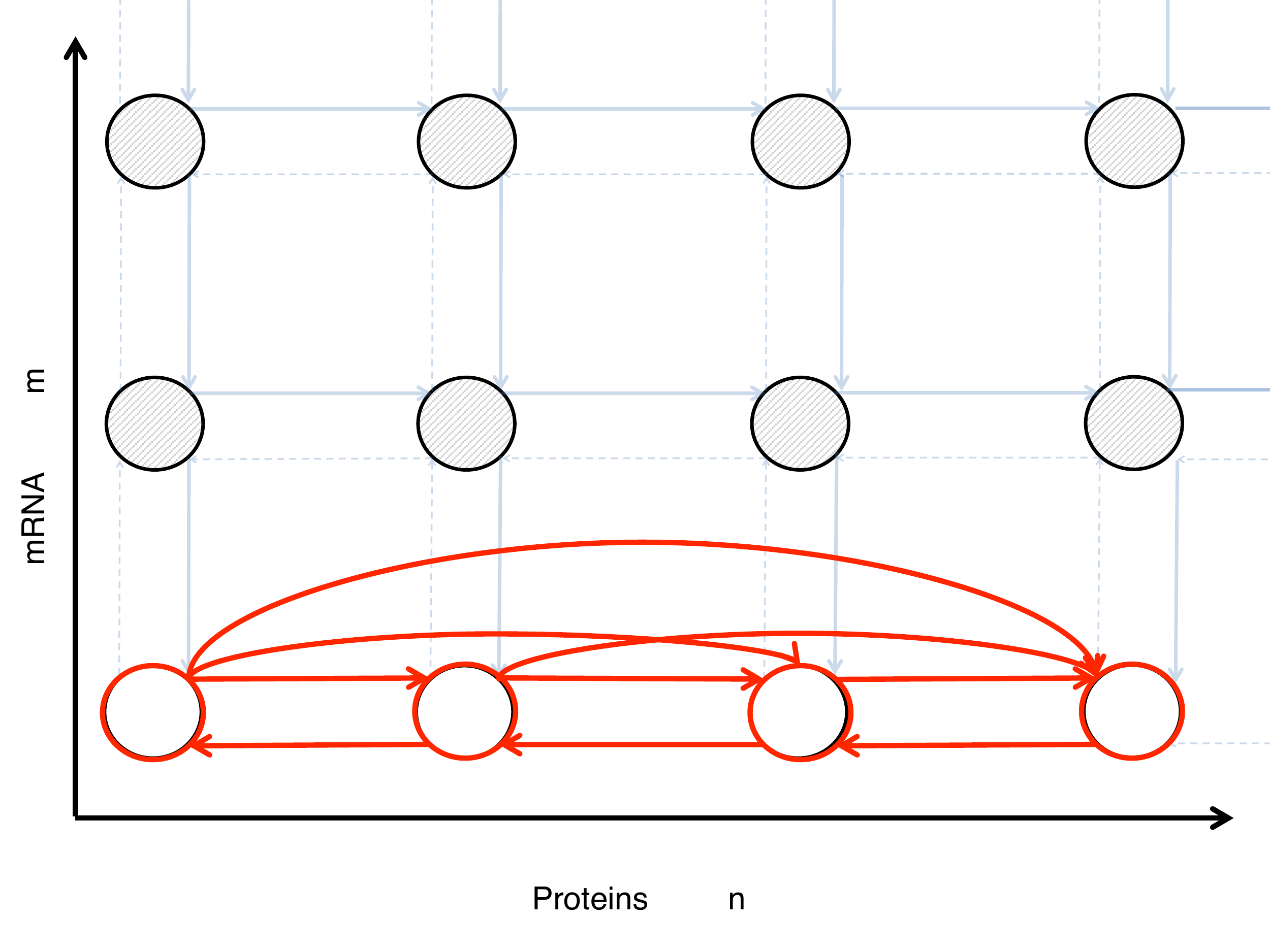}
\caption{\label{fig:prot_eff} Effective dynamics representing protein production and mRNA synthesis 
 according to eq.~(\ref{eq:prot_eff}).
The states having a non zero number of mRNA (shaded circles) have been decimated. The effective transitions (red arrows)
now connect states that were not originally linked by single transitions.}    
\end{figure}
The left nullspace of $\mathsf{M}$ is spanned by $\hat{p}^{(0)}=\sum_{i=0}^\infty\hat{\phi}_iv^i$
where $v^i=\left(v^{i_1},\,v^{i_2} \right)$ with $v^{i_1}$ being a row vector with entries $k$  \begin{eqnarray}
v^{i_1}_k=
 \left\{
 \begin{array}{c c}
 q_+^{i-k}q_s &k\le i\\
0
&  k>i
 \end{array}
 \right.
\end{eqnarray}
where $q_+=\frac{c_1}{d_0+c_1}$ is the probability that the next transition from a mRNA is translation into a protein
and $q_s=\frac{d_0}{d_0+c_1}$ that of mRNA decay.
$v^{i_2}={w^{i_2}}^T$ {  ensures} biorthonormality. We then have
\begin{eqnarray}
\hat{p}^{(0)}&=&\hat{\phi}_0
 \left(\underbrace{q_s,\,0,\,0, \ldots 0}_{m=1}\, \vline\,\underbrace{1,\,0,\,0,\,\ldots,0}_{m=0}\right)+\hat{\phi}_1
 \left(\underbrace{q_sq_+,\,q_s, \,0,\ldots 0}_{m=1}\, \vline\,\underbrace{0,\,1,\,0,\,\ldots,0}_{m=0}\right)\\\nonumber
 &+&\hat{\phi}_2
 \left(\underbrace{q_sq_+^2,\, q_sq_+,\,q_s, ,\ldots 0}_{m=1}\, \vline\,\underbrace{0,\,0,\,1,\,\ldots,0}_{m=0}\right)+\ldots
\end{eqnarray}
 The matrix of the slow transitions can be written as:
 \begin{equation}
 \mathsf{L}^{(0)}=
\left(
\begin{array}{c|c}
  \mathsf{L}^{(0)}_a &   \mathsf{L}^{(0)}_c \\
\hline
0&   \mathsf{L}^{(0)}_a
\end{array}
\right)
\end{equation}
where $\mathsf{L}^{(0)}_a$ represents the slow transitions that do not change the number of mRNA (protein degradation with rate $nd_1$)
 and the exit rates $c_0+nd_1$. It is a band diagonal matrix {  with} entries 
 $ \mathsf{L^{(0)}}^{ij}_a=-\left(c_0+id_1\right)\delta^{i,j}+(i+1)d_1\delta^{i+1,j}$ .
  $\mathsf{L}^{(0)}_c$ instead, accounts for the transitions changing the number of mRNA (RNA synthesis) and is a diagonal matrix with
  all entries equal to $c_0$.  
  To derive the effective equation, following eq.~(\ref{eq:gen_eff}), we need to evaluate
 {  
  \begin{equation}
  \mathsf{L}^{(0)}p^{(0)}=
  \left(\begin{array}{c }
\mathsf{L}^{(0)}_c p^{(0)}_b\\
\mathsf{L}^{(0)}_a p^{(0)}_b
 \end{array}
 \right)\,.
  \end{equation}
  }
  The effective evolution of the surviving state $i$ is then obtained by multiplying by $v^i$ and gives (as depicted in fig.~\ref{fig:prot_eff}):
{  
\begin{equation}\label{eq:prot_eff}
\frac{d\phi_i}{dt}=c_0q_s\sum_{l=0}^{i} q_+^l\phi_{i-l}+(i+1)d_1\phi_{i+1}-\left(c_0+id_1\right)\phi_i
\end{equation}
}
which, modulo a time rescaling, is the result found in \cite{Shahrezaei2008} expressed in their
equation (14). The decimation procedure allowed to decrease the number of states in the effective
dynamics, restricting it to the number of proteins modeled. At the same time, new connections between the surviving
states appeared. Jumps in protein numbers no longer need to be to the nearest state but can be larger, reflecting bursty protein synthesis. This biologically observed phenomenon simply results from the fact that several proteins can be produced
during the short lifetime of a single mRNA molecule.
{   
\paragraph{Summary} 
For slow proteins degradation and slow mRNA transcription, the bursty kinetics of protein synthesis from a constitutive active gene can be derived by decimation of the intermediate states involving mRNA translation and degradation. }
 \subsubsection{Block diagonal: Two-component systems }\label{sssec:signal}
\begin{figure}[h]
 \includegraphics[width=0.9\textwidth]{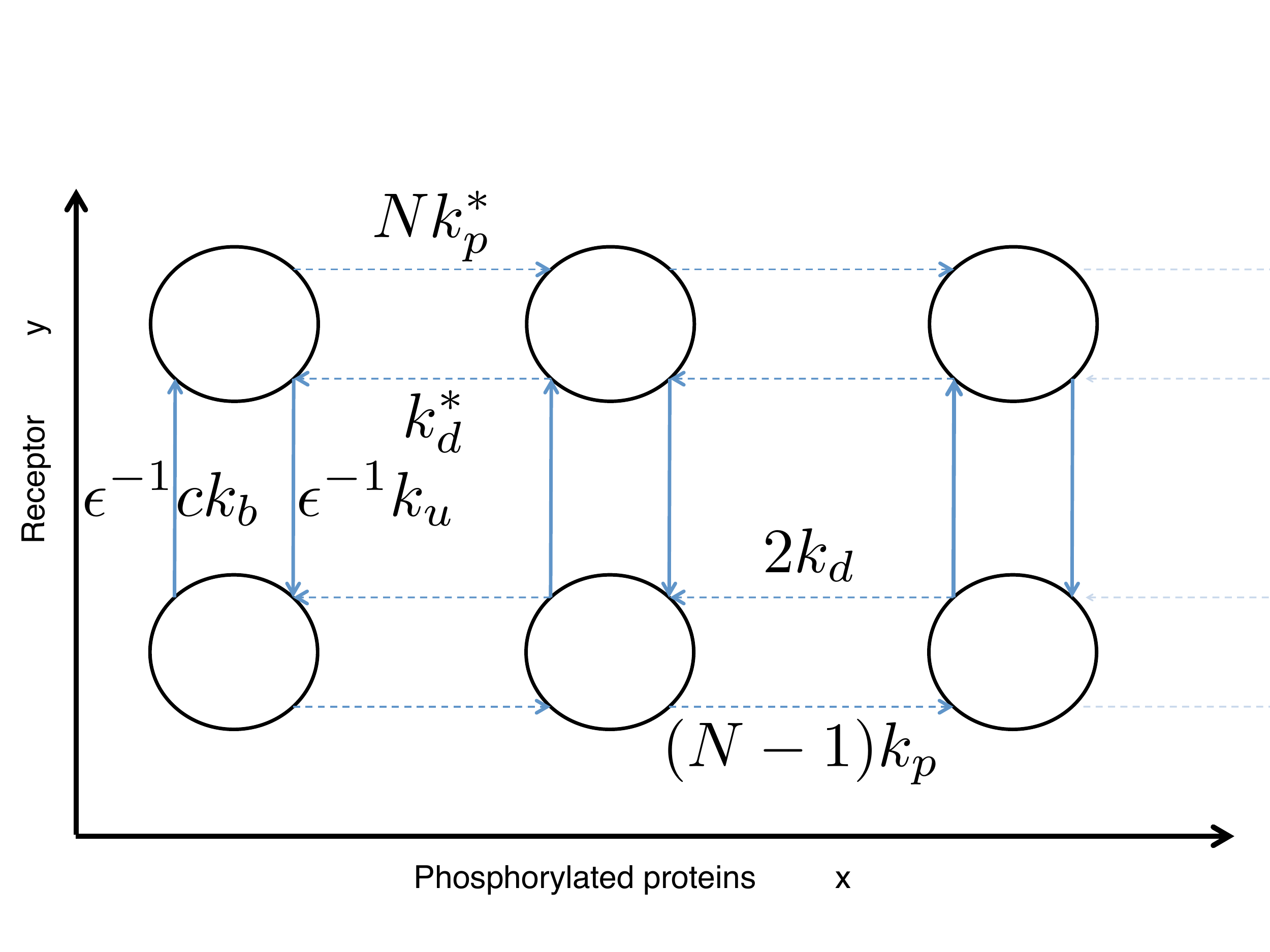}
\caption{Schematic view of two chemical reactions implementing a sensing device as described in eq. (\ref{eq:master_multiscale}). 
The variable $y$ refers to the activity state of the receptor. The phosphorylated proteins are labeled by $x$ and can take values from $0$ to the total number of available proteins $N$.
Each state is defined by the activity state of the receptor and the number phosphorylated proteins, i.e. the pair $(x,y)$.
Vertical transitions correspond to receptors activation and deactivation and occur, respectively, with rates $\epsilon^{-1}k_b c$ and $\epsilon^{-1}k_u$.
The (de-)phosphorylation reactions result in horizontal transitions.}
\label{fig:rec_prot}
\end{figure}

To illustrate an application to the case of block-diagonal fast transitions let us consider the example of a simple biochemical network mimicking the sensing process of a cell \cite{West2001}.
The system is an extension of the one introduced in \cite{mehta2012energetic} which was further studied in
\cite{bo2015thermodynamic}.   We consider the sensing mechanisms to consist of a single receptor\footnote{The extension to an arbitrary number of receptors is straightforward} that undergoes binding and unbinding with an external ligand at concentration $c$.
We assume that a receptor bound to a ligand is in the active form and that an unbound one is not.
The receptor, according to its activation state, acts on some downstream cytoplasmic proteins.
We model this {   pathway} by considering $N$ proteins which can be
phosphorylated or dephosphorylated with rates that depend on the activation state of the receptor as shown in figure \ref{fig:rec_prot}. 
The activation dynamics of the receptor is influenced by the concentration
of the external ligand and it is independent of the phosphorylation state of the proteins.
Denoting the receptor as $Y$  and the ligand concentration as $c$ its dynamics follows
   \begin{equation}
      \ce{Y <=>[{k_b}c][k_u] \overline{Y}}
   \end{equation}
where in this reaction and the following ones appear the specific probability rate constants. The number of active (bound) receptors $\overline{Y}$ is $y=0,1$. 
The receptor can reversibly activate (e.g. phosphorylate) the proteins
\begin{equation}
\ce{\overline{Y} + X <=>[k_p^*][k_d^*] \overline{Y} + X^*}
\end{equation}
\begin{equation}
\ce{Y + X <=>[k_p][k_d] Y + X^*}
\end{equation}
The number of phosphorylated proteins is $x=0,...,N$. 
We do not consider spontaneous (non-mediated by the receptor) activation and deactivation.
The phosphorylation reactions mediated by the active receptor are accompanied by ATP hydrolysis.
We describe this system as a Markov chain in which a state is defined by the number of phosphorylated proteins 
and the corresponding activation state of the receptor as shown in figure \ref{fig:rec_prot}.
We consider the case 
in which the binding and unbinding of the receptor is much faster  
than the phosphorylation dynamics of the cytoplasmic proteins. This time-scale separation is observed, for instance, in bacterial chemoreceptors. To make explicit such scaling we write 
\begin{equation*}
 ck_b\to \epsilon^{-1}ck_b\qquad \qquad  k_u\to \epsilon^{-1}k_u\;.
\end{equation*}
We can then write the master equation for the  probability of being in the state $y,x$ as 

\begin{eqnarray}\label{eq:master_multiscale} 
\frac{dp(y=0,x)}{dt} &=&\epsilon^{-1}k_u p(y=1,x)+(x+1)k_dp(y=0,x+1)+(N-x+1)k_pp(y=0,x-1)\\\nonumber
&-&\left[xk_d+(N-x)k_p+\epsilon^{-1}ck_b\right]p(y=0,x)\\
\frac{dp(y=1,x)}{dt} &=&\epsilon^{-1}ck_b p(y=0,x)+(x+1)k_d^*p(y=1,x+1)+(N-x+1)k_p^*p(y=1,x-1)\\\nonumber
&-&\left[xk_d^*+(N-x)k_p^*+\epsilon^{-1}k_u\right]p(y=0,x)\;.
\end{eqnarray} 
Notice that actual transition rates increase with the total number of proteins so 
that care must be taken when considering large numbers of proteins. We here restrict to the case in which
$N\ll \epsilon^{-1}$. Even for this simple system featuring only two chemical reactions it is non-trivial to 
find an analytic solution of the steady state probability distribution (the authors of \cite{mehta2012energetic} derived it for an infinite pool of proteins).
By ordering  states as $\ldots,(n,y=0),\,(n,y=1);\, (n',y=0),\,(n',y=1);\,\ldots$ the fast transitions, involving the receptor, are arranged in a block-diagonal manner.
As shown in the general section, this simplifies the elimination procedure. 
 At order $\epsilon^{-1}$   the equation
features only the receptor transitions and, after an initial transient relaxes to
\begin{equation}
 p(y=1)=p(ON)=\frac{ck_b}{k_u+ck_b} \qquad p(y=0)= p(OFF)=1-P(ON)=\frac{k_u}{k_u+ck_b}
\end{equation}
where $p(ON)$ is the probability of the receptor being active.
The first order solution then reads:
  \begin{equation}\label{eq:w_Rec}
p^{(0)}=\phi_0
 \left(
 \begin{array}{c }
 p(OFF)	\\
 p(ON)	\\ 
 0	 \\
  0	 \\
  0\\
  \vdots
 \end{array}
 \right)
+\phi_1
 \left(
 \begin{array}{c }
 0	 \\
  0	 \\
  p(OFF)	\\
 p(ON)	\\ 
  0	\\
\vdots
 \end{array}
 \right)
 +\dots
 \end{equation}
 where $\phi_i$ denotes the first order probability of having $i$ proteins phosphorylated.
 \begin{figure}[h]
 \includegraphics[width=0.9\textwidth]{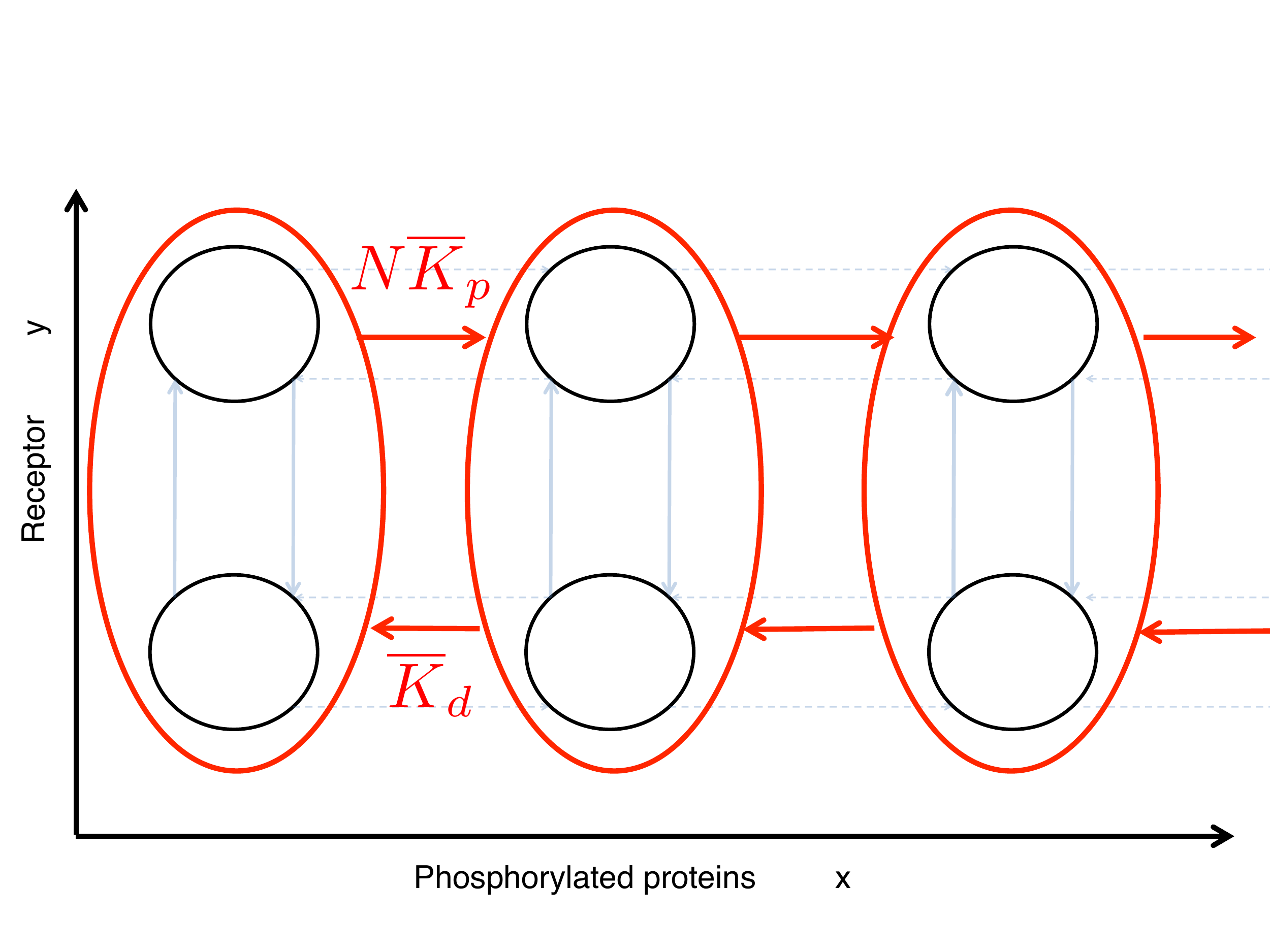}
\caption{Effective dynamics after the averaging over receptor states as described in eq. (\ref{eq:eff_general}). }
\label{fig:rec_prot_eff}
\end{figure}
 Given the block-diagonal structure of the fast operator, its left nullspace  is spanned by:
\begin{equation}
\hat{p}^{(0)}=\hat{\phi}_0\left(1,\,1,\,0,\,0,\,0,\ldots,0\right)+\hat{\phi}_1\left(0,\,0,\,1,\,1,\,0,\ldots,0\right)+\ldots
\end{equation}
so that we have the following effective equation for the protein dynamics
\begin{equation}\label{eq:eff_general}
 \frac{d\phi_x}{dt}=(x+1) \eff[K]_d\phi_{x+1}+(N-x+1) \eff[K]_p\phi_{x-1}-\left(x \eff[K]_d+(N-x) \eff[K]_p\right)\phi_x
\end{equation}
{  where  the effective rates are} averages over the equilibrium distribution of receptor states:
{  
\begin{eqnarray}\label{eq:gen_overline}
  \eff[K]_p=\overline{K}_p=p(ON) k_p^*+p(OFF)k_p   =\frac{ck_bk_p^*+k_uk_p}{k_u+ck_b}
\end{eqnarray}
and 
\begin{equation}
 \eff[K]_d=\overline{K}_d=p(ON) k_d^*+p(OFF)k_d  .
\end{equation}
}
As shown in figure~\ref{fig:rec_prot_eff}, after the averaging of the receptor states, the network has become one dimensional and the steady state solution of equation 
(\ref{eq:eff_general}) can be easily found to be
the binomial distribution
\begin{equation}
 \phi^{ss}_{x}=a^xb^{N-x}{N\choose x}
\end{equation}
where
\begin{equation}
 a=\frac{ \eff[K]_p}{ \eff[K]_p+ \eff[K]_d} \qquad \qquad b=\frac{ \eff[K]_d}{ \eff[K]_p+ \eff[K]_d}\;.
\end{equation}
Exploiting the time-scale separation by means of the asymptotic techniques presented in the previous section we 
have managed to derive the steady state distribution of the system up to corrections of order $\epsilon$.  We note that a very similar Markov network to the one we have considered (with different
interpretation of the reactions giving rise to the transition rates) can be used to model
receptor methylation as done by the authors of \cite{Wang2015}.
Under analogous time-scale separation assumptions they were able to derive solvable effective dynamics.
In closing we remark that the same procedure could have been applied to a system with $R$ independent receptors obeying
\begin{eqnarray}
 &&\frac{dp(y,x)}{dt} =\epsilon^{-1}(y+1)k_u p(y+1,x)+\epsilon^{-1}(R-y+1)ck_bp(y-1,x)+\\\nonumber
&&(x+1)\left(y k_d^*+(R-y)k_d\right)p(y,x+1)+(N-x+1)\left(y k_p^*+(R-y)k_p\right)p(y,x-1)+
 \\\nonumber
 &&-\left[x\left(y k_d^*+(R-y)k_d\right) +(N-x)\left(yk_p^*+(R-y)k_p\right)+\epsilon^{-1}yk_u+\epsilon^{-1}(R-y)ck_b\right]p(y,x)
\end{eqnarray}
and would have given similar results with the equilibrium distribution of the receptors being a binomial and with
effective rates:
\begin{eqnarray}\label{eq:gen_overline_R}
  \eff[K]_p^R=\langle y \rangle k_p^*+(R-\langle y\rangle)k_p =R\left(p(ON) k_p^*+p(OFF)k_p  \right)\qquad  \eff[K]_d^R=R \left(p(ON) k_d^*+p(OFF)k_d  \right)\;.
\end{eqnarray}
{   
\paragraph{Summary}
A two-component system where the receptor states evolve rapidly admits a block-diagonal structure.
Averaging over the receptor states gives an effective Markovian kinetics between blocks identified by the number of phosphorylated proteins.}
 \subsubsection{ Stochastic  Michaelis-Menten.}\label{sec:MM_main}
\begin{figure}[!h]
  \includegraphics[width=0.9\textwidth]{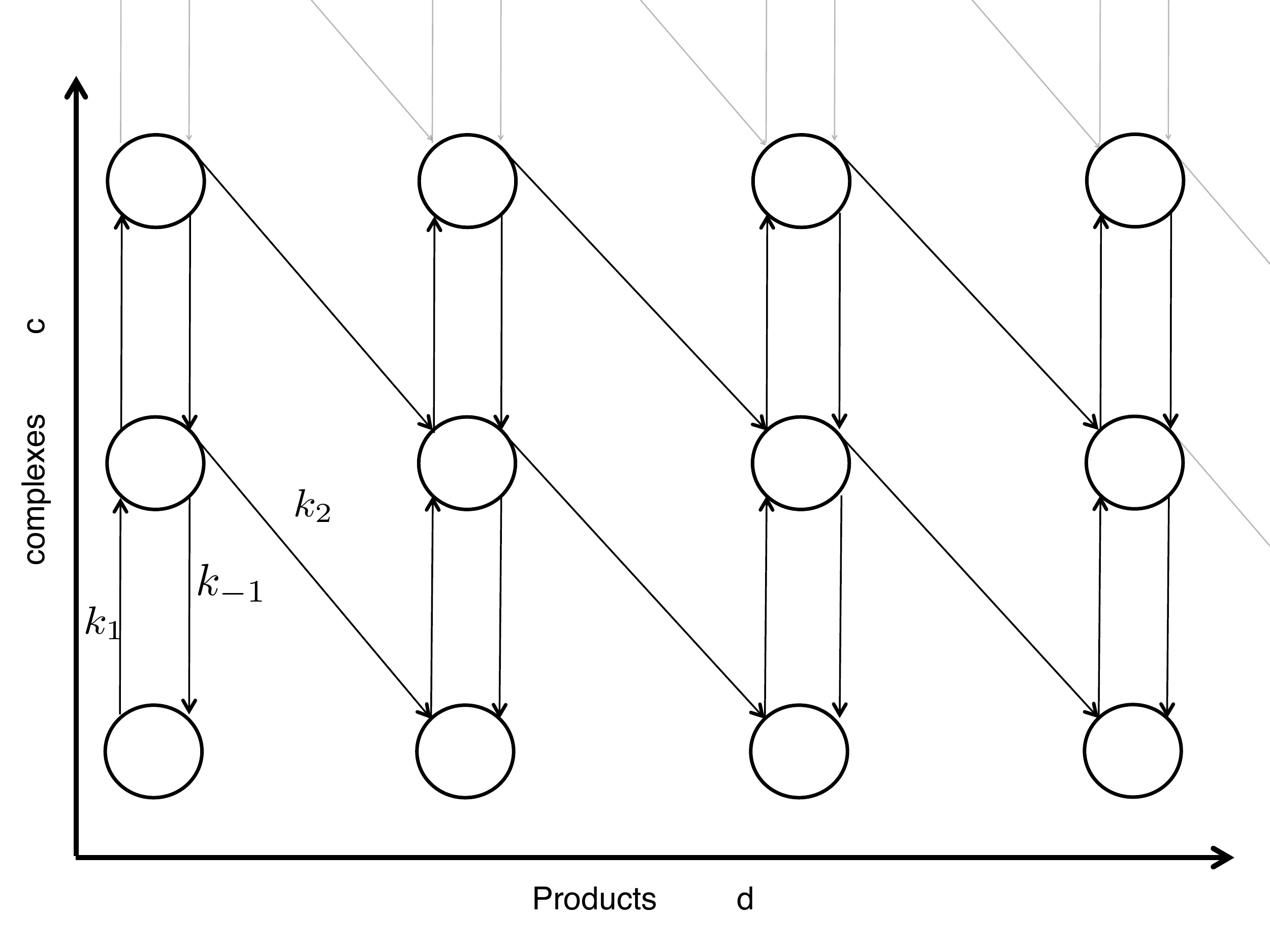}
\caption{\label{fig:mm}      Schematic view of the Markov chain representing stochastic enzyme kinetics {  according} to eq.~\ref{eq:mm}.}
\end{figure}

 Let us consider the enzymatic reaction
\begin{equation}\label{eq:MM1}
\ce{E + S <=>T[$k_1$][$k_{-1}$] C  ->T[$k_2$] E + P}
\end{equation}
converting a substrate $S$ into a product $P$ with the help of a catalytic enzyme $E$, via the intermediate state involving the
complex $C$.
For macroscopic chemical kinetics, various approximations lead to a simplified model known as 
Michaelis-Menten kinetics describing a hyperbolic dependence of the rate of product formation on the substrate
concentration $c_s$ (see e.g.~\cite{Murray2002}):
\[
\frac{dc_p}{dt}=V_{max}\frac{c_s}{K_M+c_s}
\] 
where $c_p$ is the product concentration and $K_M$ is the Michaelis constant. The specific expression of $K_M$ depends on the approximation exploited.
The two most common approximations are the {\it quasi-equilibrium} and the {\it quasi-steady-state}.
When the number of enzymes involved is small, their fluctuations become relevant and it is important to {   adopt a stochastic description that will take the form depicted} in fig.~\ref{fig:mm}. Several authors have studied how the approximations affect the stochastic
version of such enzymatic kinetics~\cite{rao2003stochastic,Gillespie2009,Sanft2011,Gopich2006,Mastny2007,sinitsyn2009adiabatic}.
Here we present how to investigate the issue within the general averaging and decimation procedure discussed
in the general section.
For the sake of simplicity we assume the substrate to be abundant with a fixed concentration $c_s$ and
we consider the stochastic dynamics of the number of complexes $c$ and product molecules $d$.
The general case of a fluctuating substrate is considered in appendix \ref{app:MM}.
The master equation governing the probability of being in a state $p(c,d)$ is:
\begin{equation}\label{eq:mm}
\frac{d}{dt} p(c,d) = k_1 c_s p(c-1,d)
 + k_{-1} (c+1) p(c+1,d)
+ k_2 (c+1)p(c+1,d-1) - \left(k_1c_s+k_{-1} c+k_2 c\right) p(c,d)
\end{equation}
where the enzyme switches between the free state  and the complex  with rate
 $c_sk_1$ and back with $k_{-1}$. Also when a product is formed (rate $k_2$) the complex involved in the reaction is back to the free enzyme form.
It is important to notice that,
when the three rates are comparable in magnitude, it is not possible to simplify the original equations and obtain an effective Markovian description for the dynamics of product formation.
One could define blocks of states identified by the same number of product molecules but
the waiting time for transitions across blocks would not be distributed exponentially~\cite{Mastny2007,Gillespie2009,Sanft2011,Gopich2006,sinitsyn2009adiabatic}.
This occurs because the transitions between complex and free enzyme  (within a block) take place with comparable rates
to the one of product formation (across blocks).
A systematic way of appreciating {   this point} is to study the statistics of product formation by analyzing the moment generating function 
 of the number of product synthesis events {  (}as done for instance by \cite{Gopich2006,sinitsyn2009adiabatic}{  )}. 
 In the minimal case, one can consider the dynamics of a single enzyme, jumping to the complex state with a rate $c_sk_1$ and returning to the free state either via unbinding from the substrate (rate $k_{-1}$) or upon product formation (rate $k_2$).     The moment generating function 
 of the number of product formation events $n_2$ in a time $t$ is:
  \begin{equation}
G_2(s,t) =\sum_{n_2=0}^{\infty}p(n_2,t)e^{-sn_2}\,.
 \end{equation}
 where $p(n_2,t)$ is the probability of observing a number $n_2$ of product syntheses in a time $t$. 
  Such quantity obeys the equation
 \begin{equation}
 \frac{d G_2(s,t)}{dt}=\mathsf{M}_sG_2(s,t)
 \end{equation}
 with the deformed fast matrix 
 \[
 \mathsf{M}_s=
 \left(
 \begin{array}{c c }
  -c_sk_1 & k_2e^{-s}+k_{-1} \\
 c_sk_1&  -k_2 -k_{-1}
 \end{array}
 \right)\;.
\]
At long times, the evolution of the generating function is determined 
by the largest eigenvalues of the modified 
 transition matrix: $G_2(s,t)\sim e^{ \lambda_+(s)t}$ with
 \begin{eqnarray}
 \lambda_+(s)=\frac{-(c_sk_1+ k_2+k_{-1})+\sqrt{(c_sk_1+ k_2+k_{-1})^2-4(1-e^{-s})c_sk_1k_2}}{2}
 \end{eqnarray}
showing that the average rate of production reads:
\begin{equation}
-\frac{1}{t}\frac{\partial G_2(s,t)}{\partial s}\biggr|_{s=0}=c_sk_1k_2/(c_sk_1+ k_2+k_{-1})\;.
\end{equation}
This average is consistent with the Michaelis-Menten rate under the quasi-steady-state assumption.
However, the other moments do not follow a Poissonian distribution, in general.
 Indeed, recalling that the moment generating function of a Poissonian variable obeys
 $G_{Poisson}(s,t)=\exp\left(\mu(t)(e^{-s}-1)\right)$
 we see that product formation can be approximately Poissonian if
 \begin{equation}\label{eq:mm_cond}
 c_sk_1k_2\ll (c_sk_1+ k_2+k_{-1})\;.
 \end{equation}
 The same condition can be derived by studying the first passage time between neighbouring product states 
 yielding eq.~(34) in \cite{Sanft2011}. 
 Other approximations relying on the large abundance of products are possible
 (see e.g.  \cite{rao2003stochastic}) but they cannot be cast in terms of the systematic procedure
 presented here and they do not justify an exponential waiting time between individual transitions involving product states.
 \paragraph{Quasi-equilibrium stochastic Michaelis-Menten}
\begin{figure}[!h]
  \includegraphics[width=0.9\textwidth]{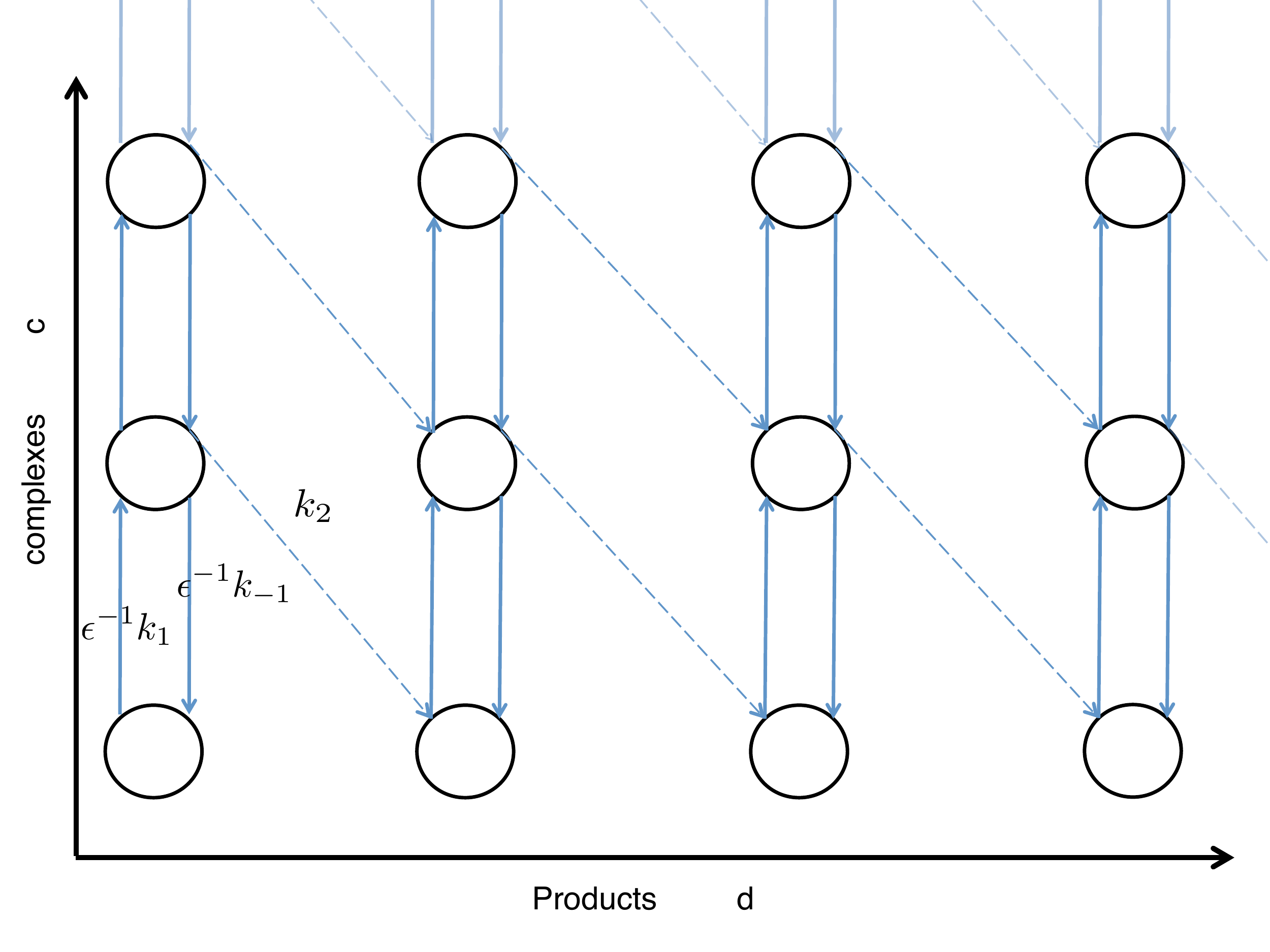}
\caption{\label{fig:mm_qe}     Quasi-equilibrium Michaelis Menten.}
\end{figure}
The quasi-equilibrium approximation is valid when the dynamics of complex formation is much  faster than 
product synthesis: $c_s k_1,k_{-1}\gg k_2$. Then, enzyme and complex reach an equilibrium between any product formation event. We make this scaling explicit by writing $c_s k_1\to \epsilon^{-1}c_sk_1$ and 
$k_{-1}\to \epsilon^{-1}k_{-1}$.  By examining figure~\ref{fig:mm_qe}, it is possible to identify blocks of rapidly connected states. The blocks are characterized by the same number of products and different complexes.
From eq.~(\ref{eq:mm_cond}) we expect the transition between blocks to follow a Poisson process. 
Since the generator of the fast transitions can  be cast in a block-diagonal way, the effective dynamics between different
blocks (number of products) can be obtained {  as  shown in section \ref{sec:dyn_block} (as also done} in the previous example in section \ref{sssec:signal}).
The effective rate of product formation, once the complex state has been averaged out, will be, following eq.~(\ref{eq:eff_Rate}),
\begin{equation}\label{eq:mm_qe}
\overline{K_p}=\sum_{c=0}^{e_t} w_dp(c) c k_2=\langle c\rangle_d k_2=e_nk_2\frac{c_s}{k_{-1}/k_1+ c_s}
\end{equation}
where $e_n$ is the number of total enzymes available, $w_d(c)$ is the first order approximation to the equilibrium probability of having $c$ complexes when having $d$ products and $\langle c\rangle_d$ is its average. 
In this case, it does not depend on the number of present products $d$ and is simply the binomial distribution: $w_d(c)=B(e_n,\frac{c_s}{k_{-1}/k_1+ c_s})$.
We have then an effective dynamics between blocks as depicted in fig.~\ref{fig:mm_qe_eff}.
 The expression in eq.~(\ref{eq:mm_qe}) reproduces the result of the quasi-equilibrium approximation for macroscopic chemical kinetics which gives $K^{eq}_M=k_{-1}/k_1$.
\begin{figure}[!h]
  \includegraphics[width=0.9\textwidth]{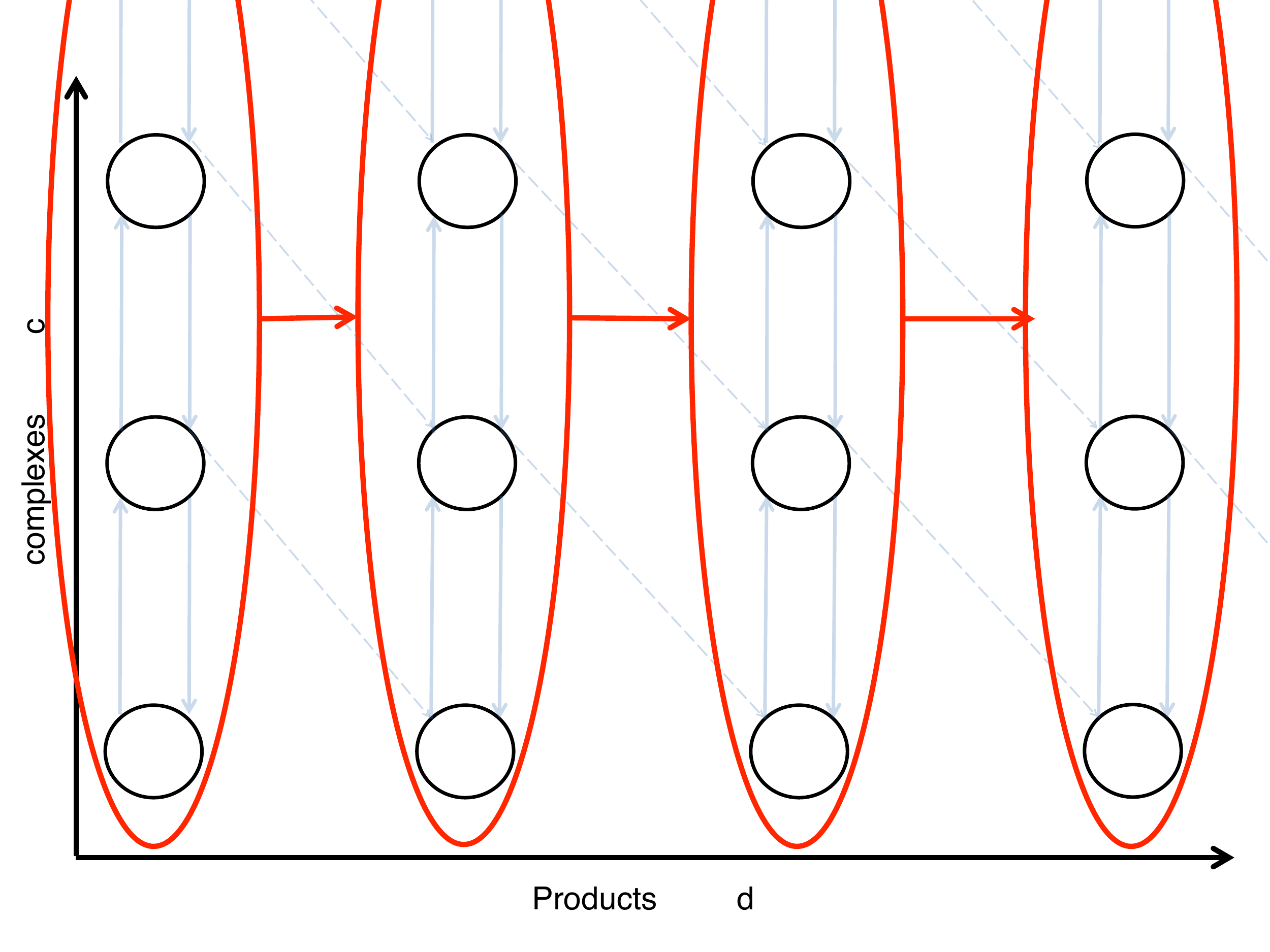}
\caption{\label{fig:mm_qe_eff}     Effective dynamics for the quasi-equilibrium Michaelis Menten.}
\end{figure}
\paragraph{Quasi-steady-state stochastic Michaelis-Menten with slow complex formation}
\begin{figure}[!h]
  \includegraphics[width=0.9\textwidth]{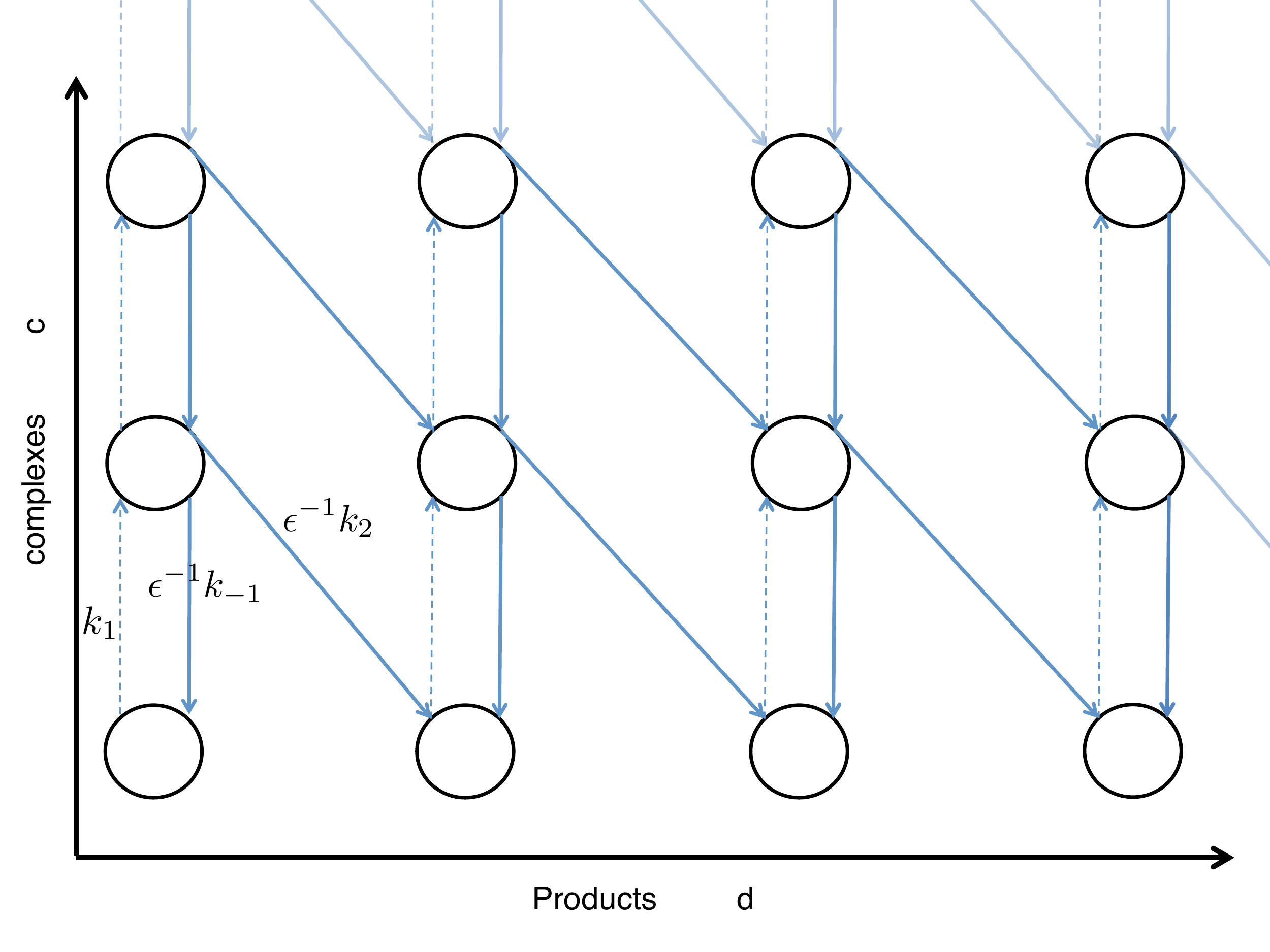}
\caption{\label{fig:mm_qss}  Quasi-steady state Michaelis Menten with slow complex formation.    
}
\end{figure}
As shown in eq.~(\ref{eq:mm_cond}), when complex  formation is slow compared to product synthesis and complex dissociation ($c_sk_1\ll k_{-1},k_2$, see fig.~\ref{fig:mm_qss}) the transitions between product states will approximately follow a Poisson process. 
At variance with the quasi-equilibrium case, we will obtain a Markovian dynamics between different product numbers not by averaging over complex states but by decimating
the states with $c>0$  (as done, e.g., by the authors of \cite{Mastny2007}).
{   Broadly} speaking, the exponentially distributed exit times are recovered since complex formation is a rate limiting step, a rare event.
With the general approach presented in this section we can obtain the effective rate of product formation when complex formation
is slower than both complex degradation and product synthesis.
Consider one total enzyme and order
the states as $\ldots,\,(c=1,d),\,(c=1,d+1),\,\ldots,\,(c=0,d),\,(c=0,d+1)\ldots$ so that
the fast transition matrix is block-triangular:
\begin{equation}
\mathsf{M}=
\left(
\begin{array}{c|c}
\mathsf{M}_a & 0 \\
\hline
\mathsf{M}_b & 0
\end{array}
\right)
\end{equation}
where $0$ represents a matrix with all entries equal to zero.
$\mathsf{M}_a$ contains the fast exit rate from the states with complex  $c=1$ and it is a diagonal matrix with entries : $-(k_{-1}+k_2)$. 
The fast transitions from states with a complex ($c=1$) to ones with a free enzyme ($c=0$) are included in $\mathsf{M}_b$ which is a band diagonal matrix
  $M^{ij}_b=k_{-1}\delta^{i,j}+k_2\delta^{i,j+1}$.
  The states with  $c>0$ do not conserve probability and can therefore be decimated.
  The probability relaxes on the fast scales to
 \begin{equation}
 p^{(0)}=\phi_0\left(
 \begin{array}{c }
 0	 \\
 \vdots\\
  0	 \\
  \hline
  1\\
0\\
0\\
0\\
\vdots\\
0
 \end{array}
 \right)+
 \phi_1\left(
 \begin{array}{c }
 0	 \\
 \vdots\\
  0	 \\
  \hline
  0\\
1\\
0\\
0\\
\vdots\\
0
 \end{array}
 \right)+
 \phi_2\left(
 \begin{array}{c }
 0	 \\
 \vdots\\
  0	 \\
  \hline
  0\\
0\\
1\\
0\\
\vdots\\
0
 \end{array}
 \right)+\ldots
  \end{equation}
where the upper block refers to states with $c=1$
and the lower one to those with $c=0$ and $\phi_i$ denotes the first order approximation to the marginal probability of having $i$ product molecules.

The left nullspace of $\mathsf{M}$ is spanned by 
\begin{eqnarray}
\hat{p}^{(0)}&=&\hat{\phi}_0
 \left(\underbrace{\frac{k_{-1}}{k_{-1}+k_{2}},\,0,\,0, \ldots 0}_{c=1}\, \vline\,\underbrace{1,\,0,\,0,\,\ldots,0}_{c=0}\right)+\hat{\phi}_1
 \left(\underbrace{\frac{k_{2}}{k_{-1}+k_{2}},\,\frac{k_{-1}}{k_{-1}+k_{2}}, \,0,\ldots 0}_{c=1}\, \vline\,\underbrace{0,\,1,\,0,\,\ldots,0}_{c=0}\right)\\\nonumber
 &+&\hat{\phi}_2
 \left(\underbrace{0,\frac{k_{2}}{k_{-1}+k_{2}},\,\frac{k_{-1}}{k_{-1}+k_{2}}, \,0,\ldots 0}_{c=1}\, \vline\,\underbrace{0,\,0,\,1,\,\ldots,0}_{c=0}\right)+\ldots
 \end{eqnarray}
 \begin{figure}[!h]
  \includegraphics[width=0.9\textwidth]{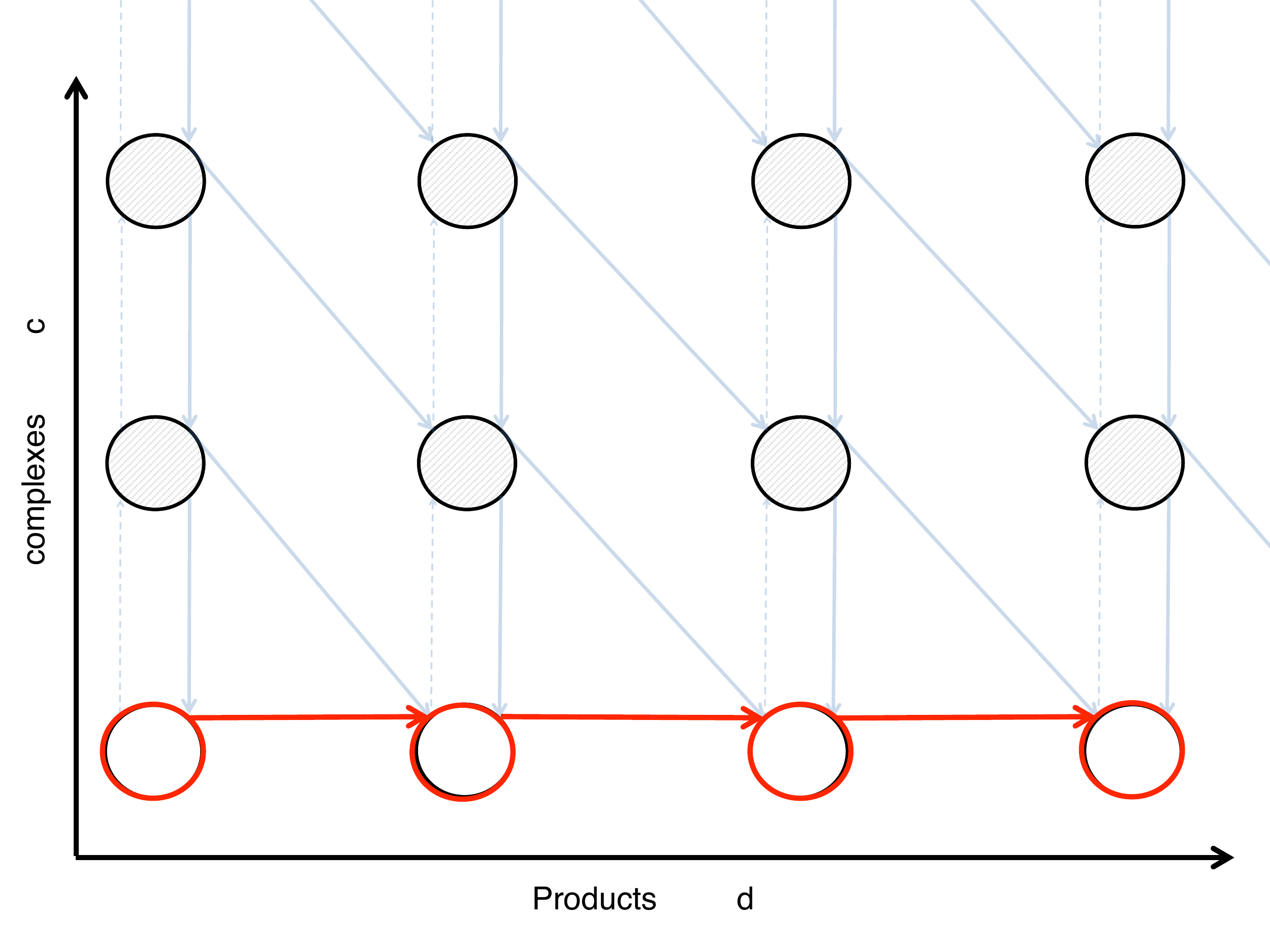}
\caption{\label{fig:mm_qss_eff}  Effective quasi-steady state Michaelis Menten with slow complex formation.   
}
\end{figure}
%
The slow transition matrix reads:
\begin{equation}
 \mathsf{L}^{(0)}=
\left(
\begin{array}{c|c}
0&  \mathsf{L}^{(0)}_c\\
\hline
0&-\mathsf{L}^{(0)}_c
\end{array}
\right)
\end{equation}
with $\mathsf{L}^{(0)}_c$ being a diagonal matrix with entries $c_sk_1$.
 The effective equation is found to be
 \begin{eqnarray}
\frac{d\phi_0}{dt}&=&-k_2\frac{c_sk_1}{k_{-1}+k_{2}}\phi_0\\
\frac{d\phi_i}{dt}&=&\left(k_2\frac{c_sk_1}{k_{-1}+k_{2}}\right)(\phi_{i-1}-\phi_i) \qquad i>0
\end{eqnarray}
and takes place on the states with no complex $(c=0)$ as depicted in fig.~\ref{fig:mm_qss_eff}.
The Michaelis constant for the quasi-steady-state approximation of macroscopic chemical kinetics is $K_M^{qss}=\frac{k_{-1}+k_{2}}{k_1}$.
For small complex formation rates the rate of product synthesis
 is 
\begin{equation}
\frac{dc^{qss}_p}{dt}=k_2\frac{c_sk_1}{k_{-1}+k_{2}+c_sk_1}\simeq 
k_2\frac{c_sk_1}{k_{-1}+k_{2}}
\end{equation}
yielding an approximately linear dependence on the substrate. The stochastic transition rate derived with the decimation procedure coincides with such macroscopic rate.
{  
\paragraph{Summary}
The stochastic Michaelis-Menten kinetics can be simplified to give an effective  kinetics for products only, under the assumptions that some
rates are much larger than others.
Specifically, the quasi-equilibrium approximation (fast substrate-enzyme binding and unbinding) has a block structure that
through averaging gives the slow evolution in terms of the blocks with different numbers of product molecules.
The case of slow binding of the substrate (but fast unbinding and product formation) is a sub-case of the quasi-steady-state
approximation and via decimation gives an effective evolution between the states with no complexes (all enzymes unbound)
and different number of products.
}
\subsubsection{The flagellar motor of {\it E. coli}}
Let us now consider a case involving three different well separated time scales
taken from the kinetics of {\it E. coli} flagellar motor \cite{berg_03}. The approach and results presented below are discussed in detail in Ref.~\cite{pac_16}.
	
\begin{figure}[h]
		\centering
		\small\def\svgwidth{.5\textwidth}
		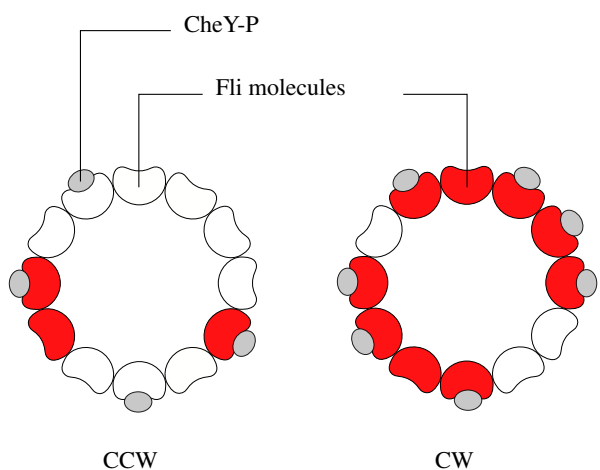
		\caption{The flagellar motor.
			The protomers (Fli molecules) are represented in white (inactive state, $I$) and red (active state, $A$), while the CheY-P regulator is the grey spot.
			The motor rotates counterclockwise when most of the protomers are in the inactive state (left) and clockwise otherwise (right).
		}\label{fig:motor-configs}
	\end{figure}

	The ring of proteins forming the motor of the {E.\,coli} flagella has been shown to be very well described by the conformational spread model \cite{db_01, dnb_01, bai_10, ma_12}.
	This model consists in $N$ identical units, or protomers{  \footnote{Not to be confused with promoters}}, each of which can appear in two different states, active ($A$) or inactive ($I$):
	a protomer in the active state increases the probability of CW rotation of the motor, and of CCW rotation in the inactive state (see Fig.\,\ref{fig:motor-configs}).
	Moreover, each protomer can also bind a {ligand}, corresponding to the CheY-P chemotactic regulator: we refer to the protomer as in the {bound} ($B$) state when a ligand is attached to it, or {unbound} ($U$) otherwise.
	Therefore, the single protomers can be in 4 different states, corresponding to all the possible {activity} and {binding} configurations.
	
	The protomers are coupled via a nearest neighbour interaction, which depends on their activity states {only}: in particular, the energy is lowered by a quantity $J$ when the neighbouring protomers are in the same activity state $A$ or $I$.
		
	Associating to each protomer a {spin} variable $\sigma_i$ taking value $+ 1$ when the protomer is active ($\alpha_i = A$), or $-1$ when it is inactive ($\alpha_i= I$), one can represent the states of the system as $s = \{(\sigma_i, \ell_i)\}_{i=1}^N$. The kinetics of the conformational spread model is then governed by the master (Kolmogorov) equation
	\begin{equation}\label{eq:CS-glauber-master}
		\frac{d}{dt} P(s,t) = \sum_{s'} \left[ P(s',t)\,K(s'\to s) - P(s,t)\, K(s\to s')\right] \comma
	\end{equation}
	where the rates are
	\begin{equation}\label{eq:CS-glauber-single}
			K(s\to s') =  \Bigg\{ \frac{\omega_f}{1-\gamma} \Big( 1 - \gamma\,\sigma_i\,\frac{\sigma_{i+1} + \sigma_{i-1}}{2} \Big)\,e^{\beta\,h(-\sigma_i,\,\ell_i)} \,\delta_{\sigma_i',\,-\sigma_i}\,\delta_{\ell_i',\,\ell_i} \\
				+     \omega_s\,e^{\beta\,h(\sigma_i,\,1-\ell_i)}\,\delta_{\sigma_i',\,\sigma_i}\,\delta_{\ell_i',\,1-\ell_i} \Bigg\}	
				\prod_{j\neq i} \delta_{\sigma'_j,\,\sigma_j}\delta_{\ell_j',\,\ell_j}
			\end{equation}
	Each term between the curly brackets in Eq.\,\eqref{eq:CS-glauber-single} is obtained from detailed balance up to multiplicative factors $\omega_f$ and $\omega_s$: these constants account for typical time scales of the flipping and binding process, respectively.
	The constant $\gamma$ in the spin-flip contribution is set by the strength of the ferromagnetic coupling, $\gamma = \tanh(\beta\,J)$.

	For a protomer in the activity state $\alpha$ (either $A$ or $I$), the rates of binding and unbinding are respectively given by:
	\begin{equation}
		\omega_s\,c\,k_b^\alpha
		\qquad \mbox{ and } \qquad
		\omega_s\,k_u^\alpha \comma
	\end{equation}
	The ratio between the rate constants $k_u^\alpha/k_b^\alpha$ is the {dissociation constant} of the binding process, $K_d^\alpha$:
	\begin{equation}
		K_d^\alpha = \frac{k_u^\alpha}{k_b^\alpha} = c_0\,e^{-\beta(\varepsilon_b^{(\alpha)}+\mu_0)}   
	\end{equation}
	We also define the constants $k_a$ and $k_i$ as 
	\begin{equation}
		k_a = \omega_f\,e^{\beta\varepsilon_I} \comma
		\qquad \mbox{ and } \qquad
		k_i = \omega_f\,e^{\beta\varepsilon_A} \comma
	\end{equation}
	which are the rate constants for activation and inactivation of the single protomer with $\ell = 0$; the rates in the case $\ell = 1$ are chosen compatibly with detailed balance condition.
	The ratio $k_i/k_a$ is called {allosteric constant}, $L$:
	\begin{equation}
		L = \frac{k_i}{k_a} = e^{\beta(\varepsilon_A - \varepsilon_I)}   
	\end{equation}

	It is worth noticing that the binding/unbinding rates at one protomer only depend on the state of the protomer itself and no other protomer in the ring:
	this assumption of independent binding is typical of allosteric models.

	\begin{figure}[t]
		\centering
		\def\svgwidth{.85\textwidth}
		\input{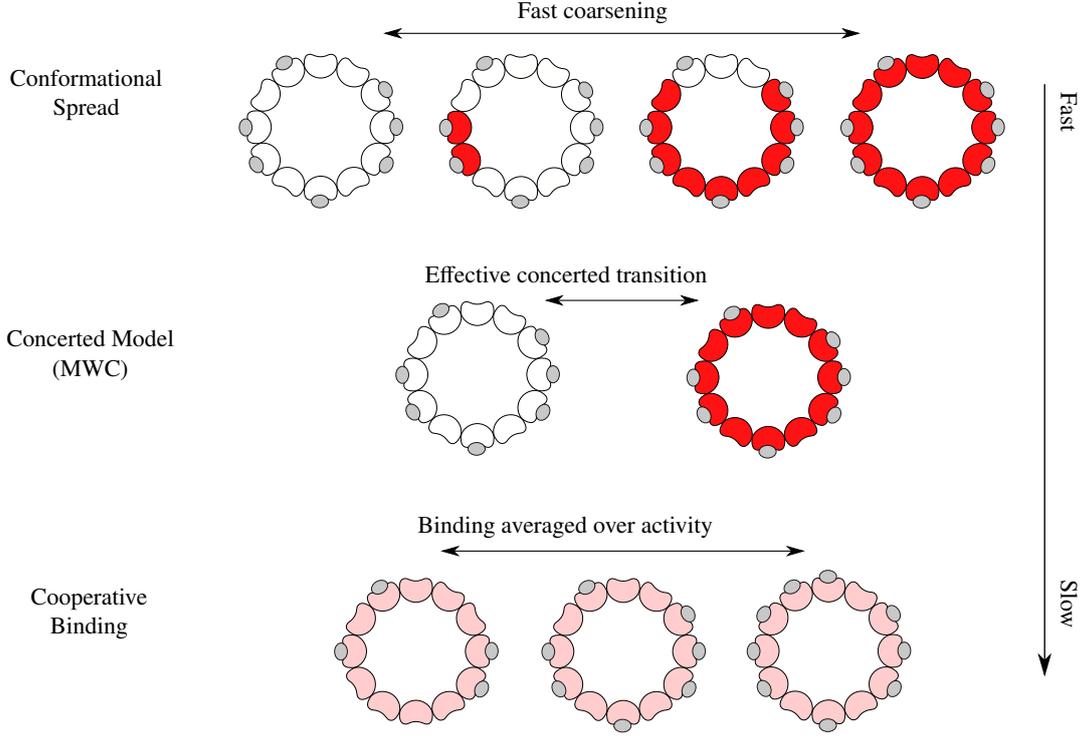}
		\caption{{Time-scale separation in the Conformational Spread Model.}
			Graphical representation of the time-scale separation scheme.
			Short-lived transient states containing domain walls are decimated in a first time-scale separation, leading from the Conformational Spread to the MWC model, while the binding dynamics is kept frozen.
			Then, over the binding time scales the activity states are averaged out, resulting into a cooperative binding model.}\label{fig:time scales}
	\end{figure}

		Experimental data provide evidence that the system has three well-separated time scales for the kinetics (see Fig.~\ref{fig:time scales}: fast coarsening of activity domains due to strong coupling $\beta J \gg 1$, slow nucleation and even slower binding/unbinding. These are summarized by the inequalities $\omega_s \ll \omega_f \ll \omega_f/(1-\gamma)$.
		
	Following Ref.\cite{pac_16} we show below how the strong coupling limit amounts to reducing the conformational spread model  to an effective Monod--Wyman--Changeux model \cite{mgp_13} and the slow-binding limit further reduces it to an effective cooperative binding model.
		
\subsubsection*{Decimation: from the Conformational Spread Model to MWC}	

The fastest rates are
	\begin{equation}\label{eq:glauber-sep1}
		K_f^{(i)}(s\to s') = \frac{1}{N}\,\frac{\omega_f}{1-\gamma}\left( 1 + \sigma_i\,\frac{\sigma_{i+1} + \sigma_{i-1}}{2} \right) \delta_{\sigma_i',\,-\sigma_i}\delta_{\ell_i',\,\ell_i}   
	\end{equation}
	The coherent configurations with all spins equal are thus the only absorbing states of the process.	The dynamics specified by $K_f$ forbids the creation of pairs of domain walls, while it allows their annihilation.
	Fast transitions are allowed within sets of states characterized by the same number of domains, and towards states with fewer domains.
	As a result, the dynamics leads to one or the other coherent configuration (all protomers active or inactive) exponentially fast, with a typical rate $\sim \omega_f/(1-\gamma)$.

	
	Therefore, on the typical time-scale of the switching dynamics, set by the rate $\omega_f$, the conformational spread model reduces to the MWC model with $N$ binding sites and two global activity states.
		While transition rates from active to inactive states depend on the arrangement of bound ligands over the $N$ sites, the equilibrium properties depend only on the total number of ligands bound. This property enables a further simplification when binding/unbinding is slower than switching.

\subsubsection*{Averaging: from MWC to a cooperative binding model}	
	For slow binding/unbinding, the activity of the ring rapidly relaxes to equilibrium at a given value of bound protomers $l$:
	\begin{eqnarray}
		P_{eq}(I|l) &= \frac{P_{eq}(l,I)}{P_{eq}(l)} = \frac{P_{eq}(l,I)}{P_{eq}(l,I) + P_{eq}(l,A)} = \frac{1}{1 + \frac{k_a}{k_i}\left(\frac{K_d^I}{K_d^A}\right)^l} \comma \label{eq:inact-given-l} \\[1ex]
		P_{eq}(A|l) &= \frac{P_{eq}(l,A)}{P_{eq}(l)} = \frac{P_{eq}(l,A)}{P_{eq}(l,I) + P_{eq}(l,A)} = \frac{1}{1 + \frac{k_i}{k_a}\left(\frac{K_d^A}{K_d^I}\right)^l} \label{eq:act-given-l}   
	\end{eqnarray}
	Then, on time scales comparable to or larger than $\omega_s^{-1}$, the relevant dynamics is essentially the slow binding/unbinding one, while the fast activation/inactivation dynamics is {averaged} over the equilibrium conditional probabilities in Eqs.\,\eqref{eq:inact-given-l}--\eqref{eq:act-given-l}, to give the effective rates $\overline{K}$ for the variable $l$:
	\begin{equation}\label{eq:eff-rates}
			\overline{K}(l \to l') = \sum_{\alpha\in\{I,A\}} P_{eq}(\alpha|l)\,K(l\to l',\,\alpha\to \alpha)   
	\end{equation}
Namely,	
	\begin{equation}\label{eq:eff-rates2}
		\overline{K}(l \to l+1) = (N-l)\,c\,\overline{k}_b^{(l)} \equiv b_l \comma \\
		\overline{K}(l \to l-1) = l\,\overline{k}_u^{(l)} \equiv u_l \comma
	\end{equation}	
	where
	\begin{equation}
		\overline{k}_{b,u}^{(l)} = \frac{k_{b,u}^A}{1 + \frac{k_i}{k_a}\left(\frac{K_d^A}{K_d^I}\right)^l} + \frac{k_{b,u}^I}{1 + \frac{k_a}{k_i}\left(\frac{K_d^I}{K_d^A}\right)^l} \;.
	\end{equation}
	The process hence obtained is a {birth-and-death process}, restricted on the set of integers between $l=0$ and $l=N$. The resulting effective cooperative binding model reproduces well the experimental observations, see \cite{pac_16}.
{  
\paragraph{Summary}
The flagellar motor of {\it E. coli} involves three widely separated timescales. By first decimating and then averaging  it is 
possible to obtain an effective kinetics in terms of a cooperative binding model.
}
 \subsubsection{Random graphs}
 \begin{figure}[!h]
  \includegraphics[width=0.9\textwidth]{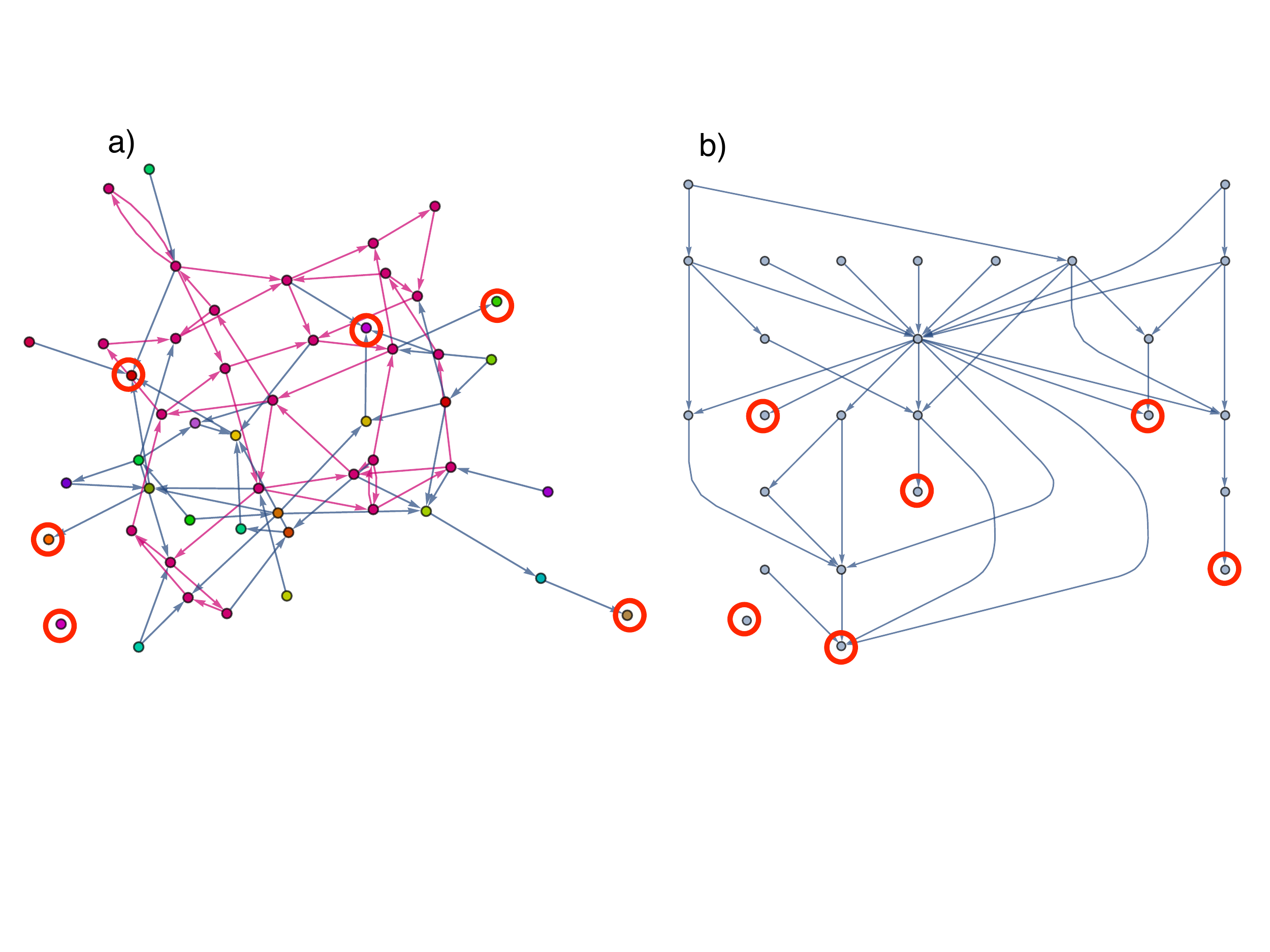}
\caption{\label{fig:rand} a) Network of the fast transitions between $50$ states  generated by drawing edges with probability $p=c/(n-1)$ with $c=1.5$. The network has one large strongly connected component involving $25$ states (connected by purple arrows) and $25$ states not strongly connected (each forming a strong component of size $1$). b) The condensation of the network is the acyclic directed graph showing transitions between the components. From its analysis it is possible to identify the states that conserve probability that are the ones with no outgoing links: the "sinks".
 The red circles identify the strongly connected components that conserve probability: i.e., the ones that "survive" the decimation procedure.}
\end{figure}
\begin{figure}[!h]
  \includegraphics[width=0.8\textwidth]{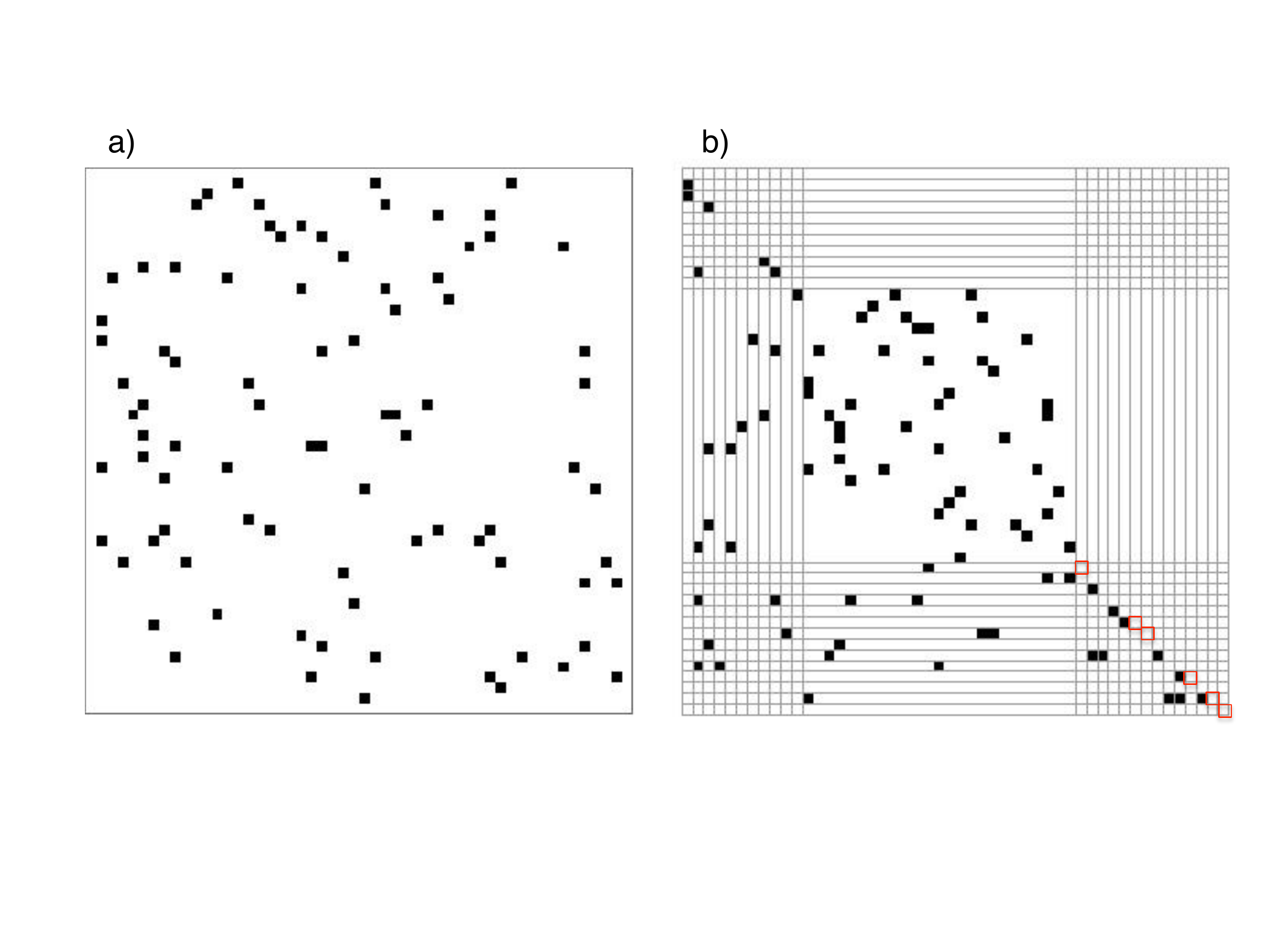}
\caption{\label{fig:rand_adj} a) Adjacency matrix of the network of the fast transitions between $50$ states  generated by drawing edges with probability $p=c/(n-1)$ with $c=1.5$. b) Rearrangement of the states to obtain a block triangular adjacency matrix. The large connected component is evident. The states in red are the ones conserving probability. }
\end{figure}

\begin{figure}[!h]
  \includegraphics[width=0.8\textwidth]{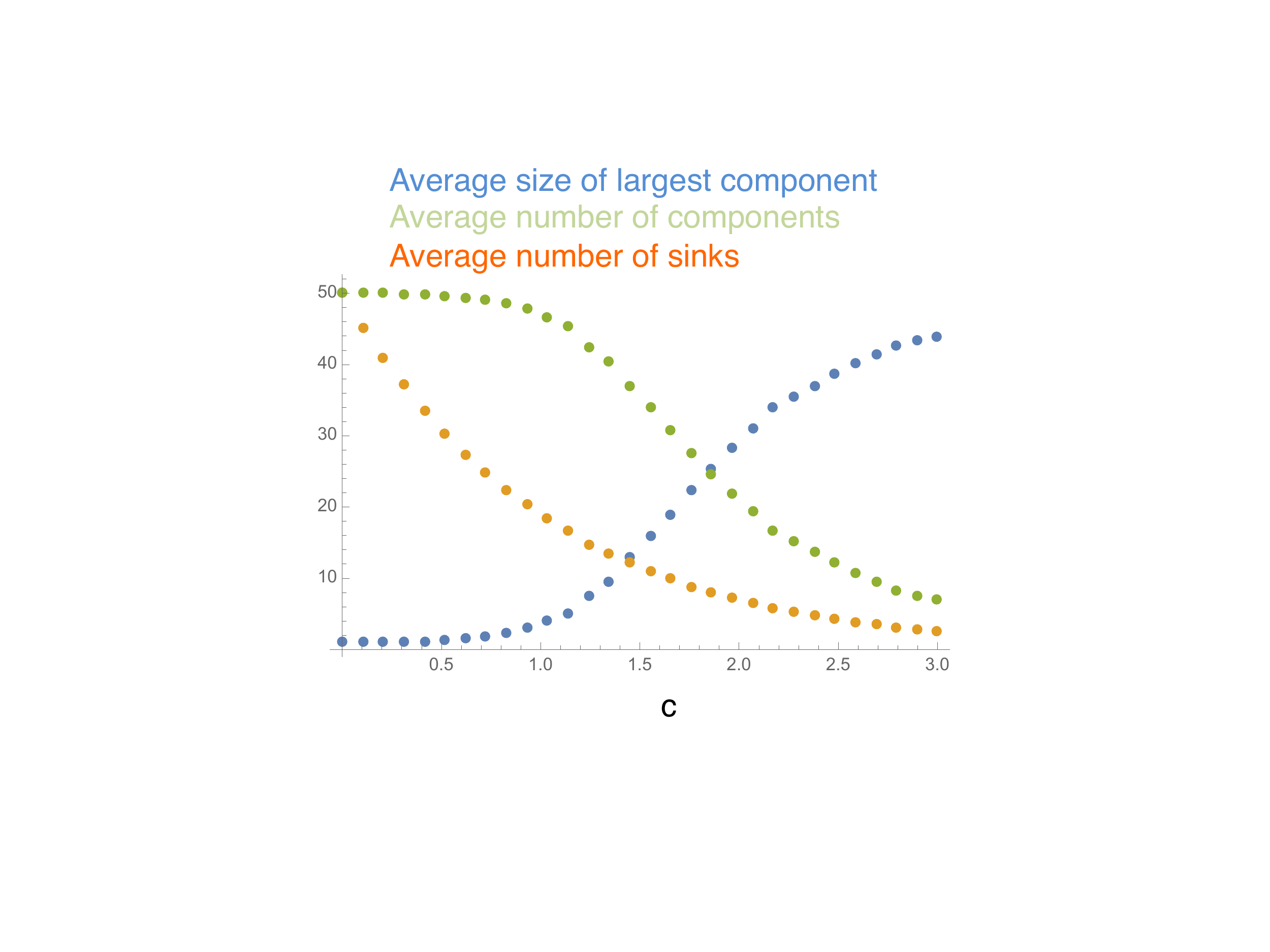}
\caption{\label{fig:rand_sim} {   Decimation and averaging of a Markov chain on a} graph with $50$ nodes: average size of largest component,
 number of components and
of sinks (components conserving probability) from a thousand random graphs in figure  as a function of the parameter $c$ which determines the probability of drawing an edge $p=c/(n-1)$. }
\end{figure}

The transitions networks discussed so far originated from chemical and biochemical systems and
had regular structures. Let us now consider how the presented approach can be applied to more general
(less structured) cases as, for instance, random directed graphs.
Take a system with $n$ states and, for the sake of simplicity, let us assume that the fast transition are given by a random Bernoulli directed graph $\vec{D}(n,p)$ which is composed of 
$n$ vertices and with directed edges randomly drawn independently with probability $p$.
An example of a network with $50$ states and an average  outgoing (and incoming) edge number per state of $1.5$ generated form $\vec{D}(n=50,\,p=1.5/49)$ is given in figure~\ref{fig:rand}. An alternative visualization is given by the adjacency matrix in figure \ref{fig:rand_adj}.
The procedure starts with the identification of the strongly connected components. For the case at hand there is
one large component involving $25$ states and $25$ components made of a single state. 
The next step is to study the transitions across such components and identify the ones conserving probability.
This can be done by studying the condensation of the graph and retaining only the sinks.
Equivalently, one can cast the adjacency matrix in the block triangular form and keep only the blocks 
that conserve probability i.e. the ones that have no off-diagonal blocks.
In this example there are only six components conserving probability which means that the effective dynamics will take place on a much smaller set of states compared to the original dynamics.
However, to obtain such effective dynamics, one needs to invert the matrix containing transitions between the $44$ transient states. 
In general the extent to which the time scale-separation can reduce the complexity of the original dynamics
depends on the structure of the fast transitions.
The relevant features are the number of strongly connected components, their size and how many
of them conserve probability.
If few components conserve probability the effective dynamics will be simpler but its derivation
will require the inversion of a larger matrix.
 For large random directed graphs  a few results about the size of the largest strongly connected component and the number of strong components are known (see e.g. \cite{Frieze2015}).
 When $c=p/n>1$ the size of the  largest strongly connected component 
 is, with high probability, a finite fraction of the number of vertices. More precisely, the fraction is $(1-x/c)^2$ where $x<1$ is defined
 by $xe^{-x}=ce^{-c}$. The other strong components are logarithmic in size.  
 For less connected graphs, when $c<1$, all strong connected components are either cycles or
 single vertices. The number of nodes in each cycle is at most $\omega$ for any $\omega(N)$ such that
 $\lim_{N\to \infty}\omega(N)=\infty$ no matter how slowly.
 To provide an intuition on how the relevant features of random direct graph depend on the connectivity we plot the average size of largest component,
 number of components and
of sinks (components conserving probability) from a thousand random graphs in figure \ref{fig:rand_sim}.
{  
\paragraph{Summary}
The problem of determining the relevant states on which the process effectively evolves on the slow time-scale can be cast in terms of the identification of 
the sinks of the graph condensation of the fast transitions. The method we have presented in the previous section applies to more complex network structures as well. Making use of established algorithms for deriving the condensation of a graph with linear complexity, one can efficiently identify which states of the original system will appear in the effective dynamics. 
}
 

  \clearpage
 
 \section{Diffusive systems}\label{sec:dyn_diff}
 {  
 \paragraph{The question} Is there a systematic way of deriving the effective, slow-scale evolution of a diffusive system displaying a fast and a slow timescale? Under what conditions
 will the effective evolution follow a diffusive process?
 }\\
 \subsection{General case}
 Let us consider continuous diffusive processes.
Such processes are widely used to model dynamics that are intrinsically of continuous
nature as for example the motion of a Brownian particle in a fluid.
In other cases, such description may  result from the approximation
of  discrete dynamics as, for instance,  the Van Kampen system size expansion \cite{van1992stochastic} of a set of chemical reactions
or the many particles limit of reaction networks giving the chemical Langevin equation \cite{gillespie2000chemical}
or the Kimura-Ohta equation in population genetics \cite{kimura1964diffusion}.
Again we will consider the case in which two well separated time scales are involved and make use of asymptotic
techniques  \cite{pavliotis2008multiscale}. As we will show in the following, continuous systems can display a 
richer temporal structure and their effective evolution may take place on time scales slower than the ones
that initially characterize the system. We will give an example from population genetics in which 
the effective equation takes place on the time scales of the original slower process, and one 
concerning  Brownian motion that evolves on a time scale slower than the initial slow one.
The general process we consider (see \cite{Bo2014}) is described by the system of It\=o  stochastic differential equations coupling the fast variables
$Y_t$ to the slow ones $X_t$ 
\begin{equation}\label{eq:cont_slow}
dX^i_t = {u}^i(X_t,Y_t,t) dt + \beta^{ij}(X_t,Y_t,t) \cdot dB^j_t\;
\end{equation}
\begin{equation}\label{eq:cont_fast}
dY^a_t = \epsilon^{-1}{z}^a(X_t,Y_t,t) dt + \epsilon^{-1/2}\sigma^{ab}(X_t,Y_t,t) \cdot d\hat{B}_t^b
\end{equation}
where $i,j=1,\ldots,n$, $a,b=1,\ldots,m$,  $B_t^j$ and $\hat{B}_t^b$ are independent Wiener processes and $\epsilon \ll 1$. 
The explicit dependence on time is only on time scales 
$O(\epsilon^0)$ or slower\footnote{In general the drift of the fast variable may also contain terms of order $O(\epsilon^0)$. However, such terms behave regularly in the limiting procedure as we will see in the example concerning a force acting on a Brownian particle
undergoing Langevin-Kramers dynamics as discussed in \cite{Celani2012} and reported in section
\ref{sec:lk}.}. We are interested in the dynamics of $X_t$ in the limit $\epsilon \to 0$.\\

Of course, one may consider different couplings and scaling between the slow and fast processes.
A relevant case with alternative coupling can be used to investigate the white noise limit of a colored noise
(see e.g.  \cite{Moon2014,wong1965convergence,hanggi1995colored,kushner1981weak,Volpe2016} 
 for the general discussion on the interpretation of the resulting noise).
The general idea there, is to separate the time-scale on which the relevant process takes place from   
the typical correlation time of the noise. An approach within the methods presented in the section 
was taken in \cite{Bo2013} following  the discussion provided in Gardiner's book \cite{gardiner1985stochastic}. More complex cases involving the interplay between the inertial time scale of
a Brownian particle and the correlation of its noise were studied in\cite{Kupferman2004,pavliotis2008multiscale}.\\

For the case we are considering, the Kolmogorov equations for the propagator $p(x,y,t|x',y',t')$ read
\footnote{Note that for diffusive processes we adhere to the usual notation {   in the mathematics literature} that $L$ is the generator of the process (backward) and $L^\dagger$ is the 
(forward) Fokker-Planck operator. For discrete processes we have defined $\mathsf{L}$ as the forward operator instead (also customary {   in the physics literature}).}
\begin{equation}
\partial_{t} p = {L}^\dagger p \;, \qquad \partial_{t'}p =  -L' p 
\end{equation}
where the $'$ indicates the dependence on the initial variables $x',y',t'$ and the generator of the diffusion process
is
\begin{equation}\label{eq:dyn_cont_gen}
L = \underbrace{{u}^i\frac{\partial}{\partial x^i} + \frac{1}{2} d^{ij} \frac{\partial^2}{\partial x^i\partial x^j}}_{L_0}
+  \epsilon^{-1}\underbrace{\left({z}^a\frac{\partial}{\partial y^a} + \frac{1}{2} g^{ab}\frac{\partial^2}{\partial y^a\partial y^b}\right)}_{M}
\end{equation}
with
\begin{equation}
d^{ij}=\beta^{ik}\beta^{jk} \qquad\qquad
g^{ab}=\sigma^{ac}\sigma^{bc}\;.
\end{equation}
We will perform the averaging procedure on the backward equation but the same approach
can be applied to the forward one (as done for example in \cite{Celani2012}). 
{  We} start by introducing the fast and slow time variables
$\theta=\epsilon^{-1}t$ and $\tilde{t}=\epsilon t$ and expand the solution as 
\begin{equation}
p=p^{(0)}+\epsilon p^{(1)}+ \epsilon^2 p^{(2)} + \ldots
\end{equation}
At order $\epsilon^{-1}$  the backward equation reads
\begin{equation}
\left(\frac{\partial}{\partial \theta}  + M\right) p^{(0)} = 0 \;.
\end{equation}
 As for the discrete case we consider the case in which the fast dynamics relaxes on fast time scales
 to  an equilibrium solution
 for any given value of the slow variable $x$ and time $t$
 \begin{equation}\label{eq:cont_equil}
 M^\dagger w_{eq} (x,y,t,\tilde{t})  = 0\;.
 \end{equation}
 In general the averaging procedure can be carried out also for the case in which the fast
 process reaches a non-equilibrium steady state. We address such general case in
 Appendix \ref{appendix:neq_diff}.
 The presence of a steady state implies that the spectrum of $M^\dagger$ has a top eigenvalue $E_0=0$
 and {   we assume} a finite gap ($E_1<0$). 
 As a consequence, after a fast transient, the solution will relax to a constant in $y$:
 \begin{equation}
 p^{(0)} = \rho(x,t,\tilde{t})\,.
 \end{equation}
 
 \paragraph{Propagator at order $\epsilon^0$}
 At order $\epsilon^0$ one has, after the initial relaxation,
 \begin{equation}
 Mp^{(1)}=-\left(\frac{\partial}{\partial t}  + L_0\right) p^{(0)}\;.
 \end{equation}
 The solvability condition requires (as in the discrete case) the RHS to be orthogonal to the nullspace of $M^\dagger$ which, according to eq.~(\ref{eq:cont_equil}) is spanned by $w_{eq}$ .
 Then, multiplying to the left by $w_{eq}$ and integrating over $y$ we have
  \begin{equation}
 0 = -\frac{\partial \rho}{\partial t} - \int dy\, w_{eq} L_0 \rho 
 \end{equation}
 where we have made use of $\int dy\, w_{eq} =1$. Therefore
 \begin{equation}\label{eq:eff0}
 \frac{\partial \rho}{\partial t} + \overline{L}_0 \rho = 0
 \end{equation}
and the generator of the slow process is
\begin{equation}\label{eq:L_cont}
\overline{L}_0 = \overline{{u}}^i \frac{\partial}{\partial x^i}  + \frac{1}{2}\overline{d}^{ij}
\frac{\partial^2}{\partial x^i \partial x^j}
 \end{equation} 
where $\overline{\cdots}=\int dy\, \cdots w_{eq}$ denotes the average over the equilibrium
distribution of the fast variables.
This concludes the elimination unless the slow drift and diffusion coefficient average to zero
 on the equilibrium distribution of fast variables. At variance with the discrete case, this is
 possible and indeed happens, for instance, in the high friction limit of Brownian motion (see section
 \ref{sec:lk}). 
 \paragraph{Propagator at order $\epsilon$}
 In the case  $\overline{{u}}^i=0$ and $\overline{{d}}^{ij}=0$ we need to proceed to
 the following order. 
 The previous solvability condition then becomes
 \begin{equation}
 \frac{\partial \rho}{\partial t} =0
 \end{equation}
 and the equation for $p^{(1)}$ reduces to
 \begin{equation}
 M p^{(1)} =- L_0 \rho = -{u}^i \frac{\partial\rho}{\partial x^i}-\frac{1}{2}d^{ij}
\frac{\partial^2\rho}{\partial x^i \partial x^j}
 \end{equation}
 with formal solution
 \begin{equation}
 p^{(1)} =- M^{-1}L_0 \rho
 + \mbox{zero modes of $M$}= - M^{-1}\left( {u}^i  \frac{\partial\rho}{\partial x^i}+\frac{1}{2}d^{ij}
\frac{\partial^2\rho}{\partial x^i \partial x^j}\right)
 + \mbox{zero modes of $M$}\;.
 \end{equation}
 The equation at order $\epsilon$ is 
 \begin{equation}
 \left(\frac{\partial}{\partial \theta}  + M\right) p^{(2)} =-\left(\frac{\partial}{\partial t}  + L_0\right) p^{(1)} - \frac{\partial}{\partial \tilde{t}} p^{(0)}\;.
 \end{equation}
From this expression it is clear that, in general, the effective equation will contain derivatives of $\rho$ of 
order higher than 2 hence describing non-Markovian effective dynamics. However, as already discussed by
\cite{risken1984fokker}, if the diffusion coefficient of the slow variable $d^{ij}$ does not depend on the fast variable 
$y$, the higher{  -}order derivatives vanish and the effective dynamics is Markovian. Since
we have required  $\overline{d}^{ij}=0$ we see that at order $\epsilon$ the dynamics is
Markovian only if there is no noise on the slow variable (i.e.  $d^{ij}=0$).
In such a case the solvability condition becomes an effective backward Kolmogorov equation
on slow time scales 
 \begin{equation}\label{eq:cont_FP_epsilon}
 \frac{\partial \rho}{\partial \tilde{t}}  + \underbrace{ \left( \overline{{u}^j \frac{\partial}{\partial x^j}  \left(- M^{-1} {u}^i \right)}  
 \right)}_{U^i} \frac{\partial\rho}{\partial x^i}  +\underbrace{ 
 \overline{{u}^i \left(- M^{-1} {u}^j\right) } 
 }_{\frac{1}{2}\tilde{D}^{ij}}\frac{\partial^2\rho}{\partial x^i\partial x^j} = 0
  \end{equation}
  where we have made use of $\overline{\partial_t M^{-1}f} =\partial_t \overline{M^{-1}f} = 0$ for any $f$ such that $\overline{f}=0$ and assumed that $w_{eq}$
 depends on $O(\epsilon)$ times only. 
 We provide the proof which is  a slight generalization of the one in Ref.~\cite{Bo2014} in appendix \ref{appendix:neq_diff}.
Relying on the fact that the fast dynamics reaches an equilibrium steady state (which amounts to say $M^\dagger w_{eq} = w_{eq} M$ by detailed balance) one can show that, for any $f$ and $g$ with $\overline{f}=\overline{g}=0$
\begin{equation}\label{eq:cont_commute}
\overline{g \left(-M^{-1} f \right)} 
= \overline{f \left(-M^{-1} g\right)} 
\end{equation}
which defines a scalar product. 
The most immediate consequence is that this ensures that the diffusion coefficient is symmetric and positive 
\begin{equation}
\frac{1}{2} D^{ij} = \overline{{u}^i \left(- M^{-1} {u}^j\right) } 
\;.
\end{equation}
{  
\paragraph{Summary}
The effective evolution of a diffusive process involving a fast and a slow scale can be found with a systematic multiple scale approach. If the fast processes reach a steady state, the effective system will follow a diffusive process itself with a generator given by the average of the generator of the slow dynamics over the stationary distribution of the fast variables. If this average is zero, one obtains a diffusive effective evolution only if  the initial system has no noise acting on the slow variables. In such a case the effective drift and diffusion coefficient 
are not simply an average of the initial slow ones but depend also on time-correlations of the slow drift at the steady state through the Green's function of the fast process.
}
\subsection{Examples for diffusive dynamics}
We shall present here two applications of the shown method.
The first one is taken from population genetics and will have an effective dynamics taking place at order 
 $\epsilon^{0}$. 
The second one is the high friction limit of Brownian motion and will require two separate time scales to obtain the relevant effective equations.
 \subsubsection{Population genetics in changing fitness landscapes}\label{sec:ko}
The stochastic equation we consider comes from population genetics and describes the evolutionary process of a population
under mutations, genetic drift and natural selection 
 (see \cite{hartl1997principles,gillespie2010population,blythe2007stochastic} for a general exposition).
A key intuition for modeling the effects of natural selection is the concept of fitness
 of a population in a certain 
environment  which describes the global ability of the population to reproduce and survive~\cite{haldane1927mathematical}. 
The different fitness of various genotypes can be effectively visualized in terms
of a fitness landscape~\cite{wright1932roles} which is usually "climbed" during the course of evolution.
Such climbing is referred to as adaptation.
Fisher's fundamental theorem of natural selection~\cite{fisher1958genetical}
states that when evolution is subject only to natural selection
in a constant environment, the fitness of a population increases
at a positive rate equal to the variance of the population.
Indeed, 
natural selection, by its very definition, is the force that favors fitter individuals,
pushing towards the genotype configuration which maximizes the global fitness.
When mutations and random drift are relevant, the stochastic nature of evolution
emerges and adaptation becomes a more complex phenomenon. 
One of the most widely accepted model for the stochastic description of the evolutionary process is the Kimura-Ohta
 equation~\cite{kimura1964diffusion} that we will describe in detail in Eq.~(\ref{eq:ko}) below.
In most natural cases the environment in which the population evolves changes in time
(see for example Refs. \cite{mustonen2008molecular,mustonen2010fitness,tanase2012fitness,rivoire2014model,Held2014,Nourmohammad2015,Melbinger2015,Desponds2016})
and genotypes that were fit under the initial condition may successively be unfavored by selection.
 Consider for instance the case of a population of bacteria shaped by natural selection to
 metabolize a certain nutrient. If such nutrient is gradually replaced by a different 
 one for which metabolic enzymes are not genetically encoded, bacteria will now be less efficient and their fitness will decrease.
 At the same time adaptation to this new environment though mutations
will start taking place and eventually lead to an increase in fitness.
We can then represent the fitness dynamics as the result of two effects: 
adaptation driven by 
natural selection and fitness changes due to environmental variations:
\begin{equation}\label{eq:fit_gen}
    \frac{d}{dt} \text{Fitness} = \text{Adaptation} + \text{Environmental changes}\,.
\end{equation}
The changes brought about by the variability of the environment give a negative contribution
countered by adaptation, 
which is, in general, positive.

We focus on the case in which the environment exerting natural selection 
on the population changes  very rapidly and in a random fashion consequently alternatively favoring the different genotypes.
Such limit of  a very rapidly changing selection goes under the name
of microevolutionary limit and was shown to give  an effective diffusive equation on time scales that are much longer than the typical
 environment variation, a result first obtained more than four decades ago (see Refs.~ \cite{Takahata1975,Takahata1979})
 We review this problem within the framework  derived in the previous section, 
  closely following Ref. \cite{bo2014adaptation}.
For the sake of simplicity we consider the case in which the population can have two different genotypes: $G$ and $G'$ and denote
by $x$  the frequency of individuals of genotype $G$ and by $1-x$ the one of $G'$.
The Kimura-Ohta equation for a population that switches between two possible genotypes is equivalent to the following 
one dimensional It\=o SDE 
\begin{equation}\label{eq:ko}
 dX_t = \biggr(s(X_t,Y_t)X_t(1-X_t)+ m(X_t)\biggr) dt +\sqrt{\frac{1}{N} X_t(1-X_t)} \cdot dB_t
\end{equation}
where $dB_t$ is a Wiener process, $m(x) = -\mu x + \nu (1-x)$ is the mutation coefficient, $\mu$ and $\nu$ are the mutation rates and $N$ is the effective population size.
   $s(x,y)$  
   is the selection coefficient, which describes the effect of natural selection
 and depends on the
  environmental state $y$ and, in the simple case we are considering, can be used to define the
 global fitness of the population $F(x,y)$ via:
\begin{equation}\label{eq:s}
  s(x,y) = \frac{\partial F(x,y)}{\partial x}\;.
\end{equation}
      The environment follows the faster dynamics which, for the sake of simplicity, we assume to be an Ornstein-Uhlenbeck process
  with constant diffusion $D$ and spring constant $K$ (of order $1$)
\begin{equation}\label{eq:env_ex}
 dY_t = -\epsilon^{-1}KY_t dt + \epsilon^{-1/2}\sqrt{2D} \cdot d\hat{B}_t
\end{equation}
where $d\hat{B}_t$ is a Wiener process and $\epsilon\ll 1$ is the parameter accounting for the time-scale separation between the
environment and the genotype evolution.
At the steady state
$\overline{y} = 0$ and $\overline{y^2} = D/K$.
For the sake of exemplification we consider the case in which the selection coefficient has
a simple linear dependence on the environment and does not depend on the genotype frequency: $s(y) = \sigma y$.
This implies that a positive environmental state favors genotype $G$ whereas
a negative one favors $G'$. 
The fitness of the population, up to an arbitrary constant, is then:
\begin{equation}\label{eq:fitness}
F(x,y) = \sigma x y\;.
\end{equation}
  In the notation of the previous section we have that the drift of the slow process is
  \begin{equation}\label{eq:ko_drift}
     u(x,y)=s(x,y)g(x) + m(x)=\sigma y x(1-x)-\mu x + \nu (1-x)
    \end{equation}

and the diffusion matrix does not depend on the fast process
\[
d(x)=\frac{1}{N}x(1-x)\;.
\]
Since the only term in the slow drift depending on the fast process is linear in $y$ and $\overline{y} = 0$  we have that the effective drift
on the slow time-scale reads 
\begin{equation}
 \overline{u}=-\mu x + \nu (1-x)
\end{equation}
and is independent of the selection coefficient. 
This concludes our elimination procedure and shows that
the effective dynamics is  described by
\begin{equation}\label{eq:ko_eff}
 dX_t =   m(X_t) dt +\sqrt{\frac{1}{N} X_t(1-X_t)} \cdot dB_t
\end{equation}
and apparently neutral: {\it i.e.}, not subject to selection.
{  
\paragraph{Summary}
Within a simple single-locus two-allele model of population genetics we have considered the evolution of a genotype in presence of  a rapidly changing 
stochastic environment.
The allele frequency follows a diffusive equation with a selection coefficient that is given by the average selection exerted by the environment.
}
 \subsubsection{From Langevin-Kramers dynamics to Brownian motion}\label{sec:lk}
We then move to an example in which the effective dynamics takes place on a time-scale which is
much slower than the one of the original system. We shall discuss the high friction limit 
of Langevin-Kramers dynamics describing the motion of a microscale particle immersed in a fluid.
This is a rather well studied system (see~\cite{bocquet1997high} for a pedagogical exposition) when
the temperature of the fluid is constant. Here we consider the case in which both the temperature
and the viscous friction of the fluid can vary smoothly in space.
The initial equations are the ones in Eq. (\ref{eq:LK}).
Here we consider also the possibility of an external force $f$ acting on the particle and,
in order to have a more concise notation, we adopt a non-dimensional expression by introducing the following rescalings:
\begin{eqnarray}\label{eq:rescale}
 X \rightarrow \frac{X}{L} \qquad & V \rightarrow V\sqrt{\frac{m}{k_BT_0}} \qquad &f \rightarrow \frac{f}{k_BT_0/L}\\
T\rightarrow \frac{T}{T_0}\qquad & t \rightarrow t\frac{\sqrt{k_BT_0/m}}{L} \qquad &\gamma  \rightarrow \frac{\gamma L}{\sqrt{k_BT_0 m}} \nonumber
\end{eqnarray}
where $T_0$ is a reference temperature, $k_B$ the Boltzmann constant, $m$ the mass of the particle and $L$ the typical length scale of the process.
We are interested in deriving the strong friction limit of the Kolmogorov equation. To keep track
of this limiting procedure we express the friction as $\gamma\to\epsilon ^{-1}\gamma $.
We then have
\begin{equation}\label{eq:LK_ex}
\begin{array}{lll}
dX^i_t &=& V^i_t dt \\
dV^i_t &=& f^i(X_t,t) dt -\epsilon^{-1}\gamma(X_t,t) V_t^i dt  + 
\epsilon^{-1/2}\sqrt{2 T(X_t,t)\gamma(X_t,t)}\, d\hat{B}^i_t 
\end{array}
\end{equation}
where $\hat{B}^i_t$ are independent Wiener processes. 
The validity and limitation of such starting model are investigated in Ref.~\cite{Falasco2016} within a fluctuating hydrodynamics approach with special attention
to the long range correlations of a fluid subject to a temperature gradient.
Here, the diffusion coefficient multiplying the noise term does not depend on velocity so, at this level, there is no need to specify the discretization prescription (It\=o, Stratonovich or others).
In the following we shall abridge the notation by omitting the explicit dependency on the
trajectory, e.g. $f^i(X_t,t) \equiv f^i_t $.
This set of equations is richer than the one discussed in the general example since the fast variable $v$  has also a slow contribution to to its drift (given by the external force $f$).
As we shall see in the following this will not impact on our elimination procedure.
%
In order to keep as close as possible to the general example discussed above we consider the backward Kolmogorov equation.
The complementary forward case is described in \cite{Celani2012}.
The backward equation  is
\begin{equation}\label{eq:FPLK}
\left(\frac{\partial }{\partial t}  +  L_{0} + \epsilon^{-1} M  \right) p_t = 0
\end{equation}
where
\begin{equation}\label{eq:LKLM}
L _0=  v_t^i \frac{\partial}{\partial x^i} +  f_t^i \frac{\partial}{\partial v^i} 
\qquad \qquad
M =\gamma_t
\biggl[ - v_t^i \frac{\partial}{\partial v^i} + T_t \frac{\partial}{\partial v^i} \frac{\partial}{\partial v^i} \biggr]  \;.
\end{equation}
 
As usual we introduce a fast time variable $\theta=\epsilon^{-1} t$, associated to frictional relaxation. We also define a very slow one 
$\tilde{t} =\epsilon t$, the time-scale of variation of temperature and friction. 
 The propagator is assumed to be a function of fast, slow and very slow times, and developed in power series 
 in $\epsilon$ as $p=p^{(0)} + \epsilon p^{(1)} +\epsilon^2 p^{(2)} \ldots$. 
\paragraph*{Fast time scales.}
 Here we show that the dynamics at fast time scales (of the order of the inverse friction)
 is ruled by the balance of thermal noise and friction, and leads to relaxation to 
 the Maxwell-Boltzmann distribution in velocity space, with the local temperature.
 After the initial relaxation, at order $\epsilon^{-1}$ the forward Kolmogorov equation reads
\begin{equation}
M^\dagger w_{eq} = 0\;.
\end{equation}
For the case we are considering the eigenfunctions of $M^\dagger$ are products of Hermite polynomials in the velocity variable multiplied by the weight
\begin{equation}\label{eq:maxwellian}
 w_{eq}=\frac{\exp\left(-\frac{v^2}{2 T}\right)}{\bigl(2\pi T\bigr)^{n/2}}
\end{equation}
i.e. the local Maxwellian equilibrium,
\begin{equation}\label{eq:M_hermite}
\psi_{k_1,\ldots,k_n} = w \prod_{i=1}^n H_{k_i}(v^i/\sqrt{T}) \qquad M \psi_{k_1,\ldots,k_n}
=-\gamma \left(\sum_{i=1}^n k_i\right) \psi_{k_1,\ldots,k_n}
\end{equation}
with $k_i=0,1,\ldots$. Since the spectrum is nonnegative, the solution relaxes exponentially fast to the zero eigenfunction for $\theta \to \infty$.
The solution of the backward equation at order $\epsilon^{-1}$ then has to be constant in the velocity variables 
\begin{equation}
p^{(0)}({x},{v},t,\tilde{t}) = \rho({x},t,\tilde{t}) 
\end{equation}
where $\rho$ is the marginal probability density in space, at lowest order.

\paragraph*{Slow time scales.}
As we know from the general formalism (see eq.~\ref{eq:eff0}) on slow time scales the dynamics is determined by the 
average of the slow operator 
$\overline{L_{0}}$. For the system under consideration 
$\overline{L_{0}}=0$ so that 
the effective equation on the slow time-scale reads
\begin{equation}
\frac{\partial \rho}{\partial t} = 0
\end{equation}
implying that the dependence on time for $\rho$ is only through the very slow time-scale $\tilde{t}$,
i.e. $p^{(0)}$ reaches a quasi-steady-state at slow time scales where the dependence on very slow time scales enters as a parameter only.

To proceed further we need to derive the explicit expression of $p^{(1)}$ from the equation on slow time scales
at order $\epsilon^0$ 
\begin{equation}
M p^{(1)} = -L_0 p^{(0)} \;.
\end{equation}

Noticing that
\begin{equation}
M v^i  = -\gamma v^i   
\end{equation}
and making use of the explicit expression of $L_0$ one obtains

\begin{equation}\label{eq:p1}
p^{(1)} =-M^{-1}L_0 p^{(0)}=
\frac{1}{\gamma}v^i\frac{\partial}{\partial x^i}\rho +r
\end{equation}
where $r=r(x,\tilde{t})$ is the contribution from the null-space of $M$.

\paragraph*{Very slow time scales.}
At these scales we obtain the overdamped dynamics in position space as follows.
At order $\epsilon^1$, after relaxation over fast variables,
the solvability condition is  obtained by integrating both sides over $w_{eq}$ and provides the description on the 
very slow dynamics
\begin{equation}\label{eq:slow_propagator}
\frac{\partial}{\partial \tilde{t}} p^{(0)}+ \overline{L_0 p^{(1)}}=0
\end{equation}
which gives
\begin{equation}
 \frac{\partial \rho}{\partial \tilde{t}} + \frac{f^i}{\gamma} \frac{\partial\rho}{\partial x^i} +
T \frac{\partial}{\partial x^i} 
\frac{1}{\gamma}  
\frac{\partial\rho}{\partial x^i}  =0\;.
\end{equation}
This corresponds to the overdamped stochastic differential equation
\begin{equation}\label{eq:lang_odamp_gen}
d X^i_t =\left( \frac{f^i}{\gamma} -\frac{1}{2\gamma} \frac{\partial T}{\partial x^i} +
 \frac{T}{2}\frac{\partial \gamma^{-1}}{\partial x^i}  \right)  dt
+ \sqrt{\frac{2 T }{\gamma} } \circ dW^i_t  \qquad \mbox{(Stratonovich)}
\end{equation}
or, in It\=o form,
\begin{equation}
dX^i_t = \left(\frac{f^i}{\gamma}+T\frac{\partial \gamma^{-1}}{\partial x^i} \right) \; dt + \sqrt{\frac{2 T}{\gamma}}\cdot dW^i_t \qquad \mbox{(It\^o)}
\end{equation}
We have then completed the averaging of the velocities degrees of freedom and obtained 
the dynamics taking place on the very slow time scales involving only the position variables. This description displays corrections of order 
$\epsilon$ and was also derived with a different approach in Ref.~\cite{Durang2015}. We note that, if friction is constant, the equation reduces to a It\=o stochastic differential equation  with drift $ f^i/\gamma$:
\begin{equation}\label{eq:lang_odamp}
 \frac{\partial \gamma}{\partial x^i}=0\qquad\qquad
dX^i_t = f^i/\gamma\; dt + \sqrt{2 T(x_t)/\gamma}\; \cdot dW^i_t \qquad \mbox{(It\=o)}
\end{equation}
in agreement with \cite{matsuo_sasa2000,yang2013brownian,Stolovitzky1998}.
When, instead,  temperature is constant we recover the result obtained
e.g. in \cite{Sancho1982} (see e.g. \cite{Volpe2016} for a review).
{  
\paragraph{Summary}
The large-friction (or small-mass) limit of a microscopic particle diffusing in a fluid can be
derived using multiple-scale techniques. When the temperature and friction are uniform in space, the result is the well-known overdamped approximation. For space-dependent temperature and friction diffusion, these techniques provide the physically correct interpretation of the multiplicative noise that appears in the overdamped approximation.
}
\clearpage

\part{Functionals}
 \section{Paths and sequential observables of discrete Markov processes}
 {  
\paragraph{The question}
We have seen how to obtain the effective slow evolution of a discrete Markov process.
Can the same procedure be applied to a functional of the stochastic trajectory such as entropy production, counting statistics, fluxes, etc...?
Will it be possible to express the evolution of a generic functional in terms of the effective states and their dynamics?}\\
\subsection{General functional of the discrete dynamics}
  \subsection*{From the master equation to the path-integral}
 For a system jumping between discrete states with given transition rates the probability density of 
observing a given path takes a rather simple expression. This is because we can decompose such probability
as the product of probabilities of each successive step. For instance the probability density of observing the sequence 
of states $\{ i \mbox{ in } (t'=\tau_0,\tau_1) ;\:  
 j \mbox{ in } (\tau_{1},\tau_{2}); \}$ is expressed as the product of the probability of not exiting state $i$
 until time $\tau_1$ times the transition rate from $i$ to $j$ $K_i^j$ times the probability of remaining in state $j$:
 \begin{equation}\label{eq:path_prob_ex}
 {\cal P}_{\tau_2,t'} = 
\exp\left[- (\tau_1-\tau_0)   e_i \right]\,
K^j_i \exp\left[- (\tau_2-\tau_1)   e_j \right]
\end{equation}
where we recall that $e_i$ is the exit rate from state $i$ as defined in section \ref{sec:dyn_discrete}.
 In general, a trajectory of the discrete Markov process is given by the sequence of states 
\[
i(\tau) \qquad  t' \le \tau \le t
\]
or, making the transitions explicit
\[
\{  i_0  \mbox{ in } (t'=\tau_0,\tau_1) ; \cdots;   i_k \mbox{ in } (\tau_{k},\tau_{k+1}); \cdots; i_{N}  \mbox{ in } (\tau_{N},\tau_{N+1}=t)  \}
\]

The probability density of having undergone exactly $N$ transitions at times $\tau_1,\ldots,\tau_N$ before time $t$,
with a sequence of states $i_1\ldots i_{N}$, starting from
initial state $i_0$, decays in time due to outbound transitions from the final state
\[
\frac{\partial}{\partial t}{\cal P}_{t,t'}^{N}(\tau_1i_1,\ldots ,\tau_N i_{N}|i_0) = - e_{i_N} {\cal P}^{N}_{t,t'}(\tau_1i_1,\ldots ,\tau_N i_{N}|i_0)
\]
for $t>\tau_N$. The initial condition (at time $t=\tau_N$) for the above equation is
\[
{\cal P}_{\tau_N,t'}^{N}(\tau_1i_1,\ldots ,\tau_N i_{N}|i_0)
= {\cal P}_{\tau_N,t'}^{N-1}(\tau_1i_1,\ldots ,\tau_{N-1} i_{N-1}|i_0) K_{i_{N-1}}^{ i_{N}}
\]
i.e. the probability of having undergone $N-1$ transitions at the specified times up to time $\tau_N$ and 
a new transition at time $\tau_N$. The density ${\cal P}^N$ has dimension $\mathrm{time}^{-N}$ and the rate $K$ has dimension $\mathrm{time}^{-1}$.

It follows that
\[
{\cal P}_{t,t'}^{N}(\tau_1i_1,\ldots ,\tau_N i_{N}|i_0) 
= \exp\left[ -\int_{\tau_N}^t e_{i_N} d\tau + \log K_{i_{N-1}}^{ i_{N}} \right] {\cal P}_{\tau_N,t'}^{N-1}(\tau_1i_1,\ldots ,\tau_{N-1} i_{N-1}|i_0) 
\]
and proceeding recursively one finally has
\[
{\cal P}^{N}_{t,t'} = 
\exp(-{\cal A}^{N}_{t,t'}) 
\]
with the action defined as
\begin{equation}\label{eq:action}
{\cal A}^N_{t,t'} =  \sum_{k=1}^{N+1} \int_{\tau_{k-1}}^{\tau_k} d\tau\, e_{i_{k-1}}
- \sum_{k=1}^{N} \log K^{i_{k}}_{i_{k-1}}\,.
\end{equation}
Recall that
this is a function of the $N$ transition times and states, given the initial state.
The normalization 
\[
\sum_{N=0}^\infty \sum_{i_1} \cdots \sum_{i_N} \int_{t'}^{t} d\tau_1 \int_{\tau_1}^{t} d\tau_2 \cdots
\int_{\tau_{N-1}}^t d\tau_N \; {\cal P}^{N}_{t,t'}  = 1
\]
follows from noticing that
\[
\sum_{N=0}^\infty \sum_{i_1} \cdots \sum_{i_{N}} \int_{t'}^{t} d\tau_1 \int_{\tau_1}^{t} d\tau_2 \cdots
\int_{\tau_{N-1}}^t d\tau_N \; {\cal P}^{N}_{t,t'} (\tau_1i_1,\ldots ,\tau_N i_ {N}|i_0)  \delta^{j}_{i_N} = p_{i_0}^{j}
\]

\subsection*{Functionals of the trajectory}
 Let us turn our attention to the study of functionals of the stochastic trajectory.
 Such functionals can depend on the time spent in a given state and on the transitions occurred.
 In general one can write
\begin{equation}
{\cal J}_{t,t'} =\sum_{k=1}^{N+1} \int_{\tau_{k-1}}^{\tau_k}  d\tau f_{i_{k-1}}  + \sum_{k=1}^{N} g^{i_k}_{i_{k-1}}
\end{equation}
where $g^a_b$ is different from zero only if a transition occurs, i.e. $a \neq b$.
Relevant instances of such a functional are the residence time in a given state $j$ ($f_i=\delta_{ij}$ and $g=0$), the transition counting statistics from $j$ to $j'$ ($f=0$ and $g_i^{i'}=\delta_j^{j'}$) as well as the action of the process
\eqref{eq:action} or the entropy production (see section \ref{sec:entro_block}). 
Notice that if $g^a_b=\phi_a-\phi_b$ then the last sum above is telescopic and it reduces 
to the evaluation of boundary terms.

The value of the functional ${\cal J}_{t'}$ undergoes both smooth changes and jumps.
The probability $p^{i}(J)$ of being in state $i$ at time $t$ with a value of the integral ${\cal J}_{t} =J$  
is given by the solution of the equation 
\begin{equation}\label{eq:pJ}
\frac{d}{dt} p^{i}(J) =-\frac{\partial}{\partial J} f_{i} p^{i}(J) + \sum_{j}\left( K_j^i p^{j }(J-g_{j}^{i}) -
K_i^j  p^{i}(J) \right)   
\end{equation}
Introducing the generating function of $J$ along trajectories that end in state $i$ at time $t$
\[
G^{i}(s)=\int_{-\infty}^\infty  dJ\, e^{-sJ}  p^{i}(J)
\]
one has
\begin{equation}\label{eq:GJ}
\frac{d}{dt} G^{i} =   
\sum_{j}\left[ K_j^{i} \exp(-s g^{i}_{j}) G^{j } -
\left(K_i^{j}+sf_{i}\delta^{j}_{i} \right)  G^{i} \right]   
\equiv\sum_{j} H^{i}_{j} G^{j}
\end{equation}
evolving under the action of the tilted generator $\mathsf{H}$.
The modified generator is given by the original one with the off-diagonal terms
 multiplied by $\exp(-s g^i_j)$ and additional $s f_i$ on the  diagonal. Setting $s=0$ one recovers the original dynamics.
 {   Notice that the long time limit of eq.~(\ref{eq:pJ}) and (\ref{eq:GJ}) contain the same information of the large deviation function of $J$ (see e.g. \cite{touchette2011basic}).
 }
\subsubsection{Averaging functionals, discrete case}
\begin{figure}[!h]
  \includegraphics[width=0.9\textwidth]{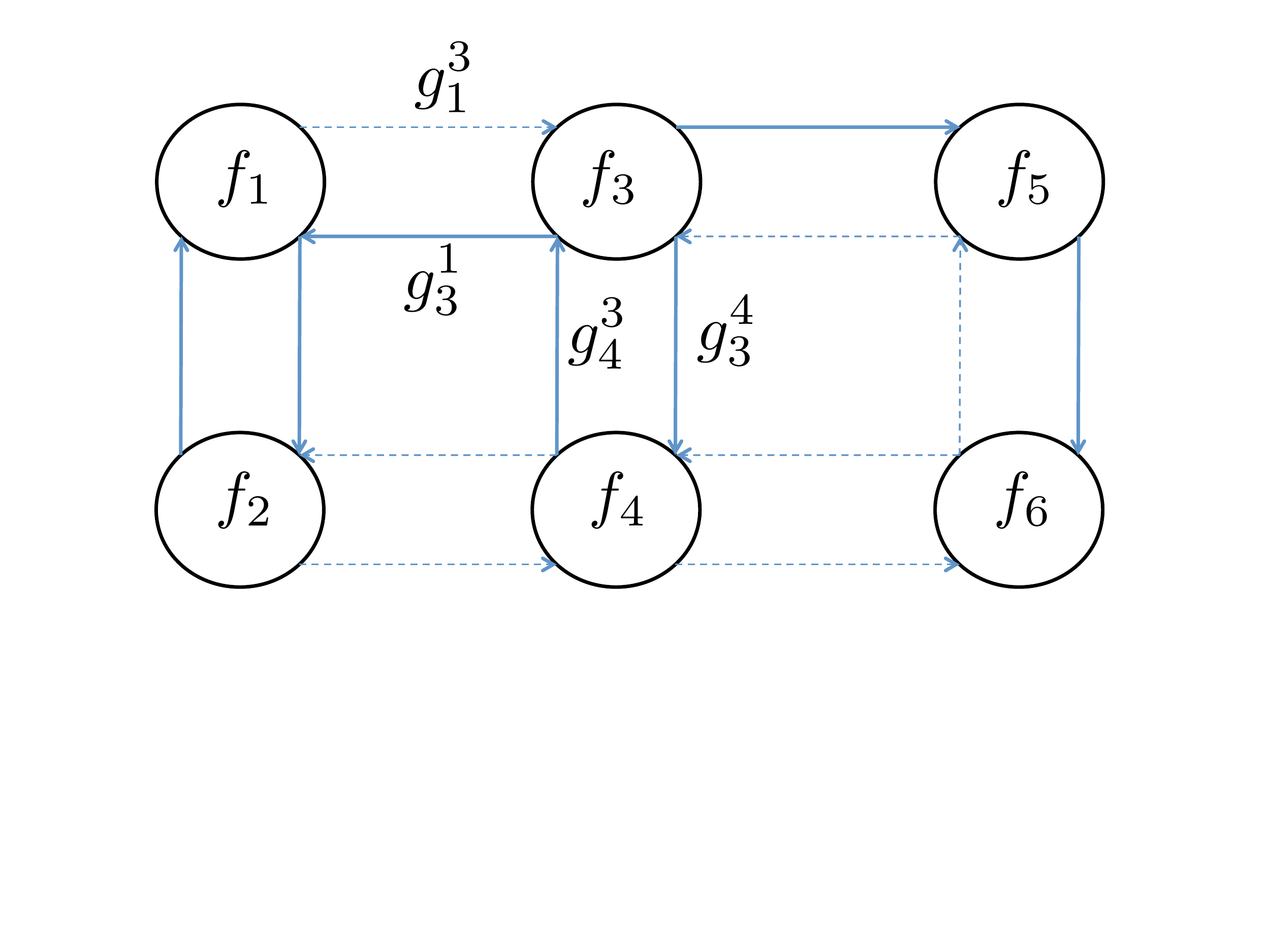}
\caption{\label{fig:dec_gen_func} Functional of the continuous-time Markov chain involving 6 states depicted in figure \ref{fig:dec_gen}. The quantities $g_i^j$ represent the change of the functional
upon transitions from state $i$ to $j$ whereas $f_i$ is the contribution picked up while residing in state $i$}
\end{figure}

As we have done for the dynamics, we shall seek an effective evolution of the generating function of the functional on the slower time scales.
As a first step we shall decompose the tilted generator in its slow and fast components
\begin{equation}\label{eq:tilted_gen_gen}
\mathsf{H} = \epsilon^{-1}\mathsf{M}_s + \mathsf{H^{(0)}}\,.
\end{equation}
For the six states example discussed in section~\ref{sec:dyn_discrete}  and described in figure~\ref{fig:dec_gen},
the fast tilted generator reads
\begin{equation}\label{eq:Mtilted}
  \tilde{M}_s=
 \left(
 \begin{array}{c c  c  c  c  | c}
 -\left(F^1_3+F^4_3+F^5_3\right)&F^3_4e^{-sg^3_4}	&\multicolumn{4}{|c }{}	\\
 F^4_3e^{-sg^4_3}	&-F^3_4	&\multicolumn{4}{|c}{  
 \text{\huge0}
   }	\\ 
 \cline{1-3}
  F^5_3e^{-sg^5_3}&0  &\multicolumn{1}{|c |}{-F^6_5}	&\multicolumn{3}{|c }{}\\
  \cline{3-5}
 F^1_3e^{-sg^1_3}	&0& 0& \multicolumn{1}{|c }{-F^2_1}	& F^1_2e^{-sg^1_2} 	 & 		 \\
  0	&0  &0 &\multicolumn{1}{|c }{F^2_1e^{-sg^2_1}}	& -F^1_2 	 & 			 \\
  \cline{4-6}
  0	&0	&F^6_5e^{-sg^6_5}	&0 & 0 &0\\
 \end{array}
 \right)
 \end{equation}
 where the superscript $\, \tilde{\,}\,$ is used to recall that we have ordered the states to have a
 block triangular structure as in eq.~\eqref{eq:Mmod}, {\it i. e.}, $(3,4,5,1,2,6)$. 
 The slow modified generator is given by
  \begin{equation}
  \tilde{L}^{(0)}=
 \left(
 \begin{array}{c c  c c c c}
 -sf_3&	0	&	S^3_5e^{-sg^3_5}&		S^3_1e^{-sg^3_1}&	0&	0	\\
 0	&	-sf_4-S^2_4-S^6_4&	0&	0&	S^4_2e^{-sg^4_2}&	S^4_6e^{-sg^4_6}	\\
0	&	0&	-sf_5-S^3_5&	0&	0&	S^5_6e^{-sg^5_6}\\
0	&	0&	0&	-sf_1-S^3_1&	0&	0\\
0	&	S^2_4e^{-sg^2_4}&	0&	0&	-sf_4-S^4_2&	0\\
0	&	S^6_4e^{-sg^6_4}&	0&	0&	0&	-sf_6-S^4_6-S^5_6
  \end{array}
 \right)
 \end{equation}
 This is schematically depicted in figure~\ref{fig:dec_gen_func} where it is highlighted that
 the functional changes of an amount $g_i^j$ in a transition form $i$ to $j$ and that
 it increases while in state $i$ at a rate $f_i$.
At order $\epsilon^{-1}$ one has
\begin{equation}
\frac{d}{d\theta}\mathsf{G^{(0)}} = \mathsf{M_s G^{(0)}}\,.
\end{equation}
In order to have a stationary state we need to require the eigenvalues of $\mathsf{M_s}$
to have non-positive real part.
This was granted for the unperturbed generator of the dynamics $\mathsf{M}$. Here, 
the presence of the additional exponential terms may lift the zero eigenvalues of $\mathsf{M}$ to positive values. Indeed, for the case at hand, the largest eigenvalue
of  block $a$ (involving state $1$ and $2$), which was $0$ for the 
generator of the dynamics, takes the following form:
\begin{equation}
\lambda_a(s)=\frac{1}{2}\left[-(F^1_2+F^2_1)+\sqrt{(F^1_2+F^2_1)^2 -4F^1_2F^2_1(1-e^{-(g^2_1+g^1_2)s})}\right]
\end{equation}
so that to ensure the steady state we shall require $g^2_1=g^1_2=0$.
This corresponds to saying that, in order to perform the averaging procedure discussed for the dynamics, we must
choose functionals that do not depend on the fast transitions within the blocks that
will be part of the effective dynamics. This is not surprising since the number of such contributions before a slow transition occurs would be extremely large (order $\epsilon^{-1}$).\footnote{Note that one may allow functionals that depend on such transitions but admit a steady state. However, additional requirements on the specific form of the functional must be made. In the current example, choosing $g^2_1=-g^1_2$ would grant a steady state. More generally, the condition to be satisfied is that changes in the functional occurring while the system is within one absorbing block must always sum to zero.}
We can then proceed to evaluate the steady state and find it to be 
\begin{equation}
\mathsf{G^{(0)}}=q_a\tilde{w}_a+q_d\tilde{w}_d
\end{equation}
where $q_a$ and $q_d$ are generic positive functions and $\tilde{w}_a$, $\tilde{w}_d$ are the eigenvectors spanning the right null space of $\mathsf{M}_s$. They coincide (modulo the state reshuffling giving the block triangular structure) with the eigenvectors of the unperturbed generator $\mathsf{M}$ introduced in eq.~(\ref{eq:w}). 

At order $\epsilon^{0}$
\[
\frac{d\mathsf{G^{(1)}}}{d\theta} + \frac{d\mathsf{G^{(0)}}}{dt}
= \mathsf{M_s G^{(1)}} + \mathsf{H^{(0)} G^{(0)}} 
\]
the solvability condition is obtained by multiplying by the eigenvectors spanning the left nullspace of $\mathsf{M_s}$:
\begin{eqnarray}\label{eq:vs}
\tilde{v}^a_s= \left(\frac{F^1_3e^{-sg^1_3}}{F^1_3+(1-e^{-s(g^3_4+g^4_3)})F_3^4+F^5_3},\,e^{-sg^3_4}\frac{F^1_3e^{-sg^1_3}}{F^1_3+(1-e^{-s(g^3_4+g^4_3)})F_3^4+F^5_3},\,0,\,1,\,1,\,0\right)
\\\nonumber
\tilde{v}^d_s=
\left(\frac{F^5_3e^{-s(g^5_3+g^6_5)}}{F^1_3+(1-e^{-s(g^3_4+g^4_3)})F_3^4+F^5_3},\,e^{-sg^3_4}\frac{F^5_3e^{-s(g^5_3+g^6_5)}}{F^1_3+(1-e^{-s(g^3_4+g^4_3)})F_3^4+F^5_3},\, e^{-sg^6_5},\,0,\,0,\,1\right)
 \end{eqnarray}
 where it can be seen that, setting $s=0$, one recovers the left eigenvectors of the unperturbed generator as in eq.~(\ref{eq:v}) apart from the reordering of states.

The entries in \eqref{eq:vs} can be understood as the generating functions for the functional $J$ along the fast dynamics,
conditioned on absorption in a block.
Developing the fractions as geometric series one finds, for instance
\begin{equation}
\frac{F_3^1 e^{-s g_3^1}}{F_3^1+F_3^4(1-e^{-s(g_4^3+g_3^4)})+F_3^5} =
\frac{F_3^1}{F_3^1+F_3^4+F_3^5} \sum_{n=0}^\infty \left(\frac{F_3^4}{F_3^1+F_3^4+F_3^5} \right)^n e^{-s(g_3^1+n g_4^3 + n g_3^4)}
\end{equation}
On the right-hand side, the prefactor is the probability of exiting from 3 to 1, the factor within brackets is 
the probability of going from 3 to 4, $n$ is the number of transits from 3 to 4, which also equals the number of transits from 4 to 3 since  the probability of going from 3 to 4 is unity.  $g_3^1+n g_4^3 + n g_3^4$ is the value of the functional along a fast path that starts in 3 and ends in 1 after $n$ returns to 3.
The effective equation then reads 
\begin{eqnarray}\label{eq:generating_eff_gen}
\frac{dq_a}{dt}&=&\frac{F^1_3e^{-sg^1_3}}{F^1_3+(1-e^{-s(g^3_4+g^4_3)})F_3^4+F^5_3}
\left[\left(w^{a_1}S^3_1e^{-sg^3_1}+w^{a_2}S^4_2e^{-s(g^4_2+g^3_4)} \right)q_a+
e^{-s(g^4_6+g^3_4)}S^4_6q_d\right]\nonumber\\
&-&\left[w^{a_1}(S^3_1+sf_1)+w^{a_2}(S^4_2+sf_2) \right]q_a\nonumber\\
\frac{dq_d}{dt}&=&\frac{F^5_3e^{-s(g^5_3+g^6_5)}}{F^1_3+(1-e^{-s(g^3_4+g^4_3)})F_3^4+F^5_3}
\left[\left(w^{a_1}S^3_1e^{-sg^3_1}+w^{a_2}S^4_2e^{-s(g^4_2+g^3_4)} \right)q_a+
e^{-s(g^4_6+g^3_4)}S^4_6 q_d\right]\\ \nonumber
&+&e^{-s(g^5_6+g^6_5)}S^5_6q_d-(S^4_6+S^5_6+sf_6)q_d
\end{eqnarray}
Developing the denominator as geometric series as discussed above 
one finds that the effective equations can be written as follows
 \begin{eqnarray}\label{eq:generating_eff_gen_paths}
\frac{dq_a}{dt}&=& w^{a_1} S_1^3 e^{-s g_1^3} \sum_{\textit{path} \in A_{3\to a}} \mathrm{Prob}[path] e^{-s \Delta J[path]} q_a
+ w^{a_2} S_2^4 e^{-s g_2^4}\sum_{\textit{path} \in A_{4\to a}} \mathrm{Prob}[path] e^{-s \Delta J[path]} q_a
 \nonumber \\
& & 
+ S_6^4 e^{-s g_6^4}\sum_{\textit{path} \in A_{4\to a}} \mathrm{Prob}[path] e^{-s \Delta J[path]}  q_d
- w^{a_1} S_1^3 q_a
- w^{a_2} S_2^4  q_a - s (w^{a_1} f_1 +w^{a_2} f_2) q_a  \nonumber \\
\frac{dq_d}{dt}&=& w^{a_1} S_1^3 e^{-s g_1^3}\sum_{\textit{path} \in A_{3\to d}} \mathrm{Prob}[path] e^{-s \Delta J[path]} q_a
+ w^{a_2} S_2^4 e^{-s g_2^4}\sum_{\textit{path} \in A_{4\to d}} \mathrm{Prob}[path] e^{-s \Delta J[path]} q_a
 \nonumber \\
& & 
+ S_6^4 e^{-s g_6^4}\sum_{\textit{path} \in A_{4\to d}} \mathrm{Prob}[path] e^{-s \Delta J[path]}  q_d
+ S_6^5 e^{-s g_6^5}\sum_{\textit{path} \in A_{5\to d}} \mathrm{Prob}[path] e^{-s \Delta J[path]}  q_d
\nonumber \\
& & 
-  S_6^4 q_d
- S_6^5  q_d - s f_6 q_d
\end{eqnarray}
where by $A_{k\to a}$ we have defined the set of fast paths (i.e. sequences of states connected by fast transitions only) 
that join state $k$ to the absorbing block $a$   (resp. $d$). 

As an example, let us consider the paths that from $3$ are absorbed into $a$
\begin{equation} 
A_{3\to a} = \left\{  (3,1),(3,4,3,1),(3,4,3,4,3,1),\ldots,(\underbrace{ 3,4,\ldots,3,4}_{n\; \mathrm{times}},3,1) ,\ldots \right\}
\end{equation}
with probabilities 
\begin{equation}
\mathrm{Prob}\left[ (\underbrace{ 3,4,\ldots,3,4}_{n\; \mathrm{times}},3,1)\right] = \frac{F_3^1}{F_3^1+F_3^4+F_3^5}
 \left(\frac{F_3^4}{F_3^1+F_3^4+F_3^5} \right)^n
\end{equation}
and changes in value for the functional $J$
\begin{equation}
\Delta J \left[ (\underbrace{ 3,4,\ldots,3,4}_{n\; \mathrm{times}},3,1)\right] = n g_3^4 + n g_4^3 + g_3^1
\end{equation}
Note that
\[
\sum_{path \in A_{3\to a}} \mathrm{Prob}(path) + \sum_{path \in A_{3\to d}} \mathrm{Prob}(path)  = 1
\]

The interpretation of the effective process described by \eqref{eq:generating_eff_gen_paths} is now straightforward. 
Let us assume the process is in block $a$. With probability 
$w^{a_1} S_1^3 dt$ it can jump to visit $3$ but then it immediately reverts back to $a$ or moves to $d$.  Among all possible fast paths that connect $3$ to the absorbing states pick one with probability $\mathrm{Prob}(path)$, and add the corresponding $\Delta J[path]$ to the value of the functional. Independently, with probability $w^{a_2} S_2^4 dt$, it can jump to $4$ and then again eventually get back to $a$ or move to $c$ through the corresponding fast paths.
If there is no jump, the process stays in $a$ and adds   $(w^{a_1} f_1 +w^{a_2} f_2)dt$ to the functional. The reasoning is similar for block $d$.

\subsubsection{Algorithm for the effective process}
The procedure detailed above for a particular example in fact applies to all processes with fast and slow transitions and for general functionals $J$. Without repeating the calculations in general, here we provide an algorithm that generates the effective joint process for states and values of the functional:
\begin{enumerate}
\item[1] Identify the absorbing blocks as discussed for the dynamics in Section \ref{sec:dyn_discrete} and compute the steady-state distributions (right-nullspace of $M$). Check that the $g_j^k$ within each block are zero.
\item[2] Choose an initial condition for the effective process, say block $a$ and set $J=0$.
\item[3] Extract an exponentially distributed time $\tau$ with rate $\sum_{i,j} w^{a_i} S_{a_i}^{j}$.
\item[4] Add $\tau \sum_{i} f_{a_i} w^{a_i} $ to $J$ and move time ahead by $\tau$. 
\item[5] Single out a slow transition channel  $a_i \to j$ according to a probability $\pi_{a_i}^j = 
w_{a_i} S_{a_i}^j/\sum_{i,j} w_{a_i} S_{a_i}^j$. Add $g_{a_i}^j$ to $J$.
\item[6] If $j$ is outside $a$, pick a fast transition from $j$ to $k$ with the corresponding probability $ F_j^k/\sum_k F_j^k$. Add $g_j^k$ to $J$ and iterate this step until the path ends in an absorbing block. 
\item[7] Cycle back to 3 until desired 
\end{enumerate}
The remarkable property of this algorithm is that time advances only on slow time-scale. The only added cost for computation is the construction of the sequence of fast states from one absorbing block to another or to itself, with only one slow transition out of the initial block.
{  
\paragraph{Summary}
It is possible to derive the effective evolution of a functional on the slow scales but, in general, it cannot be written in terms of the effective rates and blocks that described the slow effective dynamics.
The characterization of the slow dynamics of a functional generally requires additional information. Namely, which specific channel was chosen to jump form one block to another and which path through eliminated states was taken.
}
 \subsection{ Functionals of block-diagonal fast dynamics}
 {  
\paragraph{The question} Which simplifications in the effective equation for a functional are brought about by a block-diagonal structure of the fast transitions?
}

   \begin{figure}[!h]
  \includegraphics[width=0.9\textwidth]{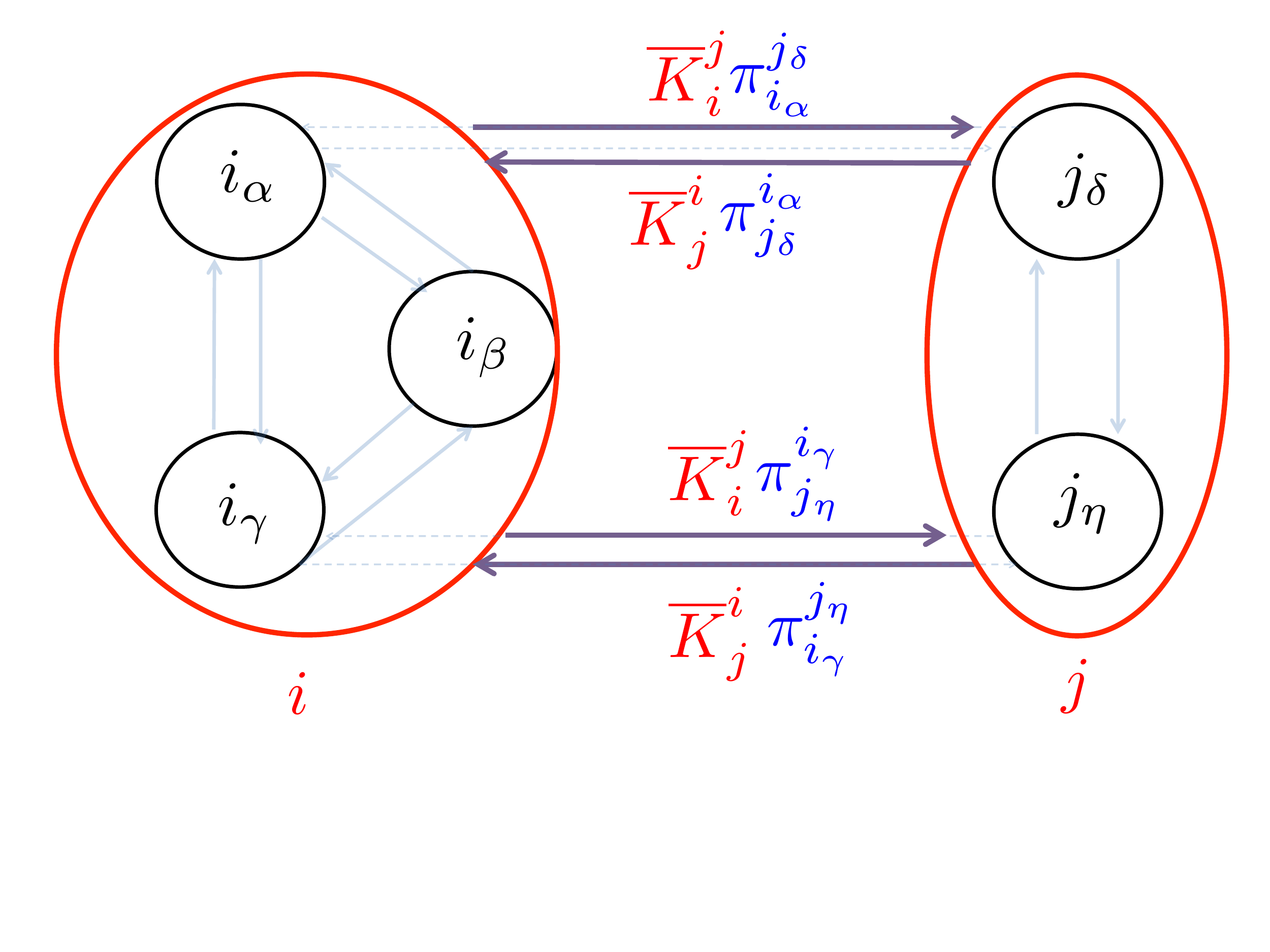}
\caption{\label{fig:blocks_func} 
Block-diagonal dynamics and transition channels. In order to correctly describe a generic functional of the trajectories the different transition channels and their probabilities must be duly accounted for.
}
\end{figure}

As we have seen when discussing the dynamics of discrete state Markov chains in section \ref{sec:dyn_block}, the presence of a block-diagonal 
structure for the fast transitions allows to identify blocks of states that are connected only
by slow transitions and simplifies the averaging procedure.
The same applies for the study of functionals of the dynamics.
We will focus on functionals that do not vary when transitions between fast
states within a block occur:
\begin{equation}
g^{i\alpha}_{i\beta} = 0 
\end{equation}
where, again, the latin indices label the block to which a state belong and the greek ones
the specific fast state within the block (see Fig.~\ref{fig:blocks_func}).
This implies that the fast part of the tilted generator describing the evolution of the generating function (eq.~\ref{eq:tilted_gen_gen}) is equal
to the generator of the dynamics
 $\mathsf{M}_s = \mathsf{M}$.
In this case, the stationary solution of the first order approximation to the generating function is
\begin{equation}
{G^{(0)}}^{i\alpha}_{m\mu} = q^i_m  w^{i\alpha} 
\end{equation}
At order $\epsilon^{0}$ the equation then reads
\begin{equation}
\frac{d\mathsf{G^{(1)}}}{d\theta} + \frac{d\mathsf{G^{(0)}}}{dt}
= \mathsf{M G^{(1)}} + \mathsf{H^{(0)} G^{(0)}}\;.
\end{equation}
Due to the simple structure of the left null-space of the fast operator in the block-diagonal case (see eq. \ref{eq:block_leftev}), the solvability condition is obtained by summing over the fast states
\begin{equation}\label{eq:block_eff_gen}
\frac{d q^i_m}{dt} 
= \sum_{j,\alpha,\beta} \overline{K}^i_j \pi^{i\alpha}_{j\beta} \exp(-s g^{i\alpha}_{j\beta}) q^j_m  - \left( \sum_k \overline{K}^k_i + s \overline{f}_{i} \right) q^i_m
\end{equation}
where 
\[
\pi^{i\alpha}_{j\beta} = \frac{S^{i\alpha}_{j\beta} 
w^{j\beta}}{\overline{K}^{i}_{j} } 
\] 
are positive normalized weights ($\sum_{\alpha,\beta} \pi^{i\alpha}_{j\beta} =1$ for all $i,j$) representing 
the {\it conditional probability}  of the specific transition from $i\alpha$ to $j\beta$ 
given that a transition from block $i$ to block $j$ has occurred, in the limit $\epsilon \to 0$.

Inverting the Laplace transform, one obtains the effective master equation for the 
probability of being in state $i$ given the initial state $m$, with a value $J$ of the functional
reads
\begin{equation}\label{eq:block_eff_mef}
\frac{d P^i_m(J)}{dt} 
= -\frac{\partial}{\partial J} \overline{f}_{i} P^{i}_{m}(J) + \sum_{j,\alpha,\beta}\left( \chi^{i\alpha}_{j\beta}  P^j_m (J-g_{j\beta}^{i\alpha}) -
\chi^{j\beta}_{i\alpha}  P^{i}_{m}(J) \right)  
\end{equation}
where the rate for a jump $g^{i\alpha}_{j\beta}$ in the value of the functional is
\[
\chi^{i\alpha}_{j\beta} = \overline{K}^i_j \pi^{i\alpha}_{j\beta}=S^{i\alpha}_{j\beta} 
w^{j\beta}
\]
Integrating over $J$ (or equivalently setting $s=0$ in the effective equation for the generating function)  one recovers the effective kinetics on slow states.\\
Expressions \eqref{eq:block_eff_gen} and \eqref{eq:block_eff_mef}
involve quantities depending on the specific fast states ($\pi^{i\alpha}_{j\beta}$, $\chi^{i\alpha}_{j\beta}$) and, hence, 
 clearly show how it is impossible, in general, to express
the effective evolution of a functional only in terms of functions of the states
of the effective dynamics (in this case the blocks).
However, this major short-coming can be overcome by considering
that for such block-diagonal systems the only additional information
required is the conditional probability of choosing a specific channel between
blocks as depicted in Figure~\ref{fig:blocks_func}.\\
More specifically, we investigate the possibility of  expressing the functional 
in terms of slow states and transitions only as
\begin{equation}
{\cal J}_{t,t'}  \xrightarrow{\epsilon \to 0} \sum_{k=1}^{N+1} \int_{\tau_{k-1}}^{\tau_k}  d\tau \overline{f}_{i_{k-1}}  + \sum_{k=1}^{N} \overline{g}^{i_k}_{i_{k-1}}
\end{equation}
with some suitable $\overline{g}$ and $\overline{f}$ depending on the block variables only.
As shown in Ref.~\cite{Bo2014}, a sufficient but obvious condition
 for a generic functional is that $g^{i_\alpha}_{j_\beta}=g^i_j$ i.e., the transition dependent part does not
depend on fast states. 

In general, however, one has the effective equation for the generating function
\begin{equation}\label{eq:disc_entro_eff}
\frac{d q^i_m}{dt} 
= \sum_{j} \overline{K}^i_j  \exp(-s\overline{g}^i_j)  q^j_m  - \sum_k \overline{K}^k_i  q^i_m 
+ \sum_j C^i_j q^j_m\
\end{equation}
where the additional contribution features the matrix
\begin{equation}\label{eq:disc_entro_an}
C^i_j = \sum_{\alpha,\beta} S^{i_\alpha}_{j_\beta} w^{j_\beta} \left[\exp(-s g^{i_\alpha}_{j_\beta}) - \exp(-s \overline{g}^i_j) \right]
 \end{equation}
which need not vanish. This means that on slow scales the effective evolution of the functional depends not only on the slow states and transitions but also on the specific fast states
at which the slow transition took place. Consequently, for a correct description of generic functionals
more details are needed than the ones available (and sufficient) from the effective evolution of the dynamics on
the slow scales (see Fig.~\ref{fig:blocks_func}). 
 {  
\paragraph{Summary}
When fast transitions are block-diagonal, there are no decimated blocks.
As a consequence, the only additional detail required to specify the evolution of a functional
is which transition channel was chosen when a jump between two blocks took place.
}
\subsubsection{Counting statistics of transitions across blocks}
An interesting functional to consider is the counting statistics of a single specified transition 
$l\lambda \to n\nu$, i.e. $f=0$ and $\hat{g}^{i\alpha}_{j\beta} = \delta_{j\beta}^{l\lambda} 
\delta^{i\alpha}_{n\nu}$ one has $\exp(-s\hat{g}^{i\alpha}_{j\beta}) = 1 +
(\exp(-s)-1)\delta_{j\beta}^{l\lambda} 
\delta^{i\alpha}_{n\nu}$ and the effective equation for the generating function in the limit
$\epsilon \to 0$ is
\[
\frac{d \hat{q}^i_m}{dt} 
= \sum_{j} \overline{K}^i_j \left(\sum_{\alpha,\beta} \pi^{i\alpha}_{j\beta} \exp(-s \hat{g}^{i\alpha}_{j\beta}) \right) \hat{q}^j_m  - \sum_j \overline{K}^j_i  \hat{q}^i_m
=  \sum_{j} \overline{K}^i_j \hat{q}^j_m  - \sum_j \overline{K}^j_i  \hat{q}^i_m
+\overline{K}^n_l (\exp(-s)-1) \pi^{n\nu}_{l\lambda} \hat{q}^l_m \delta^i_n\;.
\]
As expected, one needs to retain additional information about the
fast states.
In this case, simply knowing $\pi^{n\nu}_{l\lambda}$, {\it i. e.}, the conditional probability of choosing the channel
connecting state   $l\lambda$ to state $n\nu$ once a transition between blocks $l$ and $n$ has occurred, is sufficient to provide the correct counting statistics in the limit $\epsilon\to 0$.
For instance, the average effective transition rate ${n\nu} \to {l\lambda}$ then follows from 
\[
\frac{d}{dt} \left\langle \#^{n\nu}_{l\lambda} \right\rangle = \frac{d}{dt} \sum_i 
\left(-\frac{d\hat{q}^i_m}{ds} \right)_{s=0} = \overline{K}^n_l \pi^{n\nu}_{l\lambda} \phi^l_m 
= S_{l\lambda}^{n\nu} w^{l\lambda} \phi^l_m 
\]
and the overall average transition rate $l\to n$ is
\[
\frac{d}{dt} \left\langle \#^{n}_{l} \right\rangle =
\sum_{\lambda,\nu} S_{l\lambda}^{n\nu} w^{l\lambda} \phi^l_m =
\overline{K}^n_l  \phi^l_m 
\]

 {  
\paragraph{Summary}
For block-diagonal fast transitions, the evolution of the counting statistics of transitions across blocks can be found if, in addition of the effective rates connecting the blocks, one has access to the probability of choosing a specific transition channel.
Counting the transitions between blocks
 (irrespective of the specific states within the block)  can be fully expressed in terms of the effective rates connecting the blocks of the effective dynamics.
}

\subsection{Entropy for block-diagonal dynamics}\label{sec:entro_block}
{  
\paragraph{The question}
Is it possible to express entropy production in terms of the rates and blocks describing the effective dynamics on the slow scale?
}\\ 
Another interesting functional of  a stochastic trajectory is the entropy production.
For its definition we shall introduce the concept of the time reversed trajectory, defined  as
\[
i^*(\tau) = i(t+t'-\tau) 
\]
where
we note that the residence time spent in each state is the same as for the forward trajectory.
The entropy production of a trajectory is then defined as the log-ratio
 of the probability of observing that forward trajectory ${\cal P}$ compared to the backward one ${\cal P^*}$ (see e.g. \cite{seifert2012review}):
\begin{equation}\label{eq:entro_dys_def}
{\cal S} =\log \frac{\cal P}{\cal P^*}=-\log p^{i_t}_{t} + \log p^{i_{t'}}_{t'} +{\cal A^* -A}=  \underbrace{-\log p^{i_t}_{t} + \log p^{i_{t'}}_{t'}}_{\Delta {\cal S}_p }+\underbrace{\sum_{k=1}^{N} \log \frac{K^{i_{k}}_{i_{k-1}}}{K^{i_{k-1}}_{i_{k}}}}_{{\cal S}^{env}}
\end{equation}
where $p^{i_t}_{t}$ is the probability of starting in state $i_t$ at the initial time $t$ and
${\cal A}$ is the action as defined in eq.~\eqref{eq:action}.

The average rate of entropy production at time $t$ can be written as
\begin{equation}\label{eq:entro_aaa}
 \langle \dot{\cal S}\rangle=\sum_{i\,j}p_iK_i^j \log{\frac{p_iK_i^j}{p_jK_j^i}}\;.
\end{equation}
We see that the condition of detailed balance
\begin{equation}
 p^{eq}_iK_i^j=p_j^{eq}K_j^i
\end{equation}
corresponding to requiring that, at the steady state, each term of the sum in the 
master equation (\ref{eq:discr_me}) vanishes, implies that the system is at equilibrium and that no 
entropy is produced. Furthermore, at detailed balance and at the steady state (equilibrium) there probability fluxes between states:
\begin{equation}\label{eq:def_flux}
 J_i^j= p_iK_i^j-p_jK_j^i
\end{equation}
must be equal to zero.\\

Here we will derive the effective evolution of the entropy production on the slower scales as an illustration of the case for general functionals recovering the results of \cite{Bo2014}. Let us first recall that also
Ref.~ \cite{esposito2012stochastic}  considered the average entropy production
of  a system involving two time scales. The author showed that an entropy defined using only the effective rates would underestimate the effective entropy.
Indeed,  the effective entropy produced on the slower scales displayed a non-negative correction.
The presence of the additional term was anticipated also in \cite{santillan2011irreversible}.
The authors of Ref.~\cite{rahav2007fluctuation} showed that the effective
entropy production satisfies an approximate fluctuation relation that becomes exact for infinite time-scale separation (as $\epsilon\to 0$).
Entropy for non block-diagonal fast dynamics was investigated in
 \cite{puglisi2010entropy} that discussed  the effect of decimation and in
\cite{Altaner2012} which presented a different approach to decimation which ensured that the correct statistics of the dynamics and of its functionals are preserved.\\
 We shall present here this issue for the complete generating function (hence not restricting to the average) within the framework for general functionals highlighted in the previous section. 
For the entropy production in the environment one has
\[
{\cal S}^{env}_{t,t'} = \sum_{k=1}^{N}\left(1-\delta^{i_k}_{i_{k-1}}\right) \log \frac{S^{i_{k}\alpha_{k}}_{i_{k-1}\alpha_{k-1}}}{S^{i_{k-1}\alpha_{k-1}}_{i_{k}\alpha_{k}}} 
+ \delta^{i_k}_{i_{k-1}} \log \frac{F^{i_{k}\alpha_{k}}_{i_{k-1}\alpha_{k-1}}}{F^{i_{k-1}\alpha_{k-1}}_{i_{k}\alpha_{k}}} 
\]
that can be rewritten as
\[
{\cal S}^{env}_{t,t'} = \log w^{i_t\alpha_t} - \log w^{i_{t'}\alpha_{t'}} +
\underbrace{\sum_{k=1}^{N-1}\left(1-\delta^{i_k}_{i_{k-1}}\right) \log \frac{S^{i_{k}\alpha_{k}}_{i_{k-1}\alpha_{k-1}}w^{i_{k-1}\alpha_{k-1}}}{ S^{i_{k-1}\alpha_{k-1}}_{i_{k}\alpha_{k}}w^{i_{k}\alpha_{k}}}  
+ \delta^{i_k}_{i_{k-1}} \log \frac{F^{i_{k}\alpha_{k}}_{i_{k-1}\alpha_{k-1}}w^{i_{k-1}\alpha_{k-1}}}{ F^{i_{k-1}\alpha_{k-1}}_{i_{k}\alpha_{k}}w^{i_{k}\alpha_{k}}} 
}_{{\cal J}_{t,t'}}
 \]
 
As discussed before for generic functionals, the solvability condition at order $\epsilon^{-1}$ 
requires $g^{i\alpha}_{j\beta} = 0$, which amounts to
\[
\frac{F^{i\alpha}_{i\beta}w^{i\beta}}{ F_{i\alpha}^{i\beta}w^{i\alpha}}=1
\]
i.e. detailed balance must hold for the fast processes.
In this case
\[
{\cal S}^{env}_{t,t'} = \log w^{i_t\alpha_t} - \log w^{i_{t'}\alpha_{t'}} +
\underbrace{ \sum_{k=1}^{N-1}\left(1-\delta^{i_k}_{i_{k-1}}\right) \log \frac{S^{i_{k}\alpha_{k}}_{i_{k-1}\alpha_{k-1}}w^{i_{k-1}\alpha_{k-1}}}{ S^{i_{k-1}\alpha_{k-1}}_{i_{k}\alpha_{k}}w^{i_{k}\alpha_{k}}}  }_{{\cal J}_{t,t'}}
 \]
so that we can identify 
\begin{equation}
g^{i\alpha}_{j\beta} = \log \frac{ S^{i\alpha}_{j\beta} w^{j\beta} }{ S^{j\beta}_{i\alpha} w^{i\alpha} }
\end{equation}
(where it is understood that $g^{i\alpha}_{i\beta}=0$).
Notice that only transitions between slow states are counted as required by the solvability
condition at order $\epsilon^{-1}$.

From the 
result of the general section we do not expect to be able
to express the evolution of the effective functional in terms of only of the block variables and
know that there should be an additional term as expressed in \eqref{eq:disc_entro_an}.
To proceed to the study of such contribution we need to explicit 
 the entropy production functional that one would
define starting from the effective dynamics.
This is given by
\begin{equation}
{\cal S}_{slow} = -\log \phi^{i_t}_{t} + \log \phi^{i_{t'}}_{t'} +
\sum_{k=1}^{N} \log \frac{\overline{K}^{i_{k}}_{i_{k-1}}}{\overline{K}^{i_{k-1}}_{i_{k}}}
\end{equation}
which obeys the fluctuation theorem (see Refs. \cite{rahav2007fluctuation,Bo2014})
\begin{equation}
\langle 
e^{-{\cal S}_{slow}} \rangle =1\;.
\end{equation}
Then the related weight assigned to the slow transition reads
\begin{equation}
\overline{g}^i_j= \log \frac{\overline{K}^i_j}{\overline{K}^j_i} =
-\log \left( \sum_{\alpha\beta} \pi^{i_\alpha}_{j_\beta} \exp(-g^{i_\alpha}_{j_\beta})\right)\;.
\end{equation}
We can then read from eq.~\eqref{eq:disc_entro_an} the additional
contribution:
\[
C^i_j = \sum_{\alpha,\beta} S^{i\alpha}_{j\beta} w^{j\beta} \left[\exp(-s g^{i\alpha}_{j\beta}) - \exp(-s \overline{g}^i_j) \right]
= \overline{K}^i_j \left[\sum_{\alpha,\beta} \pi^{i\alpha}_{j\beta} 
\left(\frac{S^{j\beta}_{i\alpha} w^{i\alpha}}{S^{i\alpha}_{j\beta} w^{j\beta}}\right)^s
-\left(\frac{\overline{K}^j_i}{\overline{K}^i_j}\right)^s\right]
 \]
 \[
 = \overline{K}^i_j \left(\frac{\overline{K}^j_i}{\overline{K}^i_j}\right)^s  \left[ 
 \sum_{\alpha,\beta} \pi^{i\alpha}_{j\beta} 
 \left(\frac{\pi^{j\beta}_{i\alpha}}{\pi^{i\alpha}_{j\beta}}
 \right)^s-1
 \right]
 \]

Such additional contribution disappears ({\it i. e.}, it is possible to describe the effective evolution
of the 
the functional in terms of the effective dynamics alone)
 if and only if 
\begin{equation}\label{eq:disc_cond_reg1}
\frac{ S^{i_\alpha}_{j_\beta} w^{j_\beta} }{ S^{j_\beta}_{i_\alpha} w^{i_\alpha} }= \frac{\overline{K}^i_j}{\overline{K}^j_i}
\end{equation}
 i.e. 
\begin{equation}\label{eq:disc_cond_reg2}
\pi^{i_\alpha}_{j_\beta} = \pi^{j_\beta}_{i_\alpha}\;.
\end{equation}
A sufficient condition is that the full dynamics obeys detailed balance as shown also in \cite{santillan2011irreversible}.
Indeed, in this case
\begin{equation}
\frac{ S^{i_\alpha}_{j_\beta}}{ S^{j_\beta}_{i_\alpha} }= \frac{\rho^{i_\alpha}}{\rho^{j_\beta}}
\end{equation}
where $\rho$ is the equilibrium distribution for the full system. Then, in the limit $\epsilon \to 0$
\begin{equation}
 \frac{\rho^{i_\alpha}}{\rho^{j_\beta}} \to \frac{\varphi^{i}w^{i_\alpha}}{\varphi^{j}w^{j_\beta}} 
\end{equation}
where $\varphi$ is the equilibrium distribution of the slow effective dynamics and therefore obeys
\begin{equation}
\frac{\overline{K}^i_j}{\overline{K}^j_i} = \frac{\varphi^i}{\varphi^j}
\end{equation}
by detailed balance. Together with the two previous relations, this shows that the regularity condition
\eqref{eq:disc_cond_reg1} is satisfied. Therefore, at detailed balance, the limit of entropy production is regular and
the effective Markov process with rates $\overline{K}^i_j$ captures the
correct entropy in addition to the correct kinetics. \\
Another sufficient condition is that there exists only one allowed transition between states $j$ and $i \neq j$ and back
. 
In this case trivially $\pi^i_j=\pi^j_i=1$, and the limit of the entropy production is regular as well.

Introducing the anomalous entropy production defined as the difference
between the effective entropy production and the entropy defined from the effective dynamics equals:
\begin{equation}\label{eq:anomaly_discrete}
{\cal S}_{anom} = \lim_{\epsilon\to 0}{\cal S}-{\cal S}_{slow}=\sum_{k=1}^{N}\log \frac{ S^{i_k\alpha_k}_{i_{k-1}\alpha_{k-1}} w^{i_{k-1}\alpha_{k-1}} \overline{K}^{i_{k-1}}_{i_{k}}}{ S^{i_{k-1}\alpha_{k-1}}_{i_k\alpha_k} w^{i_k\alpha_k}\overline{K}^{i_{k}}_{i_{k-1}} }
= \sum_{k=1}^{N} \log \frac{\pi^{i_k\alpha_k}_{i_{k-1}\alpha_{k-1}}}{\pi^{i_{k-1}\alpha_{k-1}}_{i_{k}\alpha_k}}
\end{equation}
It follows that
\[
\exp\left[-{\cal S}_{anom}\right] = \frac{\prod_{k=1}^N \pi^{i_{k-1}\alpha_{k-1}}_{i_{k}\alpha_k} }
{\prod_{k=1}^N \pi^{i_k\alpha_k}_{i_{k-1}\alpha_{k-1}} }
\]
is the backward/forward ratio of conditional probabilities of fast states at a given  sequence of
slow states. The anomalous contribution, based on the probability of choosing transition channel once
a transition between block has occurred, accounts for the possible cycles that where collapsed
in the effective rates description.

The generating function of the anomalous entropy term obeys
\[
\frac{d \tilde{q}_m^i}{dt} 
= \sum_{j} \overline{K}^i_j \tilde{q}_m^j  - \sum_j \overline{K}^j_i  \tilde{q}_m^i 
+ \sum_j A^i_j \tilde{q}_m^j
\]
which gives for $s=1$ (since $A^{i}_j \bigr|_{s=1}=0$)
\[
\left\langle e^{-{\cal S}_{anom}} \right\rangle=\sum_i \left. \tilde{q}_m^i \right|_{s=1}= 1
\]
and
\begin{equation}\label{eq:disc_anom}
\frac{d}{dt} \left\langle {\cal S}_{anom} \right\rangle = 
\sum_{i,j} \phi^j_m \overline{K}^i_j D\left({\pi^{i\alpha}_{j\beta} }\Bigl|\Bigr| {\pi^{j\beta}_{i\alpha}}\right) \ge 0
\end{equation}
where $D$ is the Kullback-Leibler divergence, one recovers the results of
Ref.~\cite{esposito2012stochastic} .
We have then seen how the effective dynamics may miss relevant parts of the entropy production
thereby hiding some of the system irreversibility and dissipation.
The authors of Ref.~\cite{Wang2016} showed
that systems subject to fast drivings
display a violation of the fluctuation-response relation
and evaluated its typical shape in the space of frequencies
for large time-scale separations.
Such violation of the fluctuation response relation is known
to be linked to entropy production for Langevin systems
by the Harada-Sasa equality \cite{Harada2005}.
Investigating the connections between such violation
of the fluctuation response relation in systems
 with time-scale separation
 and the anomalous entropy production represents an interesting future perspective.
  {  
\paragraph{Summary}
When the original system does not obey detailed balance the entropy defined from the effective dynamics underestimates the full one. To recover the correct limit, it is necessary to include an additional contribution related to the different channels connecting the blocks. To compute the correct average entropy production it is sufficient to know what is the conditional probability
of choosing a specific channel once a transition between two blocks has occurred. 
}
  \subsection{Some examples of coarse-graining of stochastic systems not based on time-scale separation} 
  Before moving to some specific examples we shall briefly report on some studies
  that in a similar spirit to what we have shown here contrast 
  the limiting behavior of coarse-grained functionals to the value
  they would take if defined from the coarse grained dynamics.
   In contrast to the main focus of the present report the coarse-graining considered
   in these studies does not rely on time-scale separation.  
    In Ref.~\cite{Horowitz2015} the author considered how approximating a chemical master equation in terms of a diffusive process (either as a chemical Langevin equation~\cite{gillespie2000chemical} or as a system size expansion~\cite{van1992stochastic}) affects its entropy production. He showed that, away from equilibrium, the entropy production of the diffusive approximations is unrelated to the
  original one of the chemical master equation. The authors of Ref.~\cite{Zimmermann2015} considered the case of a colloidal probe attached to a molecular motor and derived a coarse grained description of the motor dynamics eliminating the probe. 
    Such elimination involves continuous (probe position) and discrete
     (motor states) degrees of freedom and does not rely on a time-scale separation.
    In their study, a careful definition of the effective rates ensured the possibility to
    express the average entropy production in terms of the effective rates.
    Intuitively, this can be linked to the fact that  there are no cycles in the full phase-space that are disrupted by the coarse-graining. 
\subsection{Example for entropy production in discrete systems}
 \subsubsection{Irreversibility in two-component systems}\label{sec:thermo_discrete_ex}
Let us consider the thermodynamics of the biochemical sensing system described in section \ref{sssec:signal}.
We will show how, with the results of the previous section, we are able to 
derive the leading order contribution to
the entropy production of the system when the receptor is much faster than the phosphorylation dynamics of the proteins and we shall compute its average value.
Recall that since we are considering the case of a single receptor
 and no spontaneous (de-)phosphorylation, each
transition between states of the Markov chain corresponds to a single chemical reaction.
This ensures that the entropy production of the Markov chain (defined in \ref{eq:entro_dys_def}) captures
the ones produced in the chemical reactions \cite{esposito2010three}.
The effective equation for the dynamics of the proteins derived in (\ref{eq:eff_general}) satisfies detailed balance and,  at the steady state, reaches equilibrium.
This is not surprising since the reduced network is one-dimensional and cannot display cycles.
Since the effective equation is at equilibrium, at the steady state, its associated entropy production is zero.
However, we know from the initial dynamics that the full system is out of equilibrium and produces entropy.
With the results from eq.~(\ref{eq:anomaly_discrete}) we can compute the statistics
 of this missing contribution without having to solve the full dynamics.
 The key idea is to retain information about the different channels
 connecting the effective slow states.
 We note in passing that, based on the observation that
 when concerning chemical {   systems} one should account separately for the different chemical reaction
 connecting states of the Markov chain (see \cite{esposito2010three}),
 the authors of \cite{Wang2015}, in a model formally equivalent to the one considered here, exploited
 time separation to simplify the dynamics and derived the correct effective average entropy production.
 For the sake of illustration we shall compute the average entropy production
 of the system described in  section \ref{sssec:signal}.
To
  evaluate the average anomaly we need
to write the conditional probability of undergoing a specific transition between two fast states for a given slow transition:
 \begin{equation}
\pi^{i_\alpha}_{j_\beta} = \frac{S^{i_\alpha}_{j_\beta} 
w^{j_\beta}}{\overline{K}^{i}_{j} } \,.
\end{equation} 
As shown in figure \ref{fig:rec_prot} and in eq. (\ref{eq:master_multiscale})
the only admissible transitions
 between slow states are the (de-)phosphorylation ones which connect states with the same receptor state
  \begin{eqnarray}\label{eq:pi_general}
\pi^{x,ON}_{x+1,ON} = 
\frac{p(ON)k_d^* }{\barred[K]_d}=\frac{p(ON)k_d^* }{p(ON)k_d^*+ p(OFF)k_d }\qquad
\pi^{x+1,ON}_{x,ON} = 
\frac{p(ON)k_p^* }{\barred[K]_p}=\frac{p(ON)k_p^* }{p(ON)k_p^*+ p(OFF)k_p }\\\nonumber
\pi^{x,OFF}_{x+1,OFF} =
\frac{p(OFF)k_d }{\barred[K]_d}=\frac{p(OFF)k_d }{p(ON)k_d^* +p(OFF)k_d }\qquad
\pi^{x+1,OFF}_{x,OFF} = 
\frac{p(OFF)k_p }{\barred[K]_p}=\frac{p(OFF)k_p }{p(ON)k_p^*+ p(OFF)k_p }
\end{eqnarray} 
We can now proceed to the evaluation of  the average anomalous entropy production:
\begin{eqnarray}
\frac{d}{dt} \left\langle {\cal S}_{anom} \right\rangle = \sum_{x\;x'}\phi_{x'}\barred[K]^{x}_{x'}
\sum_{y\;y'}\pi^{x,y}_{x',y'}\log\frac{\pi^{x,y}_{x',y'}}{\pi^{x',y'}_{x,y}}\;,
\end{eqnarray}
and carry out the sum over the slow states $x'$ and $y,\,y'$ which gives
\begin{eqnarray}
\frac{d}{dt} \left\langle {\cal S}_{anom} \right\rangle &=&  
 \sum_{x}\biggr[(x+1)\phi_{x+1}\barred[K]_d
  \left(\frac{k_dp(OFF)}{\barred[K]_d}\log\frac{k_d\barred[K]_p}{k_p\barred[K]_d}+
  \frac{k^*_dp(ON)}{\barred[K]_d}\log\frac{k^*_d\barred[K]_p}{k^*_p\barred[K]_d}
  \right)\\\nonumber
 &-& (N-x)\phi_{x}\barred[K]_p \left(\frac{k_pp(OFF)}{\barred[K]_p}\log\frac{k_d\barred[K]_p}{k_p\barred[K]_d}+
  \frac{k^*_pp(ON)}{\barred[K]_p}\log\frac{k^*_d\barred[K]_p}{k^*_p\barred[K]_d}
  \right)\biggr]\;.
  \end{eqnarray}
The slow effective dynamics obeys detailed balance so that at the steady state for $(x+1)\phi_{x+1}\barred[K]_d=(N-x)\barred[K]_p\phi_x$.
Exploiting this property and taking the sum over $x$ one 
gets
\begin{eqnarray}
\frac{d}{dt} \left\langle {\cal S}_{anom} \right\rangle =
 N\frac{\barred[K]_p\barred[K]_d}{\barred[K]_p+\barred[K]_d}
  \left[p(OFF)\left(\frac{k_d}{\barred[K]_d}-\frac{k_p}{\barred[K]_p}\right)\log\frac{k_d}{k_p}+
 p(ON)\left(\frac{k^*_d}{\barred[K]_d}-\frac{k^*_p}{\barred[K]_p}\right)\log\frac{k^*_d}{k^*_p}  \right]\;.
\end{eqnarray}
By recalling the explicit expression of the effective rates one can rewrite the result as:
\begin{eqnarray}
\frac{d}{dt} \left\langle {\cal S}_{anom} \right\rangle =
N \frac{\barred[K]_p\barred[K]_d}{\barred[K]_p+\barred[K]_d}
  \left[
 p(ON)\left(\frac{k^*_d}{\barred[K]_d}-\frac{k^*_p}{\barred[K]_p}\right)\log\frac{k^*_dk_p}{k^*_pk_d}  \right]
\end{eqnarray}
and finally obtain
\begin{eqnarray}
\frac{d}{dt} \left\langle {\cal S}_{anom} \right\rangle =
N \frac{p(ON)p(OFF)}{\barred[K]_p+\barred[K]_d}
  \left[
 \left(k^*_dk_p-k^*_pk_d\right)\log\frac{k^*_dk_p}{k^*_pk_d}  \right]\;.
\end{eqnarray}

This result gives the entropy production up to corrections of order $\epsilon$.
We remark that since the regular contribution to entropy vanishes in this case  if we had defined the thermodynamics from the effective dynamics we would have obtained a vanishing entropy production overlooking the dissipation connected to the eliminated receptors' states.
The only sets of parameters for which the computed entropy production vanishes are the ones where $k_p^*/k_d^* = k_p /k_d $ which correspond to the detailed balance
condition on the rates of the full system.
At the steady state this implies that the full system is at equilibrium. This confirms the result we have obtained in the general
case stating that if the full system obeys detailed balance there is no anomalous entropy production.
 {  
\paragraph{Summary}
As we have seen before, two-components systems with fast receptor dynamics admit an effective kinetics in terms of the number of phosphorylated proteins only.
The effective system, considering only the blocks (number of phosphorylated proteins), is apparently at equilibrium and  therefore has a vanishing entropy production. The original system, however,  does not obey detailed balance and at the steady state produces entropy with a finite rate. Such a rate can be correctly estimated if, in addition to the effective rates,  one also has access to the probability that, given a phosphorylation event, the reaction involved an active or an inactive receptor.
}

 \section{The diffusive case}
{  
\paragraph{The question}
We have seen under which conditions an effective diffusive dynamics can be derived for a diffusive system involving a fast and a slow scale. Is it possible to obtain an effective dynamics for a functional of the trajectories of the  initial system? Can such effective evolution be expressed in terms of the effective dynamics derived in the previous section?
}
 \subsection{Averaging for functionals. Diffusive dynamics}\label{sec:diff_func}
 We now focus on the study of functionals of the diffusive trajectory of
 a system involving two well separated time scales, described by Eqs.~\eqref{eq:cont_slow} and \eqref{eq:cont_fast}
 that we report here for convenience
 \begin{equation*}
dX^i_t = {u}^i(X_t,Y_t,t) dt + \beta^{ij}(X_t,Y_t,t) \cdot dB^j_t\;
\end{equation*}
\begin{equation*}
dY^a_t = \epsilon^{-1}{z}^a(X_t,Y_t,t) dt + \epsilon^{-1/2}\sigma^{ab}(X_t,Y_t,t) \cdot d\hat{B}_t^b
\end{equation*}

 Our interest is in eliminating the faster degrees of freedom and 
 and obtaining an effective description on the slower time scales as
 done for the dynamics.
 We will see that, as shown for the discrete case, it is not always possible 
 to express the effective evolution of a functional solely as a functional of the effective dynamics
 but that additional details of the full process must be retained. 
 
 A possible way of approaching such problem is to perform the averaging procedure 
 discussed for the dynamics in section \ref{sec:dyn_diff} on the Feynman-Kac equation
 governing the evolution
 of the generating function of the functional of interest.
 Such approach would be similar to the one we have adopted for the discrete case
 and is the one followed, for instance, in Ref.~\cite{Celani2012}.  
Here, we shall present an alternative method, generalizing the one exploited in Ref.~\cite{Bo2014}.  
Consider a general functional of the fast and slow processes:
\begin{equation}\label{eq:def_a}
Z = F(t,X_t,Y_t)-F(t',X_{t'},Y_{t'}) + \underbrace{\int_{t'}^t h(\tau,X_\tau,Y_\tau) d\tau
+ r_i(\tau,X_\tau,Y_\tau) \cdot dX^i_\tau}_{A} + 
\underbrace{\int_{t'}^t f_a(\tau,X_\tau,Y_\tau) \cdot dY^a_\tau}_{B} 
\end{equation}
where all explicit time dependencies are $O(1)$ and $A$ evolves along the slow process
$X$ and $B$ follows the fast one $Y$.
Notice that this representation is not unique. For example,
if $f_a=\frac{\partial \phi}{\partial y^a}$ the $B$ term can be removed by noticing that 
$f_a \cdot dY^a =f_a \circ dY^a -\frac{1}{2}g^{ab}\frac{\partial f_a}{\partial y_b} d\tau
 = d \phi - \frac{\partial \phi}{\partial \tau}d\tau -   \frac{\partial \phi}{\partial x^i}\circ 
 dX^i -\frac{1}{2}g^{ab}\frac{\partial^2 \phi}{\partial y_a \partial y_b} d\tau$. Therefore, in the following we shall assume that $f_a$ has a rotational part only, i.e.
 it is an $m-$dimensional curl.
 {  
 We restrict to the case in which $f$, $r$ and $h$ are of order 1. In general, they may 
have different scalings in powers of $\epsilon$, as in, for example, the large deviations study of Ref.~\cite{Bouchet2016}.
  }

The boundary terms statistics is, to lowest order,
determined by the stationary distribution of the fast variables at given slow ones 
\begin{equation}
\langle \delta(F(t,x,Y_t)- F_*) \rangle \xrightarrow{\epsilon\to 0}\int dy \, w(y) \delta(F(t,x,y)- F_*)
\end{equation}
so that we can treat them separately.
In other words, they become random variables whose distribution depends parametrically
on the slow variables.
 The key idea of the present aproach is to consider $A$ and $B$ as stochastic variables and study the generator of the joint process $(X_t,A_t,Y_t,B_t)$:
\begin{equation}\label{eq:generator}
H = \underbrace{L_0 + (h+r_i u^i)\frac{\partial}{\partial A} + 
r_i d^{ij} \frac{\partial^2}{\partial A\partial x^j} + \frac{1}{2} d^{ij} r_i r_j 
\frac{\partial^2}{\partial A^2} }_{H_0} +
\epsilon^{-1} \underbrace{\left[ M + f_a z^a \frac{\partial}{\partial B} + 
f_a g^{ab} \frac{\partial^2}{\partial B \partial y^b} +
\frac{1}{2}f_a f_b g^{ab} \frac{\partial^2}{\partial B^2} \right]}_K
\end{equation}
where $L_0$ and $M$ are the generators of the slow and fast dynamics as in eq.~(\ref{eq:cont_slow}), (\ref{eq:cont_fast}). 
The intermediate steps are provided in appendix \ref{appendix:generator}.
We can then address the problem of the averaging of the functional in terms 
of the
 adiabatic elimination of fast $(Y_t,B_t)$ variables with slow
$(X_t,A_t)$ as done in section \ref{sec:dyn_diff}.
\subsubsection{Stochastic integral at first order}
 Before proceding, we notice that,
at order $\epsilon^{-1}$, the problem admits a stationary solution only if $f=0$,
i.e. the stochastic integral can only have a gradient term depending on the fast trajectory.
Indeed, for the process $(Y_t,B_t)$ generated by $K$, at fixed slow variables, it follows from the existence of the equilibrium distribution of $Y$ ($w$) that at large times the marginal distribution of $B$ at time $t$ obeys
a diffusion equation with constant drift $\overline{f_az^a}$ and diffusion $\overline{f_af_bg^{ab}}/2$. Since the diffusion matrix of the fast process is positive, this implies that  the  stationary distribution can be obtained only if $f=0$. With this requirement $K=M$ and as in eq.~(\ref{eq:cont_equil}): $
K^\dagger w = M^\dagger w= 0 $.
Therefore
\begin{equation}
\left(\frac{\partial}{\partial \theta} + K\right) q^{(0)} = 0
\end{equation}
has a solution that is independent of the fast processes:
\begin{equation}
q^{(0)}=\eta(x,A,t)
\end{equation}

At order $\epsilon^0$ 
\begin{equation}
\left(\frac{\partial}{\partial t} + H_0\right) q^{(0)} +\left(\frac{\partial}{\partial \theta} + K \right)q^{(1)} = 0
\end{equation}
admits a stationary solution only if it is orthogonal to the nullspace of $M^\dagger$, spanned by $w$. We then find 
\begin{equation}
\left( \frac{\partial}{\partial t} + \overline{H}_0 \right) \eta =0 
\end{equation}
with the effective generator of the joint process $(X,A)$ on slow scales:
\begin{equation}
\overline{H}_0 = \overline{L}_0 +
 \overline{\left(h + r_i u^i\right)} \frac{\partial}{\partial A} 
 + \overline{r_i d^{ij}} \frac{\partial^2}{\partial A\partial x^j} + \frac{1}{2} \overline{d^{ij} r_i r_j }
\frac{\partial^2}{\partial A^2}
\end{equation}
where, as in section \ref{sec:dyn_diff}, the overbars denote averages over the
equilibrium distribution of the fast process $w(y)$.
We now convert the generator to the corresponding stochastic differential equations 
along slow trajectories for
the variables $X$, following $\overline{L}_0$ (just as in eq.~\ref{eq:L_cont}) and $A$ determined by the remaining terms.
{  
If there is no noise on the slow variables ($d=0$), the effective dynamics of eq.~(\ref{eq:L_cont}) is deterministic and the functional has the regular limit: $dA_t  =  \left(\overline{h} + \overline{r_i u^i} 
\right) dt$. If  $\overline{d}\neq0$ the effective dynamics is diffusive and}
one has that the effective dynamics for $A_t$ in the limit $\epsilon \to 0$ is given by
\begin{equation}\label{eq:effective_A_0}
dA_t  =  \left(\overline{h} + \overline{r_i u^i} - \overline{d}^{-1}_{ik} \overline{r_j d^{ij}} \overline{u}^k 
\right) dt + \overline{d}^{-1}_{ik} \overline{r_j d^{ij}} 
\cdot dX^k_t +\left(\overline{d^{ij} r_i r_j}-\overline{d}^{-1}_{ik} \overline{r_j d^{ij}}\; \overline{r_l d^{kl}}\right)^{1/2} \cdot d\xi_t
\end{equation}
along slow trajectories $X_t$ generated by $\overline{L}_0$,
where $\xi_t$ is an independent Wiener process. 
The presence of the additional noise term implies that it is not possible to express
the evolution of the statistics of the functional on the slow time scales in terms of
the slow variables only. 
We remark that the term under the square root multiplying the noise is non-negative:
{  
\begin{equation}\label{eq:nonneg_an}
\overline{d^{ij}r_i r_j} \ge\overline{d}^{-1}_{ik} \overline{d^{ij} r_j} \;\overline{d^{kl}r_l}\,.
\end{equation}
}
This follows from the non-negativity of the diffusion matrix of the slow variables $d$.
Indeed,
 $d^{ij} (v_i-r_i)(v_j-r_j) \ge 0$ for all vectors $v$ implies
$\overline{d}^{ij} v_i v_j - 2\overline{d^{ij} r_j} v_i + \overline{d^{ij}r_i r_j} \ge 0$  which in turn,
for the specific choice $v_i = \overline{d}^{-1}_{ik}\overline{d^{kl}r_l}$, gives eq.~(\ref{eq:nonneg_an})  as required.
The equality holds only when $v_i=\overline{d}^{-1}_{ik}\overline{d^{kl}r_l}=r^i$, i.e.
$r_i$ does not depend on $y$. 
Then, only when $r$ is independent of the fast variables, it is possible to
express the effective evolution of the functional as a stochastic integral over slow trajectories giving
\begin{equation}
\frac{\partial r_i}{\partial y_j}=0,\; \forall i,j  \qquad \longrightarrow  \qquad 
dA_t =  \overline{h}\, dt + r_k \cdot dX^k_t\;.
\end{equation}

It is worth  noticing that even in the case when neither $u$ nor $d$ depend on $y$, the
dynamics of $A$ will require the additional term if $r$ depends on $y$. Indeed, in this case $dA_t = \overline{h}\,dt + \overline{r}_k \cdot dX^k_t +
\left[d^{ij}\left(\overline{r_i r_j}-\overline{r}_i \overline{r}_j\right)\right]^{1/2} \cdot d\xi_t$.
Also,
when $d$ is independent of $y$ but $u$ is not , one has
$dA_t = \left(\overline{h} +\overline{r_i u^i} - \overline{r}_i \overline{u}^i \right)\,dt + \overline{r}_k \cdot dX^k_t +
\left[d^{ij}\left(\overline{r_i r_j}-\overline{r}_i \overline{r}_j\right)\right]^{1/2} \cdot d\xi_t$.
 \subsubsection{Stochastic integral at order $\epsilon$}\label{sec:funct_diff_eps}
 As we have seen for the dynamics in section \ref{sec:dyn_diff}, if $\overline{u}^i = 0$ and $d^{ij}=0$ the slow dynamics will be on scales $O(\epsilon)$.
 The averaging of the functional becomes then more involved.
 First,  for the dynamics of the functional to be on  
scales $O(\epsilon)$ we need to require
\begin{equation}
\overline{h} + \overline{r_i u^i} = 0
\end{equation}
which gives
\begin{equation}
\frac{\partial \eta}{ \partial t} = 0
\end{equation}
Under this condition, the solution is
\begin{equation}
q^{(1)} = 
- \left(M^{-1} {u}^i \right) \frac{\partial\eta}{\partial x^i} -
\left(M^{-1} \left(h + r_i{u}^i \right) \right)  \frac{\partial\eta}{\partial A} 
\end{equation}
apart from zero modes of $M$.

At order $\epsilon$ one has
\begin{equation}
 \left(\frac{\partial}{\partial \theta}  + K\right) q^{(2)} =-\left(\frac{\partial}{\partial t}  + H_0\right) q^{(1)} - \frac{\partial}{\partial \tilde{t}} q^{(0)}
\end{equation}
with solvability condition
\begin{equation}
 \frac{\partial \eta}{\partial \tilde{t}}  - \overline{
\left( \frac{\partial}{\partial t} + u^i \frac{\partial}{\partial x^i} + (h +r_i u^i) \frac{\partial}{\partial A} \right)
\left(\left(M^{-1} {u}^i \right) \frac{\partial}{\partial x^i} +
\left(M^{-1} \left(h + r_i{u}^i \right) \right)  \frac{\partial}{\partial A} \right) 
 }\eta
 = 0 \;.
 \end{equation} 
 Making use of the commutativity condition \eqref{eq:cont_commute} ensured by
 the detailed balance of the fast variables (see Appendix~\ref{appendix:neq_fun} for the non-equilibrium case)
and defining $\alpha = h+r_k u^k$  we have 
 \begin{eqnarray}\label{eq:effective_functional}
 && \frac{\partial \eta}{\partial \tilde{t}} + U^i \frac{\partial \eta }{\partial x^i} 
  + \frac{1}{2} D^{ij} \frac{\partial^2 \eta }{\partial {x}^i \partial {x}^j } +
 \\\nonumber&&
 + \overline{ u^i \frac{\partial}{\partial x^i} \left( - M^{-1} \right) \alpha }
 \frac{\partial \eta}{\partial A}
 + 2\overline{ \alpha  (- M^{-1}) u^j } \frac{\partial^2 \eta}{\partial x^j\partial A}
 + \overline{ \alpha (-M^{-1}) \alpha } \frac{\partial^2\eta}{\partial {A}^2} = 0
  \end{eqnarray}
  where $U$ and $D$ are the drift and diffusion matrix of the effective dynamics as defined in eq. (\ref{eq:cont_FP_epsilon}).
Let us now compare this effective evolution with the one which can be defined in terms of the 
slow variables alone. 
  If we require $ A_t$ to be expressed as a stochastic integral along slow trajectories 
$\tilde{X}^i$ at order $\epsilon$
\begin{equation}
d\tilde{X}^i_t = U^i dt  + \zeta^{ij} \cdot dW^j_t \qquad\qquad
\tilde{A}_t = \int_{t'}^t \tilde{h} d\tau + \tilde{f}_i \cdot d\tilde{X}^i
\end{equation} 
with
\begin{equation}
\zeta^{ik}\zeta^{jk} = D^{ij}
\end{equation}
the process $(\tilde{X}_t,\tilde{A}_t)$ is generated by
\begin{equation}\label{eq:gen_wish}
U^i \frac{\partial}{\partial \tilde{x}^i} + \frac{1}{2} D^{ij} \frac{\partial^2}{\partial \tilde{x}^i \partial \tilde{x}^j } + \left(\tilde{h} + \tilde{f}_i U^i \right) \frac{\partial}{\partial \tilde{A}} 
+ \tilde{f}_i D^{ij}  \frac{\partial^2}{\partial \tilde{x}^j \partial \tilde{A}} 
+  \frac{1}{2} \tilde{f}_i \tilde{f}_j D^{ij}  \frac{\partial^2}{\partial \tilde{A}^2} 
\end{equation}
If we want eq.~(\ref{eq:gen_wish}) to be consistent with the result of the elimination of the fast variables (\ref{eq:effective_functional})
we need to impose
\begin{equation}\label{eq:f}
\tilde{f}_i = 2\left(D^{-1}\right)_{ij}
 \overline{ \alpha  (- M^{-1}) u^j }
 \end{equation}
which consequently requires
\begin{equation}\label{eq:cont_condition2}
2 \left(D^{-1}\right)_{ij}
 \overline{ \alpha  (- M^{-1}) u^j }\; \overline{ \alpha  (- M^{-1}) u^i }=- \overline{ \alpha M^{-1} \alpha } \,.
\end{equation}
If the latter equality is met, using the definition for $\tilde{f}_i$ and the freedom in choosing 
$\tilde{h}$, a closed expression for $A_t$ in terms of slow paths only can be obtained.
 As shown in appendix \ref{appendix:cont_reg_epsilon},
condition~(\ref{eq:cont_condition2}) is satisfied if and only if $\alpha$ belongs to the subspace
spanned by the vectors $\{u_k, k=1,\ldots,n  \}$, or in other words, $\alpha$ is
a linear combination of the components of the slow drift $u_k$ (i.e. with coefficients independent of the fast variables)\footnote{Recall that the 
the slow drift $u$  depends in general on the fast variable $y$ as shown in eq.~(\ref{eq:cont_slow}).}
or, equivalently, must be amenable to the form
 \begin{equation}
 h=0 \qquad \mbox{and} \qquad \mbox{$r_k$ depends on slow variables only}.
 \end{equation}
When $\alpha$ is not a linear combination of the components of the slow drift $u_k$ 
 eq.~(\ref{eq:gen_wish}) and (\ref{eq:effective_functional}) cannot be equal and it is then 
 not possible to express the effective evolution of the functional only in terms
 of functions of the effective dynamics.
 To highlight the role played by the regular component of $\alpha$ (parallel to $u$) 
 in the
following we shall decompose $\alpha$ as
\begin{equation}
\alpha = \alpha^{\perp} + {\alpha^{\parallel}}_j u^j 
\end{equation}
with $\overline{\alpha^{\perp} M^{-1} u^j} = 0$ for all $j$, which gives
\begin{equation}\label{eq:cont_regpart_f}
\tilde{f}_i  =2 \left(D^{-1}\right)_{ij} \overline{\alpha (-M^{-1} u^j) } =
\left(D^{-1}\right)_{ij} {\alpha^{\parallel}}_k D^{jk} = {\alpha^{\parallel}}_i\;.
\end{equation}
We remark that parallel and perpendicular have to be intended with respect
to the product defined in (\ref{eq:cont_commute}).
Making use of
\begin{equation}
M^{-1}\alpha
= M^{-1}\alpha^{\perp} + {\alpha^{\parallel}}_k M^{-1} u^k
\end{equation}
and of the definitions of $U$ and $D$ given in eq.~(\ref{eq:cont_FP_epsilon}) we can express 
the effective equation for $\eta$ (\ref{eq:effective_functional}) as
\begin{eqnarray}\label{eq:detbal_effective_functional_alpha}
  \frac{\partial \eta}{\partial \tilde{t}} + U^i \frac{\partial \eta }{\partial x^i} 
  + \frac{1}{2} D^{ij} \frac{\partial^2 \eta }{\partial {x}^i \partial {x}^j }  
 +\left( \overline{u^i \frac{\partial}{\partial x^i} \left( - M^{-1}\right) \alpha^{\perp}  } 
  +  \frac{1}{2} \left(D^{kj} \right) \frac{\partial \hat{\alpha}_k}{\partial x^j}+
  + \alpha^{\parallel}_k U_k  \right)\frac{\partial \eta}{\partial A}
\\\nonumber
  + D^{jk} \alpha^{\parallel}_k \frac{\partial^2 \eta}{\partial x^j\partial A}
 +\frac{1}{2} D^{jk} \alpha^{\parallel}_j \alpha^{\parallel}_k \frac{\partial^2\eta}{\partial {A}^2} 
 + \overline{\alpha^{\perp}(-M^{-1})\alpha^{\perp}} \frac{\partial^2\eta}{\partial {A}^2} = 0
\end{eqnarray}
 This corresponds to having that the functional $A_t$ in the limit $\epsilon \to 0$ becomes then
a stochastic integral along slow trajectories plus an additional contribution
\begin{eqnarray}\label{eq:functional_Result_0}
\lim_{\epsilon\to0}A_t= \int_{t'}^t  \alpha^{\parallel}_k\circ dX^k_\tau + 
\int_{t'}^t \overline{u^i \frac{\partial}{\partial x^i}\left( - M^{-1}\right) \alpha^\perp}
\;d\tau + 
\left(2 \overline{\alpha^{\perp}(-M^{-1})\alpha^{\perp}} \right)^{1/2}dW'_\tau
\end{eqnarray}
where $W'_t$ is an independent Wiener process. 
Notice that,
since $\overline{ u^iM^{-1}\alpha^{\perp} } = 0$ also $\frac{\partial}{\partial x^i}\overline{ u^iM^{-1}\alpha^{\perp} } = 0$ and we then have that
  \begin{equation}\label{eq:entr_parts}
\overline{u^i \frac{\partial}{\partial x^i}\left( - M^{-1}\right) \alpha^{\perp}  }
= - \overline{ \left(\frac{\partial u^i }{\partial x^i} +u^i \frac{\partial \log w_{eq}}{\partial x^i}\right)  \left( - M^{-1}\right) \alpha^{\perp}}
\end{equation}
which will prove important when discussing the averaging of entropy production (see \cite{Bo2014}).
To conclude we   introduce the projection operator onto the subspace spanned by $\{u_k, k=1,\ldots,n\}$, defined
as
 \begin{equation}\label{eq:projector}
 \Pi(\bullet) = 2 u^j \left(D^{-1}\right)_{jk} \overline{\bullet (-M^{-1}) u^k}
 \end{equation}
which shall be useful for the discussion of specific examples.
It follows that a generic $\alpha=h+r_i u^i$ with $\overline{\alpha}=0$ can be decomposed 
in its parallel an perpendicular components by means of the projector:
\begin{eqnarray}
h+r_i u^i = \underbrace{2\left(D^{-1}\right)_{jk} \overline{(h+r_i u^i) (-M^{-1}) u^k}}_{\alpha^{\parallel}_j} u^j +\nonumber +
\underbrace{h + \left[r_j- 2\left(D^{-1}\right)_{jk} \overline{(h+r_i u^i) (-M^{-1}) u^k}\right] u^j}_{\alpha^{\perp}}
\end{eqnarray}
{  
\paragraph{Summary}
It is possible to derive the effective evolution of a functional of the trajectories on the slow scales. However, in general, such functional cannot be written solely in terms of the effective trajectories.
}
 \subsection{Averaging of the entropy production of diffusive dynamics: the anomalous entropy}\label{sec:diff_ent}
 {  
\paragraph{The question}
Consider entropy production in the environment as a specific functional.
Under what conditions can it be expressed in terms of the effective diffusion process?
Which additional processes must be taken into consideration in order to recover the correct functional?}\\
 A functional of the stochastic trajectories that has attracted considerable attention is the entropy production \cite{seifert2012review,chetrite2008fluctuation}.
The averaging of entropy production for general diffusive processes was
studied in detail in Ref.~\cite{Bo2014}.
In general, the entropy production along a stochastic trajectory measures the irreversibility of a system and can be defined as the log ratio of the path probability forward and backward in time as shown in \eqref{eq:entro_dys_def}
 \begin{equation*}\label{eq:def_entro}
{\cal S}_{tot}\equiv\log \frac{\cal P}{\cal P^*}  
\end{equation*}
and can be split as the sum of two contributions :
\begin{equation}\label{eq:total_entropy_def}
 d{\cal S}_{\mathit{tot}}=d{\cal S}_p+ \delta {\cal S}_{\mathit{env}}
\end{equation}
where, the state function
\begin{equation}\label{eq:Sp}
{\cal S}_{p}({\bm x},{\bm y},t) = -\ln p ({\bm x},{\bm y},t)
\end{equation}
 is  defined as the logarithm of the probability density $p$  at time $t$ 
 (the solution
 of the Fokker-Planck equation associated to Eq.~\eqref{eq:cont_slow}).
 For diffusive processes, the time-reversed process must be defined with additional care.
 For instance, in the presence of dissipative 
forces (as for example friction) a naive
inversion may lead to anti-dissipative dynamics (see Ref.~\cite{chetrite2008fluctuation} for detailed definitions, proofs and examples).
To overcome this issue one should
split the drift into a dissipative $(+)$ and a conservative $(-)$ part that transform as vector and pseudo vector fields under time-reversal 
\[
u = u_+ + u_- \qquad u_+ \longrightarrow u_+
\qquad u_- \longrightarrow - u_-
\]
\[
z = z_+ + z_- \qquad z_+ \longrightarrow z_+
\qquad z_- \longrightarrow - z_-
\]
For the study of entropy production on different time scales 
we shall consider the case in which the fast processes are completely dissipative, i.e. no conservative part of the drift $z_{-}=0$ . This is consistent with the requirement that the fast degrees of freedom are at equilibrium as required below.
For a system obeying equations (\ref{eq:cont_slow}) and (\ref{eq:cont_fast}), the entropy produced in the environment for the full system of fast and slow variables reads
\begin{eqnarray}\label{eq:def_sent}
{\cal S}_{env} =  \int_{t'}^t  2 g^{-1}_{ab} \hat{z}^b \circ dY^a_{\tau} 
+2 d^{-1}_{ij} \hat{u}_+^i \circ dX^j_\tau  
- 2 d^{-1}_{ij} \hat{u}_+^i u_{-}^j d\tau 
- \frac{\partial u_-^i}{\partial x^i} d\tau  
\end{eqnarray} 
where $\hat{z}^b = z^b - \frac{1}{2} \frac{\partial g^{ab}}{\partial y^a}$
and $\hat{u}^i = u^i -\frac{1}{2} \frac{\partial d^{ij}}{\partial x^j}$.
As already discussed for the discrete case, in order to have a finite entropy production
we must consider the case in which the fast degrees of freedom relax to an equilibrium steady-state $w_{eq}$ as defined in eq. (\ref{eq:cont_equil}).
By considering the Fokker-Planck equation associated to (\ref{eq:cont_fast}) one can see that
requiring equilibrium corresponds to imposing
\begin{equation}\label{eq:cont_fast_eq}
2 g_{ab}^{-1} \hat{z}^b = \frac{\partial}{\partial y_a} \log w_{eq}
\end{equation}
 which ensures that the basic condition that $f_a$ in eq.~(\ref{eq:def_a}) be a gradient is met.
This leads to the following expression
\begin{equation}
S_{env} = \log w_{eq}(X_t,Y_t,t)-\log w_{eq}(X_{t'},Y_{t'},t') +
\end{equation}
\begin{equation}
+
\underbrace{\int_{t'}^t \underbrace{\left(2 d^{-1}_{ij} \hat{u}_+^j  - \frac{\partial \log w}{\partial x^i} \right)}_{r_i} \cdot dX^i_\tau  + \underbrace{\left( - \frac{\partial \log w}{\partial t} - \frac{1}{2} d^{ij} \frac{\partial^2 \log w}{\partial x^i \partial x^j}+
d^{jk} \frac{\partial}{\partial x^k} \left(d^{-1}_{ij} \hat{u}_+^i \right)
- 2 d^{-1}_{ij} \hat{u}_+^i u_{-}^j 
- \frac{\partial u_-^i}{\partial x^i} \right)}_h d\tau  }_{A}
\end{equation}
to which we can directly apply the results of the previous section.
 \subsection*{Entropy at first order}
Noticing that
\begin{equation}
\overline{\hat{u}}^i  = \int w \left(u^i - \frac{1}{2} \frac{\partial d^{ij}}{\partial x^j} \right) dy 
= \overline{u}^i - \frac{1}{2} \frac{\partial \overline{d}^{ij}}{\partial x^j} +
\frac{1}{2} \overline{d^{ij}\frac{\partial \log w}{\partial x^j}} = \hat{\overline{u}}^i +\frac{1}{2} \overline{d^{ij}\frac{\partial \log w}{\partial x^j}}
\end{equation}
where we have introduced: 
\begin{equation}
 \hat{\overline{u}}^i= \overline{u}^i - \frac{1}{2} \frac{\partial \overline{d}^{ij}}{\partial x^j} \;.
\end{equation}
Then,
one has
\begin{equation}
\overline{d^{ij} r_j} = 2 \left(\overline{\hat{u}}^i_+ - \frac{1}{2}\overline{d^{ij}\frac{\partial \log w}{\partial x^j}}\right)\equiv 2 \hat{\overline{u}}_+^i
\end{equation}
\begin{equation}
\overline{d^{ij}r_i r_j} = 4 \overline{d^{-1}_{ij} \left(\hat{u}^i_+ - \frac{1}{2}d^{ik}\frac{\partial \log w}{\partial x^k}\right)\left(\hat{u}^j_+ - \frac{1}{2}d^{jl}\frac{\partial \log w}{\partial x^l}\right)} = 4 \left( \overline{d^{-1}_{ij}\hat{u}^i_+\hat{u}^j_+} - \overline{\hat{u}^i_+\frac{\partial \log w}{\partial x^i}} + \frac{1}{4} \overline{d^{ij} \frac{\partial \log w}{\partial x^i}\frac{\partial \log w}{\partial x^j}} \right)
\end{equation}
and finally, making use of $\overline{\partial_t \log w}=0$ the asymptotic slow dynamics 
%
\begin{equation}\label{eq:cont_entro1}
dA_t=\underbrace{
2 \overline{d}^{-1}_{ij} \hat{\overline{u}}_+^i \circ dX^j_\tau  
- 2 \overline{d}^{-1}_{ij} \hat{\overline{u}}_+^i \overline{u}_{-}^j d\tau 
- \frac{\partial \overline{u}_-^i}{\partial x^i} d\tau }_{d{\cal S}_{env}^{slow}}+d{\cal S}_{anom}\\
\end{equation}
along slow trajectories $X_t$ generated by $\overline{L}_0$. 
The first term coincides with the definition of entropy one would have starting from
the effective diffusive dynamics of eq.~(\ref{eq:eff0}).
The additional anomalous contribution reads
\begin{eqnarray}\label{eq:cont_anomaly1}
d{\cal S}_{anom}=\frac{1}{2}l\, d\tau+l^{1/2} d\xi_t
\end{eqnarray} 
where $l$ is defined as
\begin{eqnarray}\label{eq:cont_anomaly1_bis}
l=4\left( \overline{d^{-1}_{ij} \left(\hat{u}^i_+ - \frac{1}{2}d^{ik}\frac{\partial \log w}{\partial x^k}\right)\left(\hat{u}^j_+ 
- \frac{1}{2}d^{jl}\frac{\partial \log w}{\partial x^l}\right)} -\overline{d}^{-1}_{ij} \hat{\overline{u}}_+^i\hat{\overline{u}}_+^j
  \right) \ge0
\end{eqnarray} 
and $\xi_t$ is yet another independent Wiener process.
Interestingly, it can be shown that
\begin{equation}\label{eq:cont_ft1.2}
\langle e^{-{\cal S}_{anom}} \rangle = 1\;.
\end{equation}
If the full system is at equilibrium,
the anomalous contribution can be shown to vanish.
 \subsection*{Entropy at order $\epsilon$}
As seen for the general functional, in case  $\overline{{u}}^i=0$ and ${d}^{ij}=0$ the solution is to be sought
at order $\epsilon$. 
In such case, the slow dynamics is deterministic and this implies that
the slow drift will be conservative: 
$u_+=0$ (see \cite{chetrite2008fluctuation}).

This simplifies the expression of (\ref{eq:def_sent}) to

\begin{equation}\label{eq:cont_entro_simplepsilon}
 {\cal S}_{env} =  \int_{t'}^t  2 g^{-1}_{ab} \hat{z}^b \circ dY^a_{\tau} - \frac{\partial u^i}{\partial x^i} d\tau 
\end{equation}
which, exploiting the equilibrium condition on fast variables
(\ref{eq:cont_fast_eq}) and rearranging the terms can be expressed as

\begin{equation}\label{eq:cont_entro_simplepsilon_boundaries}
 {\cal S}_{env}
= \log w_{eq}(X_t,Y_t,t)-\log w_{eq}(X_{t'},Y_{t'},t') \underbrace{- 
\int_{t'}^t w_{eq}^{-1} \frac{\partial}{\partial x^i} \left( u^i  w_{eq} 
\right) d\tau}_{{ A}_t}
\end{equation}
under the assumption that $w_{eq}$ depends on $O(\epsilon)$ times only. 
We can now 
make use of the results of the previous section.
Identifying $\alpha=w_{eq}^{-1} \frac{\partial}{\partial x^i} \left( u^i  w_{eq}\right) $ and applying the projector defined in eq.~(\ref{eq:projector}) we can obtain the component
parallel to the drift
\begin{equation}
\Pi \left( w_{eq}^{-1} \frac{\partial}{\partial x^i} \left( u^i  w_{eq} 
\right) 
\right) = 2 u^j \left(D^{-1}\right)_{jk} \overline{w_{eq}^{-1} 
\left( \frac{\partial}{\partial x^i} \left( u^i  w_{eq} 
\right) \right)  (-M^{-1}) u^k}
= 2 u^j \left(D^{-1}\right)_{jk}  \left( \frac{1}{2}\frac{\partial D^{ik}}{\partial x^i} - U^k \right)
\end{equation}
\begin{equation}
= -\underbrace{ 2 \left(D^{-1}\right)_{jk} \hat{U}^k}_{\alpha^\parallel_j} u^j 
\end{equation}
and the regular part of the entropy in the limit $\epsilon \to 0$ is
\begin{equation}
S^{\epsilon \to 0}_{reg} = \int_{t'}^t 2 \left(D^{-1}\right)_{jk} \hat{U}^k \circ dX_\tau^j
\end{equation}
As for the additional contribution one has
\begin{equation}\label{eq:add_exa}
\overline{u^i \frac{\partial}{\partial x^i}\left( - M^{-1}\right) \alpha^\perp }
= - \overline{w_{eq}^{-1} \left(\frac{\partial}{\partial x^i}\left(u^i w_{eq}\right) \right)
\left(-M^{-1}\right) \alpha^\perp} = \overline{\alpha (-M^{-1}) \alpha^\perp} = 
\overline{\alpha^\perp(-M^{-1}) \alpha^\perp} 
\end{equation}
and 
\begin{equation}\label{eq:entro_correction}
dS_{anom} = \overline{\alpha^\perp(-M^{-1})\alpha^\perp} d\tau + \left(2 \overline{\alpha^\perp(-M^{-1})\alpha^\perp} \right)^{1/2}
dW'_t
\end{equation}
which implies for the averages along trajectories and $W'_t$
\begin{equation}
d\langle S_{anom} \rangle \ge 0
\end{equation}
and
\begin{equation}
d \langle e^{-S_{anom}} \rangle = - \langle e^{-S_{anom}} dS_{anom} \rangle + \langle \overline{\alpha^\perp(-M^{-1})\alpha^\perp} e^{-S_{anom}} \rangle dt = 0
\end{equation}
with
\begin{equation}
\alpha^\perp= - w_{eq}^{-1} \frac{\partial}{\partial x^i} \left(u^i w_{eq}\right)
 - 2 \left(D^{-1}\right)_{jk} \hat{U}^k u^j\;.
\end{equation}
{  
\paragraph{Summary}
We derived under what conditions on the system drift and diffusion coefficient the limit of the entropy
production of the full system coincides with the one of the effective system.
When such conditions are not met, an anomalous entropy production is present.
It is positive on average and to describe its evolution it is necessary to introduce and additional independent white-noise term.  
}
\subsection{Examples for functionals of diffusive trajectories}
  \subsubsection{Adaptation in microevolution}
Let us now consider an example where the elimination procedure stops at order $\epsilon^0$.
As we have seen in section \ref{sec:ko} the effective equation for the 
evolution of a genotype in a rapidly fluctuating environment that does not favor 
on average any genotype seems to be neutral (see eq.~\ref{eq:ko_eff}).

We have seen in eq.~(\ref{eq:fit_gen}) that the global fitness of a population in a varying environment changes accordingly with positive contributions given by adaptation and negative ones due to the unsteadiness of the environment. 
We can now consider it more in detail and write
\begin{eqnarray}\label{eq:dfit}
  d F(X_t,Y_t) =& \underbrace{s(X_t,Y_t) \circ dX_t}_{\mbox{adaptation}}+\underbrace{\frac{\partial F}{\partial y} \circ dY_t}_
  {\mbox{env. changes}} 
\end{eqnarray}
 
where $X_t$ and $Y_t$ are, respectively, the first genotype frequency and the environmental 
state at time $t$.
The first term on the right hand side 
accounts for the changes in fitness caused by changes in the genotype frequencies. It is the fitness flux at time $t$ introduced
 in Ref.\cite{mustonen2010fitness}
\begin{equation}\label{eq:ff}
  d\phi(t) = s(X_t,Y_t) \circ dX_t=\frac{\partial F}{\partial x}\circ dX_t\;.
\end{equation}
It provides a  measure of the adaptation driven by natural selection in the spirit of Fisher's  derivation of his fundamental
theorem (see ~\cite{Price1972} for a general discussion). 
Intuitively, this can be understood by taking the case in which the first genotype has a higher fitness; {\it i. e.}, positive selection coefficien $s>0$.
When the first genotype increases its frequency, the population is adapting to the environment and this gives a positive  fitness flux.\\
Generally speaking, the process of adaptation to an environment during the course of time
is not time symmetric. It shows a temporal
direction and is therefore an irreversible process.
A precise connection between  population genetics and out-of-equilibrium, irreversible stochastic systems can be made (see e.g. Ref. \cite{Sella2005,mustonen2010fitness}).
With the general results of the previous section we can investigate whether
the limit of the fitness flux coincides with the one defined from the effective
slow dynamics (in which case it should vanish).
We find that, 
despite the fact that the genotype follows a neutral effective evolution, adaptation is  continuously
taking place yielding a non-zero fitness flux.
Such result was obtained in Ref.~\cite{bo2014adaptation} to which we refer for a step by step derivation. 
By use of eq.~(\ref{eq:effective_A_0}),
identifying $r=s(x,y)$, $h=0$,   and  recalling that $u=s(x,y)x(1-x)+m(x)$ and that the diffusion coefficient of the
slow system $d=x(1-x)/N$ does not depend on the fast variables we have 

\begin{equation}
\lim_{\epsilon\to 0}d\phi_t  =
\left[\left(\overline{s^2}-\overline{s} \;\overline{s}\right)x(1-x)/N\right]dt+  \overline{s} 
\cdot dX_t +\left((\overline{s^2}-\overline{s}\; \overline{s})x(1-x)/N\right)^{1/2} \cdot d\xi_t
\end{equation}
%
As a consequence, in the microevolutionary limit the average value of the fitness flux is always greater 
than what one would obtain by simply using
the definition (\ref{eq:ff}) with
the effective dynamics. 
     In order to understand the implications of this result, consider the  case 
 in which the average selection vanishes i.e. no specific genotype
 is favored in the long term. In this case, the effective Kimura-Ohta
 equation is not subject to natural selection and we would expect no adaptation
 as, for example shown in eq.~(\ref{eq:ko_eff}).
However, if we compute the limiting value of fitness flux 
we obtain a finite positive value testifying that
adaptation is continuously taking place. 
 This is what happens for example if the environment follows an Ornstein-Uhlenbeck process
      discussed in section \ref{sec:ko} where
$\overline{s} = 0$ and $\overline{s^2} = \sigma^2D/K$.
At the stationary state, the regular term of the fitness flux $\langle \phi_{\mathit{eff}} \rangle$
is equal to zero because $\overline{s}=0$ and the anomalous one gives
\begin{equation}\label{eq_adapt_res}
 \lim_{\epsilon\to 0}\langle d\phi(t)\rangle=\langle d\phi_{\mathit{anom}}\rangle=\frac{D}{K} \sigma^2 \langle g \rangle \langle x(1-x) \rangle/N
\end{equation}
which is positive on average.
%
Let us interpret such result. The population is evolving in a rapidly fluctuating environment and consequently it is subject to an erratic selection so that, if one studies the evolution of the population on long time scales it may seem neutral
because the random contributions of the selection exerted by the succession of environments tend to cancel 
off.
However, this apparent
neutrality does not mean that the population has not been subject to selection and that it
has not adapted to the environments it has encountered. Indeed,  despite the fast environmental switches, the population manages to adapt on 
the fast time scales yielding a finite positive contribution 
to the fitness which is given by eq. (\ref{eq_adapt_res}).
{  
\paragraph{Summary}
We considered the evolution of a population under the influence of a rapidly changing stochastic environment. On a slow scale, the population dynamics is set by the average selection exerted by the environment. However, if one measures how much the population has adapted to the environment one finds an additional positive contribution to what would be estimated from the effective dynamics alone. Most notably, for an environment that is on average neutral, one finds a finite rate of adaptation. 
}

\subsubsection{Thermodynamics of a Brownian particle in a temperature gradient}\label{section:brown_thermo}
As we have discussed in section \ref{sec:lk} an example of a multiscale continuous system with effective
dynamics at the order $\epsilon$ is the Langevin-Kramers
dynamics in the strong friction limit which describes the Brownian motion of a particle immersed in a fluid 
with temperature $T$.  We 
study here the averaging of entropy production of a Brownian particle in a temperature gradient. We cast the results of \cite{Celani2012} whitin the framework of the general procedure for the homegenization of stochastic functionals
 provided in section \ref{sec:diff_func}.
 For Langevin-Kramers dynamics the entropy produced by the particle in the environment (the fluid) along 
a path is given by the integral of the released heat  divided 
by temperature
\begin{equation}\label{eq:Senv_def_lk}
{\cal S}_{\mathit{env}} = \int^t_{t'}   \left( \frac{f^i(X_\tau,\tau) V^i_\tau}{T(X_\tau,\tau)} d\tau - \frac{V^i_\tau}{T(X_\tau,\tau)} \circ dV^i_\tau \right)\;.
\end{equation}
as detailed in Refs.~\cite{seifert2012review,sekimoto2010stochastic,chetrite2008fluctuation}

The equations describing the process are (\ref{eq:LK_ex}) and, for the sake of compactness we do not consider
the external force ($f=0$) and {   we restrict to a} constant friction coefficient. 
In the formalism of equations (\ref{eq:cont_slow}) and (\ref{eq:cont_fast}) we have that the slow drift
$u^i=V^i_t$ and that there is no noise on the slow variable.
 
 We recall that
 the fast drift $z^a=-\gamma v^a$ is given by the friction term and the diffusion
matrix for the fast variables is $g^{ab}=2T\gamma\delta^{ab}$. With these choices, we have that
the eigenfunctions of $M$ are the Hermite polynomials (see \ref{eq:M_hermite})
and the equilibrium distribution corresponds to the local Maxwellian (see \ref{eq:maxwellian}).
Since there is no noise on the slow variable and the slow drift has zero average
with respect to the  equilibrium distribution, $\overline{L}=0$ and the effective dynamics takes place on
time scales of order $\epsilon$.
According to the definition of equation (\ref{eq:cont_FP_epsilon}) 
the effective drift is the $U=0$ and the diffusion matrix $D^{ij}=\frac{T}{\gamma}\delta^{ij}$.
When temperature does not depend on space, we see from \eqref{eq:Senv_def_lk}
that the integral can be performed immediately and we are left we boundary terms only:
${\cal S}_{\mathit{env}} = (V^2_{t'}-V^2_t)/2T$. 
In this case then $\alpha=0$.
On the contrary, if temperature does depend on space we can rewrite:
\begin{equation}
{\cal S}_{\mathit{env}} = \frac{V^j_{t'}V^j_{t'}}{T(X_{t'})}-\frac{V^j_{t}V^j_{t}}{T(X_{t})}-\int^t_{t'}  \frac{\partial T(X_\tau)}{\partial x^i}\frac{V_\tau^i}{2T^2(X_\tau)}V_\tau^jV^j_\tau d\tau 
\end{equation}
where we have made use of the fact that $dX^i=V^idt$. We then have that
\begin{equation}\alpha=-\frac{\partial T}{\partial x^i}\frac{v^i}{2T^2}v^jv^j\;.
\end{equation}
We can split it into two components, 
a part parallel (with respect to the product defined in (\ref{eq:cont_commute}))
 to the drift of the slow variables $u^i=v^i$ and an orthogonal one.
 Recalling that for this generator we have that the Hermite polynomials are eigenfunctions 
 \begin{eqnarray}
 M(\underbrace{v^i}_{H_1})=-\gamma \underbrace{v^i}_{H_1} \qquad \qquad
  M\underbrace{\left(v^i\left(v^jv^j-(n+2)T\right)\right)}_{H_3} = -3\gamma\underbrace{\ v^i\left(v^jv^j-(n+2)T\right)}_{H_3}
  \end{eqnarray}
and provide an orthogonal base:
 \begin{equation}
 \int \left(v^i\left(v^jv^j-(n+2)T\right)\right)v^iw(y)dy= \int H_3 H_1w(y)dy=0\;,
 \end{equation}
 we can then decompose
 \begin{equation}
 \alpha=-\frac{\partial T}{\partial x^i}\frac{1}{2T^2}\left(H_3+(n+2)TH_1\right)
 \end{equation}
 where the component parallel to the drift reads:
 \begin{equation}
  {\alpha^{\parallel}}^iv^i=-\frac{\partial T}{\partial x^i}\frac{(n+2)}{2T}v^i
   \end{equation}
  and the orthogonal one gives 
   \begin{equation} 
   \alpha^{\perp}=\frac{\partial T}{\partial x^i}\frac{v^i}{2T^2}\left((n+2)T-v^jv^j\right)\;.
   \end{equation}
Hence, the regular part of the entropy production reads
 \begin{equation}
   {\cal S}^{\epsilon \to 0}_{reg} = -\int_{t'}^t \frac{(n+2)}{2T}\frac{\partial T}{\partial x^i} \circ dX_\tau^j
   \end{equation}
and the anomalous contribution {   is} given by 
 \begin{equation}
d{\cal S}_{anom} = \frac{\partial T}{\partial x^i}\frac{\partial T}{\partial x^i}\frac{(n+2)}{6\gamma T} d\tau + 
\left(\frac{\partial T}{\partial x^i}\frac{\partial T}{\partial x^i}\frac{(n+2)}{3\gamma T}\right)^{1/2}
dW'_\tau\;.
\end{equation}
Notice that the fact that 
\begin{eqnarray}
 \overline{u^i \frac{\partial}{\partial x^i}\left( - M^{-1}\right) \alpha^\perp}=
 \overline{\alpha^{\perp}(-M^{-1})\alpha^{\perp}} 
\end{eqnarray}
is not accidental but completely general for the averaging of entropy production
(see eq.~\ref{eq:add_exa} and \cite{Bo2014}).
Finally, by separating the boundary terms $n d\log T$ that will combine with the other boundary terms $d (v^2/T)$ to give the Maxwellian, one  
recovers the results of Ref.~\cite{Celani2012}.
The presence of the positive correction to the average entropy production
was also identified in \cite{Spinney2012}.
Some recent studies have extended the analysis of the anomalous entropy to the rotational 
degrees of freedom of the brownian particle \cite{Lan2015,Marino2016}.\\
The symmetry breaking in the velocity distribution introduced by the presence of temperature gradients is responsible for the presence of the anomaly and may affect also other functionals. 
In the following we shall consider some other thermodynamic functionals and 
systematically compare the {\it exact asymptotic} expression for vanishing inertia to the overdamped {\it approximation}.
We will see that some of them are sensitive to
these effects (which are neglected by the overdamped approximation) and will therefore pick up anomalous contributions
whereas others are not affected by such corrections and will be simply given by the corresponding definition in the overdamped approximation.

\paragraph{Heat}
The case of the heat exchange between a Brownian particle and the
environment in presence of a temperature gradient was considered in
Ref.~\cite{Bo2013b}.
When the thermal environment is inhomogeneous it is important to consider separately the heat exchanges that take
place at different temperatures.
It is then useful to define the average rate of heat release to the thermostat at {   a specific} temperature. 
As shown in the supplementary material of \cite{Bo2013b} 
the asymptotic expressions for the heat released at a given temperature in the
limit of vanishingly small inertia is not regular. This is at the core of
the entropic anomaly.\\
On the contrary, the overall heat exchange (irrespective of the 
thermostat) has a regular behaviour, its definition coincides
with the one given for the overdamped effective dynamics {  as shown also in Ref.~\cite{Ge2014}}.
\paragraph{Excess and housekeeping entropy}
The excess entropy is the contribution to entropy production due to unsteadiness 
of the environment through time-changes of  protocol, temperature, friction, etc. 
It measures the additional dissipation due to external driving modifying the
non-equilibrium steady state. If such driving takes place on slow time scales the limit is then regular.
The regularity of excess entropy upon elimination of variables in the form of insensitivity to coarse graining has been discussed 
in Refs.~\cite{Nakayama2015,Ford2015}.
The anomalous entropy therefore contributes
to the housekeeping entropy which 
is the contribution to entropy production due to the maintenance of an out-of-equilibrium state. 
In conclusion
the limit of the housekeeping entropy is anomalous.
{  
\paragraph{Summary}
Consider diffusion in a fluid subject to a  temperature gradient.
For large friction (small inertia) the dynamics can be described in terms
of the positional degrees of freedom alone (overdamped approximation), averaging over velocities.
The rate of  entropy production in the environment in the overdamped approximation 
underestimates the actual contribution.
The missing dissipation is linked to the fact that the overdamped system 
assumes a symmetric velocity distribution whereas the full system does not have this symmetry, even in the limit of
vanishing inertia. The anomalous entropy production takes its name from this
irreversibility linked to the asymmetries that are overlooked by the overdamped approximation.
}

  \subsection{Homogenization and information loss}
  {  
\paragraph{The question}
Is it possible to give an interpretation of the entropic anomaly in terms of information loss?
}\\
 Intuitively, averaging over some degrees of freedom and reducing the amount of details used in the description of a system  is linked to information loss.
    Such intuition can be substantiated by considering our analysis on the averaging of functionals of the stochastic trajectories. The study of entropy production offers a rather transparent interpretation 
in terms of information theory and the anomalous contributions missed by the effective dynamics can be
directly linked to missing information. \\
By its very definition entropy production has an informational meaning and can be seen as the log-likelihood of the forward process versus the backward one (see eq.~\ref{eq:entro_dys_def}). 
As such, it can be employed to test the hypothesis that the process is the forward one versus the one that it is the backward one {\it i.e.} to test the direction of the arrow of time \cite{Jarzynski2011,Roldan2015}. 
Then,  the average entropy production  is the Kullback-Leibler divergence between the forward and the backward process:
\begin{equation}
 D_{KL}\left({\cal P}||{\cal P}^*\right)\equiv \langle {\cal S}
 (x) \rangle\,.
\end{equation}
It can be  interpreted in terms of the expected discrimination information between the forward and backward evolution hypotheses. 
Indeed, according to the Stein lemma, the Kullback-Leibler divergence sets the rate at which the probability 
of a false negative decision (type II error)
 exponentially decreases with repeated measurements~\cite{cover2012elements,Lexa2004}. 
  The  Kullback-Leibler divergence, similarly to mutual information, obeys a data processing inequality \cite{Kullback1951,Lexa2004} 
 so that, loosely speaking, acting  on the original variables (here a trajectory), will not increase the
 divergence of the two probabilities.
    We can consider the averaging procedure presented in the previous sections as the transformation
    we are applying to our variables:  
 a  projection from the set of complete paths 
to the space of the effective paths.
Then, according to the data processing inequality, the  difference between the Kullback-Leibler divergence of the full process and the one of the effective dynamics must be non negative and, in the limit of large time scales separation,
defines the average anomalous entropy production
\begin{equation}
 \lim_{\epsilon\to 0}D_{KL}\left({\cal P}||{\cal P}^*\right)-D_{KL}\left({\cal P}_{eff}||{\cal P}^*_{eff}\right)
 =\langle {\cal S}_{anom}(x) \rangle\ge0\,.
\end{equation}
Then one sees how the average anomalous entropy production quantifies the loss in the ability to discriminate
the forward and backward trajectory when the comparison is carried out by means of the projected effective dynamics instead of the full one.
An interesting corollary of the data processing inequality is that the Kullback-Leibler of the
projected process is equal to the original one only in case the effective dynamics is {  a} sufficient statistic
for the direction of the arrow of time.
We know that the anomalous entropy production vanishes when the rates obey detailed balance.
This means that, in this case, the trajectories of the effective system contain as much information
on the time irreversibility of the system as the original ones. 

An interesting question to address in the future is to consider the influence of the averaging procedure 
on the ability to discriminate between dynamics transformation that are not just time reversal but related to different symmetries such as rotations or general isometries, as the ones considered
in \cite{Lacoste2014,Lacoste2015}.

Additionally, as shown in Ref.~\cite{Roldan2015},  the steady state average entropy production rate can be shown to be inversely proportional to the minimal time needed to decide on the direction of the arrow of time. In this framework, 
missing some entropy production implies needing longer observation times before being able to 
decide if the system is evolving forward in time.
The average anomalous entropy production represents how much more time one needs to wait if one
has access only to the effective dynamics of the system. 
{  
\paragraph{Summary}
The entropy that is missed by the effective dynamics (anomaly) can be interpreted as a loss in the ability of discriminating the direction of the arrow of time.
}
  \section{Conclusion and discussion}
In the present report we have considered stochastic systems involving well separated time scales. 
The existence of such time-scale separation motivates and justifies the attempt to derive effective
dynamics taking place on the slower time scales resulting from decimation and coarse-graining  of the faster processes. The reduced effective process usually involves fewer degrees of freedom (varying on the slower time scales) and is, consequently, more amenable to analytical treatments and numerical simulation.
We focused on continuous-time discrete-state and diffusive Markov processes and 
used a systematic averaging procedure to obtain the effective dynamics.
Such procedure allowed us to highlight what are the conditions under which
the averaged dynamics are Markovian and to what extent the original system can be simplified.
With the same tools we addressed the behavior of functionals of the stochastic trajectory upon averaging the faster processes.
Our central attention was devoted to the question: is it possible to express the effective evolution
of a generic functional on the slower time scales as a functional of the reduced effective dynamics?
We have found that under general conditions this is not possible.
In other words, the effective description of a functional requires more details than the ones sufficient to describe the dynamics. We have studied what are the necessary additional details 
that must be retained to provide a consistent representation of the functional and shown how to account for them in the effective description.

The first message that we intend to convey takes the form of a warning.
Often, one studies a system starting with equations that result form the elimination of some faster degrees of freedom
but are capable of capturing the relevant features of the dynamics taking place at the chosen scale.
Our findings suggest that additional care must be taken when studying functionals.
Such issue is best captured by the example of entropy production, a measure of irreversibility.
For such functional, the eliminated
variables, though having no impact on the dynamics, are crucial to characterize irreversibility and hence entropy production. An effective dynamics that apparently seems  at equilibrium may, in fact, hide some underlying nonequilibrium processes that give rise to a finite entropy production and dissipation.

On the other hand, we have highlighted what are the additional details that are needed for
a consistent description on the slow scales, for general functionals and for entropy production in particular.
In this perspective our results provide a tool for deriving the correct effective
descriptions without the necessity of studying the full system involving fast and slow processes.
We have provided an application of such method to give the effective dynamics of the counting statistics of
a given transition for a discrete Markov process and to correctly compute the 
entropy production of stochastic systems.
The applicability of these general results have been illustrated by specific examples ranging from the entropy production
 of a Brownian particle in an inhomogeneous temperature environment
to the effective 
thermodynamics of simple biochemical network involving fast and slow reactions and
to measures of adaptation in a model of population genetics
 in a rapidly changing environment.

In summary, the elimination of fast transitions is a procedure that can be in general performed in stochastic systems, to some degree. Due attention must be paid to the fact that some information about fast processes must be retained. Asymptotic methods allow to identify this information and provide a way to derive effective equations and develop efficient algorithms  that correctly describe both dynamics and functionals of trajectories on slow time-scales.
{  
\section*{Acknowledgements}
The authors wish to thank Erik Aurell and Ralf Eichhorn for stimulating discussions and Ralf Eichhorn for  helpful comments on an earlier version of the manuscript.
Furthermore, they would like to thank an anonymous referee for their valuable suggestions. SB acknowledges ICTP for hospitality.  AC thanks Nordita for hospitality during the scientific program: "Stochastic thermodynamics in biology" in September 2015.}
\newpage
\appendix
\renewcommand*{\thesection}{\Alph{section}}
\section{Quasi equilibrium Michaelis-Menten with low substrate abundance.}\label{app:MM}
Let us consider the case of a Michaelis-Menten enzymatic reaction at low substrate concentration where we allow reversible product formation. This is an extension of the model discussed in section \ref{sec:MM_main}.
\begin{equation}\label{eq:MM1_gen}
\ce{E + S <=>T[$k_1$][$k_{-1}$] C <=>T[$k_2$][$k_{-2}$] E + P}
\end{equation}
where the $k$ are the specific probability rate constants, i.e. the rates for single molecule reactions.
The system is in a state specified by the number of free enzymes $e$,
substrate molecules
 $s$, enzyme-substrate complexes $c$, and products $d$. However, the conservation laws impose
\[
e  + c =e_n \qquad\qquad s + c + d = m
\]
where the total number of enzymes (free or in complex) $e_n$ and the  $m$ are fixed by the initial state. 

Choosing $d$ and $c$ as independent state variables, taking values in the ranges $0\le c \le \min(m,e_n)$ and $0 \le d +c \le m$, 
the master equation reads
\[
\frac{d}{dt} p(c,d) = k_1 (e_n-c+1)(m-d-c+1) p(c-1,d)- k_1(e_n-c)(m-d-c) p(c,d) 
 + k_{-1} (c+1) p(c+1,d) -  k_{-1} c p(c,d) 
 \]
\[
+ k_2 (c+1)p(c+1,d-1) - k_2 c p(c,d)
+k_{-2} (e_n-c+1)(d+1) p(c-1,d+1)
-k_{-2} (e_n-c)d p(c,d)
\]
or, in compact form
\[
\frac{d}{dt} P^{cd} = \sum_{c'd'} K^{cd}_{c'd'} P^{c'd'} - K^{c'd'}_{cd} P^{cd}
\]
with 
\[
K^{cd}_{c'd'} = k_1(e_n-c')(m-d'-c') \delta^{c-1}_{c'}\delta^d_{d'} + k_{-1} c' \delta^{c+1}_{c'}\delta^d_{d'}
+ k_2  c' \delta^{c+1}_{c'}\delta^{d-1}_{d'} + k_{-2} (e_n-c')d' \delta^{c-1}_{c'}\delta^{d+1}_{d'}
\]
Let us consider the quasi equilibrium case: {\it i. e.},  the reactions that involve the product are far slower than the other ones.
In this case
\[
K^{cd}_{c'd'} = \epsilon^{-1} F^{cd}_{c'd'} \delta^d_{d'} + S^{cd}_{c'd'} (1-\delta^d_{d'})
\]
where
\[
\epsilon^{-1} F^{cd}_{c'd} = k_1(e_n-c')(m-d-c') \delta^{c-1}_{c'}+ k_{-1} c' \delta^{c+1}_{c'} \qquad\qquad
S^{cd}_{c'd'} = k_2  c' \delta^{c+1}_{c'}\delta^{d-1}_{d'} + k_{-2} (e_n-c')d' \delta^{c-1}_{c'}\delta^{d+1}_{d'}
\]
In this case, the transition matrix of the fast dynamics is block diagonal and we can
use the results of section \ref{sec:dyn_block}.
In the limit $\epsilon\to 0$, the slow states, labeled by $d$, evolve according to the effective Markov process
\[
\frac{d}{dt} P^{d} = \sum_{c'd'} \overline{K}^{d}_{d'} P^{d'} - \overline{K}^{d'}_{d} P^{d}
\]
where, according to eq.~\ref{eq:eff_Rate},
\[
\overline{K}^d_{d'} = \sum_{c,c'} S^{cd}_{c'd'} w^{c'd'}
\]
and $w^{c'd'}$ is the steady state of the fast dynamics at fixed $d'$, i.e. the solution of
\[
k_1 (e_n-c+1)(m-d-c+1) w^{c-1,d}- k_1(e_n-c)(m-d-c) w^{c,d} 
 + k_{-1} (c+1) w^{c+1,d} -  k_{-1} c w^{c,d}  = 0
\]
If detailed balance holds for the fast system, then the equilibrium solution has to obey
\[
\left(k_1(e_n-c')(m-d-c') \delta^{c-1}_{c'}+ k_{-1} c' \delta^{c+1}_{c'}\right) w_{eq}^{c'd} =  \left(k_1(e_n-c)(m-d-c) \delta^{c'-1}_{c}+ k_{-1} c \delta^{c'+1}_{c}\right) w_{eq}^{cd}
\]
i.e.
\[
k_1(e_n-c+1)(m-d-c+1)w_{eq}^{c-1,d}=k_{-1} c w_{eq}^{cd}
\]
which is easily solved to give
\[
w_{eq}^{cd}=
w^{0d}_{e_nm} \left(\frac{k_1}{k_{-1}} \right)^c \frac{e_n! (m-d)!}{(e_n-c)!(m-d-c)! c!} 
\]
The constant is found by imposing normalization
\[
\frac{1}{w^{0d}_{e_nm}} = \sum_{c=0}^{\min(e_n,m-d)} \left(\frac{k_1}{k_{-1}} \right)^c \frac{e_n! (m-d)!}{(e_n-c)!(m-d-c)! c!} 
\]
One has therefore
\[
\overline{K}^d_{d'} =  w_{e_nm}^{0d'} \sum_{c=0}^{\min(e_n,m-d)} \sum_{c'=0}^{\min(e_n,m-d')} \left(k_2  c' \delta^{c+1}_{c'}\delta^{d-1}_{d'} + k_{-2} (e_n-c')d' \delta^{c-1}_{c'}\delta^{d+1}_{d'}\right)\left(\frac{k_1}{k_{-1}} \right)^{c'} \frac{e_n! (m-d')!}{(e_n-c')!(m-d'-c')! c'!} 
\]
\[
= k_2 w^{0d'}_{e_nm} \delta^{d-1}_{d'}  \sum_{c=0}^{\min(e_n,m-d)} \sum_{c'=0}^{\min(e_n,m-d')} c' \delta^{c+1}_{c'} \left(\frac{k_1}{k_{-1}} \right)^{c'} \frac{e_n! (m-d')!}{(e_n-c')!(m-d'-c')! c'!} 
\]
\[
+ k_{-2} w^{0d'}_{e_nm} \delta^{d+1}_{d'} \sum_{c=0}^{\min(e_n,m-d)} \sum_{c'=0}^{\min(e_n,m-d')} (e_n-c')d' \delta^{c-1}_{c'} \left(\frac{k_1}{k_{-1}} \right)^{c'} \frac{e_n! (m-d')!}{(e_n-c')!(m-d'-c')! c'!} 
\]
\[
= k_2 w^{0d'}_{e_nm} \delta^{d-1}_{d'}  \left(\frac{k_1}{k_{-1}} \right) e_n(m-d')\sum_{c'=1}^{\min(e_n,m-d')} \left(\frac{k_1}{k_{-1}} \right)^{c'-1} \frac{(e_n-1)! (m-d'-1)!}{(e_n-1-(c'-1))!(m-d'-1-(c'-1))! (c'-1)!} 
\]
\[
+ k_{-2} w^{0d'}_{e_nm} \delta^{d+1}_{d'} e_n d' \sum_{c'=0}^{\min(e_n-1,m-d')}\left(\frac{k_1}{k_{-1}} \right)^{c'} \frac{(e_n-1)! (m-d')!}{(e_n-c'-1)!(m-d'-c')! c'!} 
\]
Renominating $c'-1 \to c'$ in the first sum, using $\min(e_n,m-d')-1=\min(e_n-1,m-1-d')$ and resumming one obtains
\[
\overline{K}^d_{d'} = k_2  e_n(m-d')  \left(\frac{k_1}{k_{-1}} \right) \frac{w^{0d'}_{e_nm}}{w^{0d'}_{e_n-1,m-1}}  \delta^{d-1}_{d'} 
+ k_{-2} e_n d' \frac{w^{0d'}_{e_nm}}{w^{0d'}_{e_n-1,m}}  \delta^{d+1}_{d'} 
\]
Notice that the effective rate can also be written as
\[
\overline{K}^d_{d'} = k_2 \left\langle c | e_nmd'\right\rangle_{eq} \delta^{d-1}_{d'}  + k_{-2}d' \left( e_n - \left\langle c | e_nmd'\right\rangle_{eq} \right) \delta^{d+1}_{d'}  
\]
where 
\[
 \left\langle c | e_nmd\right\rangle_{eq}=\sum_{c=0}^{\min(e_n,m-d)} c w^{cd}_{eq}
\]

\subsection{Thermodynamics of quasi equilibrium Michaelis-Menten}
As we have seen in section \ref{sec:entro_block}, in the case of block-diagonal fast dynamics the 
additional  anomalous entropy to the one expressed in terms of the effective rates involves  the conditional probabilities
\[
\pi^{cd}_{c'd'} = \frac{S^{cd}_{c'd'}w^{c'd'}}{\overline{K}^d_{d'}} = 
 \delta^{d-1}_{d'} \delta^{c+1}_{c'} \frac{c' w_{e_nm}^{c'd'}}{\left\langle c | e_nmd'\right\rangle_{eq}} 
+\delta^{d+1}_{d'} \delta^{c-1}_{c'} \frac{(e_n-c') w_{e_nm}^{c'd'}}{e_n-\left\langle c | e_nmd'\right\rangle_{eq}} 
\]
The only pair that survives in the present case is
\[
\pi^{cd}_{c+1,d-1} = \frac{(c+1)w^{c+1,d-1}_{e_nm}}{\left\langle c | e_n,m,d-1\right\rangle_{eq}}
\]
\[
\pi^{c+1,d-1}_{cd} = \frac{(e_n-c)w^{c,d}_{e_nm}}{e_n-\left\langle c | e_n,m,d\right\rangle_{eq}}
\]
However, since the complete system obeys detailed balance, we know from section \ref{sec:entro_block}
that there will be no anomalous entropy production.
\subsubsection{Single enzyme Michaelis-Menten}
For a single enzyme ($e_n=1$) it is simple to evaluate the fast equilibrium probabilities:
\[
w_{eq}^{cd} = \left\{
\begin{array}{lll}
\frac{1}{\left(1+ (m-d) \frac{k_1}{k_{-1}} \right)} & \mbox{for} & c=0 \\
\frac{ (m-d) \frac{k_1}{k_{-1}}}{\left(1+ (m-d) \frac{k_1}{k_{-1}} \right)}& \mbox{for} & c=1 \\
\end{array}
\right.
\]
\[
\left\langle c | 1md\right\rangle_{eq}= \frac{ (m-d)}{\left( \frac{k_{-1}}{k_{1}} + (m-d)\right)}
\]
and the effective rate from substrate to product:
\[
\overline{K}_{d}^{d+1} = k_2 \left\langle c | 1md\right\rangle_{eq} = \frac{ k_2 (m-d) }{\left( \frac{k_{-1}}{k_{1}} + (m-d)\right)}\
\]
which is equivalent to the rate appearing in macroscopic chemical kinetics
and in eq.~(\ref{eq:mm_qe}).
In this case, the conditional probability of choosing a specific channel for going from
substrate to product are trivial since it is possible to form a product only form
the state with one complex $c=1$
\[
\pi^{0d}_{1,d-1} = \frac{w^{1,d-1}_{eq}}{\left\langle c | 1,m,d-1\right\rangle_{eq}} =1 
\]
and the reverse one
\[
\pi^{1,d-1}_{0d} = \frac{w^{0,d}_{eq}}{1-\left\langle c | 1,m,d\right\rangle_{eq}} = 1\;.
\]
We then clearly see that, as proven in general for systems obeying detailed balance, there
is no anomalous entropy production. 
\subsubsection{Two enzymes Michaelis-Menten}
When $e_n=2$

\[
w_{eq}^{cd} = \left\{
\begin{array}{lll}
\frac{1}{\left(1+ 2(m-d) \frac{k_1}{k_{-1}} +(m-d)(m-d-1)  \left(\frac{k_1}{k_{-1}}\right)^2 \right)} & \mbox{for} & c=0 \\
\frac{2(m-d) \frac{k_1}{k_{-1}} }{\left(1+ 2(m-d) \frac{k_1}{k_{-1}} +(m-d)(m-d-1)  \left(\frac{k_1}{k_{-1}}\right)^2 \right)} & \mbox{for} & c=1 \\
\frac{(m-d)(m-d-1)  \left(\frac{k_1}{k_{-1}}\right)^2 }{\left(1+ 2(m-d) \frac{k_1}{k_{-1}} +(m-d)(m-d-1)  \left(\frac{k_1}{k_{-1}}\right)^2 \right)} & \mbox{for} & c=2 
\end{array}
\right.
\]
and the average number of complexes reads:
\[
\left\langle c | 2md\right\rangle_{eq}= 2(m-d) \left(\frac{k_1}{k_{-1}}\right) \frac{ 1 + (m-d-1)  \left(\frac{k_1}{k_{-1}}\right) }{\left(1+ 2(m-d) \frac{k_1}{k_{-1}} +(m-d)(m-d-1)  \left(\frac{k_1}{k_{-1}}\right)^2 \right)}\;.
\]
The effective rate of product formation is obtained by $k_2 \left\langle c | 2md\right\rangle_{eq}$. One can see that it reduces to the 
macroscopic
Michaelis Menten equilibrium approximation in the case of large substrate 
abundance which grants that $m-d\simeq m-d-1$.
The conditional probability for different transition channels 
read:
\[
\pi^{0d}_{1,d-1} = \frac{w^{1,d-1}_{2m}}{\left\langle c | 2,m,d-1\right\rangle_{eq}} = \frac{1}{1+(m-d)\left(\frac{k_1}{k_{-1}}\right)}
\]
\[
\pi^{1d}_{2,d-1} = \frac{2w^{2,d-1}_{2m}}{\left\langle c | 2,m,d-1\right\rangle_{eq}} =\frac{(m-d)\left(\frac{k_1}{k_{-1}}\right)}{1+(m-d)\left(\frac{k_1}{k_{-1}}\right)}
\]
\[
\pi^{1,d-1}_{0d} = \frac{2 w^{0,d}_{2m}}{2-\left\langle c | 2,m,d\right\rangle_{eq}} = \frac{1}{1+(m-d)\left(\frac{k_1}{k_{-1}}\right)}
\]
\[
\pi^{2,d-1}_{1d} = \frac{w^{1,d}_{2m}}{2-\left\langle c | 2,m,d\right\rangle_{eq}}=\frac{(m-d)\left(\frac{k_1}{k_{-1}}\right)}{1+(m-d)\left(\frac{k_1}{k_{-1}}\right)}
\]
showing that, as expected from the detailed balance condition,  there is no anomalous entropy production.

\section{Proofs for the effective dynamics of diffusive systems at order $\epsilon$ and extensions to non-equilibrium fast diffusive processes}\label{appendix:neq_diff}
We here provide the proof that  $
 \overline{M^{-1}f} = 0$ for any $f$ such that $\overline{f}=0$. 
We also show that when the fast dynamics reaches an equilibrium steady state, eq.~(\ref{eq:cont_commute}) holds: {\it i. e.}, for any $f$ and $g$ with $\overline{f}=\overline{g}=0$ we have that
$\overline{g \left(-M^{-1} f \right)} 
= \overline{f \left(-M^{-1} g\right)}$.
We start by considering fast dynamics that reaches a steady state on the
short time scales characterized by a steady state distribution $w(y)$ which is not necessarily
an equilibrium one.
For any $\overline{f}=0$, one can write for a generic variable $y$
 \[
  f (y')=\int dy \delta(y-y')f(y)-\underbrace{\int dy w(y)f(y)}_{=0}
 \]
which can be expressed in terms of the propagator $W$  of the fast process generated by $M$ by recalling that
\[
 W(y,t|y',t)=\delta(y-y')\qquad\qquad w(y)=W(y,t|y',-\infty)
\]
so that 
\begin{equation}\label{eq:fy1}
 f (y')=\int dy \left[\int_{-\infty}^{t} dt' \frac{\partial}{\partial t'}W(y,t|y',t')\right] f(y)\;.
\end{equation}
The fast propagator obeys
\begin{equation}
\left(\frac{\partial}{\partial t'} + M' \right) W(y,t|y',t') =0
\qquad\qquad
\left(\frac{\partial}{\partial t} - M^\dagger \right) W(y,t|y',t') =0
\end{equation}
and admits a biorthogonal decomposition as 
\begin{eqnarray}
W(y,t|y',t') = \sum_{E} \phi_E(y')\phi_E^+(y) e^{-E(t-t')} \nonumber \\ 
M \phi_E = -E \phi_E \qquad\qquad M^\dagger \phi^+_E = -E \phi^+_E\;.
\end{eqnarray}
 In general, combining this expression with eq.~(\ref{eq:fy1}) we have that 
\begin{equation*}
 f (y')=-\int dy \int_{-\infty}^{t} dt'\left[M'W(y,t|y',t')\right] f(y)
 \end{equation*}
and that consequently
\begin{eqnarray}\label{eq:m-}
(-M^{-1} f) (y')&=&\int dy \int_{-\infty}^t dt'\, W(y,t|y',t') f(y)= \nonumber\\
&=& \int dy \int_{-\infty}^t dt'\, \sum_{E} \phi_E(y')\phi_E^+(y) e^{-E(t-t')} f(y)=\nonumber\\
&=&\left(\sum_{E>0} E^{-1} \phi_E(y') \int dy\, \phi_E^+(y) f(y)\right)+ \phi_0(y')\int_{-\infty}^t dt'\overline{f}=\nonumber\\
&=&\sum_{E>0} E^{-1} \phi_E(y') \int dy\, \phi_E^+(y) f(y)\;.
\end{eqnarray}
Since the steady state is the right eigenfunction associate with the $0$ eigenvalue ($w=\phi_0^+$) it is orthogonal to the eigenfunctions spanning the other eigenspaces and we
finally obtain:
 \begin{equation}
\overline{M^{-1} f} = 0\;.
 \end{equation}

With eq.~(\ref{eq:m-}) we can express the effective diffusion coefficient  appearing in eq.~(\ref{eq:cont_FP_epsilon}), as
\begin{equation}\label{eq:diff}
D^{ij}=2 \overline{{u}^i \left(- M^{-1} {u}^j\right) } =2 \sum_{E>0} E^{-1} \int dz'\,\phi_E(z')w(z') u^i(z') \int dz\, \phi_E^+(z) u^j(z)=2 \sum_{E>0} E^{-1} \overline{\phi_E u^i} \int dz\, \phi_E^+(z) u^j(z)
\end{equation}
This also implies the Green-Kubo-Taylor formula for the diffusion coefficient
\begin{equation}\label{eq:green-kubo}
\frac{1}{2} D^{ij} =-\overline{{u}^i \left(M^{-1} {u}^j \right)} = 
\int dy\,dz \int_{-\infty}^t dt'\, w(y) {u}^i(y,t)  W(z,t|y,t') {u}^j(z,t) 
=\int_{-\infty}^t  dt'\langle {u}^i(X_t,Y_t,t) {u}^j(X_t,Y_{t'},t) \rangle_{\mathit{ss,fast}}
\end{equation}
We remark that for non-equilibrium fast dynamics, 
the diffusion matrix in general is not symmetric. It can however, be shown to be positive definite by proving that 
\begin{equation}\label{eq:norm}
 \overline{f(-M^{-1})f}>0
\end{equation}
for any function $f$ satisfying $\overline{f}=0$.
Since the process is stationary the autocorrelation depends only on the difference between
the times $\tau=t-t'$ and we can express the product
\[
 \overline{f(-M^{-1})f}=\int^{\infty}_0  d\tau\langle {f}^i(X_t,Y_\tau,\tau) {f}^j(X_t,Y_{0},0) \rangle_{\mathit{ss,fast}}
\]

The Wiener Kinchin theorem states that
\[
 \langle\hat{f}(k)\hat{f}^*(q) \rangle=\frac{1}{2\pi}\delta(k-q)\int_{-\infty}^{\infty} d\tau \langle f(\tau)f(0) \rangle e^{ik\tau} 
\]
where $\hat{f}(k)$ denotes the Fourier transform. By setting $q=k=0$ we have
\[
  \frac{1}{\pi}\overline{f(-M^{-1})f}=\frac{1}{\pi}\int^{\infty}_0  d\tau\langle {f}^i(X_t,Y_\tau,\tau) {f}^j(X_t,Y_{0},0) 
=\langle|\hat{f}(0)|^2 \rangle \ge 0
\]
that proves \eqref{eq:norm} for all functions such that $\hat{f}(0)=\int_{-\infty}^{\infty} d\tau f(\tau)\neq 0  $.\\
The asymmetry of the diffusion matrix \eqref{eq:diff} is not relevant for the propagator since the it is traced with the space derivatives.
We can therefore define a symmetrized diffusion matrix without altering the form of propagator:
\begin{equation}
 \tilde{D}^{ij}=\left(\frac{D^{ij}+D^{ji}}{2}\right)=\overline{{u}^i \left(- M^{-1} {u}^j\right) }+\overline{{u}^j \left(- M^{-1} {u}^i\right) }
\end{equation}
 where clearly $ D^{ij}\frac{\partial^2\rho}{\partial x^i\partial x^j}=\tilde{D}^{ij}\frac{\partial^2\rho}{\partial x^i\partial x^j} $.
\subsection{Equilibrium dynamics}
If the steady state of the fast variable is an equilibrium one 
\begin{equation}
\phi^+_E = w_{eq} \phi_E 
\end{equation}
 by detailed balance and, by use of equation (\ref{eq:m-}) one has that, for any
 $f$ and $g$ with $\overline{f}=\overline{g}=0$
\begin{equation}\label{eq:detbalproducts}
\overline{g \left(-M^{-1} f \right)} 
= \sum_{E>0} E^{-1} \overline{f\phi_E}\;\overline{g\phi_E}
= \overline{f \left(-M^{-1} g\right)} 
\end{equation}

The diffusion coefficient is therefore symmetric and positive 
\begin{equation}\label{eq:detbaldiffusion}
\frac{1}{2} D^{ij} = \overline{{u}^i \left(- M^{-1} {u}^j\right) } 
= \sum_{E>0} E^{-1} \overline{\phi_E u^i } \; \overline{\phi_E u^j } 
\end{equation}

\subsection{Functionals for non-equilibrium fast dynamics}\label{appendix:neq_fun}
In this appendix we generalize the discussion of section \ref{sec:funct_diff_eps}
to the case in which the fast dynamics reach a non-equilbrium steady state.
The main difference is that in the non-equilibrium case is that, at variance with \eqref{eq:cont_commute}, in general,
\begin{equation}\label{eq:cont_non_commute}
\overline{g \left(-M^{-1} f \right)} 
\neq \overline{f \left(-M^{-1} g\right)} 
\end{equation}
which implies that eq.~(\ref{eq:effective_functional})
becomes

 \begin{eqnarray}\label{eq:effective_functional_neq}
 && \frac{\partial \eta}{\partial \tilde{t}} + U^i \frac{\partial \eta }{\partial x^i} 
  + \frac{1}{2} D^{ij} \frac{\partial^2 \eta }{\partial {x}^i \partial {x}^j } +
 \\\nonumber&&
 + \overline{ u^i \frac{\partial}{\partial x^i} \left( - M^{-1} \right) \alpha }
 \frac{\partial \eta}{\partial A}
 + \left( \overline{ \alpha  (- M^{-1}) u^j }+
\overline{ u^j (- M^{-1})   \alpha  }\right) \frac{\partial^2 \eta}{\partial x^j\partial A}
 + \overline{ \alpha (-M^{-1}) \alpha } \frac{\partial^2\eta}{\partial {A}^2} = 0
  \end{eqnarray}
If we want eq.~(\ref{eq:gen_wish}) to be consistent with the result of the elimination of the fast variables (\ref{eq:effective_functional_neq})
we need to impose
\begin{equation}\label{eq:f_neq}
\tilde{f}_i = \left(D^{-1}\right)_{ij}
\left( \overline{ \alpha  (- M^{-1}) u^j }+
\overline{ u^j (- M^{-1})   \alpha  }\right)
\end{equation}
which consequently requires
\begin{equation}\label{eq:cont_condition2_neq}
\frac{1}{2} \left(D^{-1}\right)_{ij}
\left( \overline{ \alpha  (- M^{-1}) u^j }+
\overline{ u^j (- M^{-1})   \alpha  }\right)
\left( \overline{ \alpha  (- M^{-1}) u^i }+\overline{ u^i (- M^{-1})   \alpha  }\right)
=- \overline{ \alpha M^{-1} \alpha } 
\end{equation}

If the latter equality is met, using the definition for $\tilde{f}_i$ and the freedom in choosing 
$\tilde{h}$, a closed expression for $A_t$ in terms of slow paths only can be obtained.

\subsubsection{Conditions for the regularity of a functional of a diffusive trajectory.}
\label{appendix:cont_reg_epsilon}
We show here that condition~(\ref{eq:cont_condition2_neq}) is satisfied if and only if $\alpha$ belongs to the subspace
spanned by the vectors $\{u_k, k=1,\ldots,n  \}$.
The same steps can be applied to the simpler case of equilibrium fast dynamics 
and condition~(\ref{eq:cont_condition2})

Defining 
 $\beta_i=\left(\zeta^{-1}\right)_{ij} u^j$ equation~(\ref{eq:cont_condition2_neq}) takes the form

\begin{equation}
\left( \overline{ \alpha  (- M^{-1}) \beta_k }+
\overline{ \beta_k(- M^{-1})   \alpha  }\right)
\left( \overline{ \alpha  (- M^{-1}) \beta_k}+\overline{ \beta_k (- M^{-1})   \alpha  }\right)
=- 2\overline{ \alpha M^{-1} \alpha }
\end{equation}

%
%

This means that $\alpha$ must be in the subspace
spanned by the vectors $\{\beta_k, k=1,\ldots,n  \}$, or in other words, $\alpha$ is
a linear combination of the functions $\beta_k$ (i.e. with coefficients independent of the fast variables). Indeed, noticing that
\begin{equation}
- \overline{\beta_k(M^{-1} \beta_l)}- \overline{\beta_l(M^{-1} \beta_k)}=
\left(\zeta^{-1}\right)_{ik}D^{ij} \left(\zeta^{-1}\right)_{jl}=\delta_{kl}
\end{equation}
%

and decomposing $\alpha$ in components within the subspace and orthogonal to it

\begin{equation}\label{eq:alphadecomp}
\alpha=\alpha^{\perp}+\tilde{\alpha}_i \beta_i \qquad\mbox{where} \qquad 
-\overline{  \alpha^{\perp} M^{-1}\beta_j}-\overline{\beta_j   M^{-1}\alpha^{\perp}}  
 =0 \qquad \mbox{for all $j=1,\ldots,n$}
\end{equation}
we have
\begin{equation}
 \overline{\alpha^{\perp}(-M^{-1})\alpha^{\perp}}=\sum_{E>0} E^{-1} \overline{\alpha^{\perp} \phi_E} \int \alpha^{\perp}(z) \phi_E^+(z)\,dz=0
\end{equation}

 which, given \eqref{eq:norm}, implies  $\alpha^{\perp}=0$ (the component along $\phi_0$ is absent since $\overline{\alpha^{\perp}}=0$).

\subsubsection{Corrections to the regular terms}

 When $\alpha$ is not a linear combination of the components of the slow drift $u_k$ 
 eq.~(\ref{eq:gen_wish}) and (\ref{eq:effective_functional}) cannot be equal and it is then 
 not possible to express the effective evolution of the functional only in terms
 of functions of the effective dynamics.
 To highlight the role played by the regular component of $\alpha$ (parallel to $u$) 
 in the
following we shall decompose $\alpha$ as
\begin{equation}
\alpha = \alpha^{\perp} + {\alpha^{\parallel}}_j u^j 
\end{equation}
with $\overline{\alpha^{\perp} M^{-1} u^j} = 0$ for all $j$, which gives
\begin{equation}\label{eq:cont_regpart_f_neq}
\tilde{f}_i  =2 \left(D^{-1}\right)_{ij} \overline{\alpha (-M^{-1} u^j) } =
\left(D^{-1}\right)_{ij} {\alpha^{\parallel}}_k D^{jk} = {\alpha^{\parallel}}_i\;.
\end{equation}
Making use of
\begin{equation}
M^{-1}\alpha
= M^{-1}\alpha^{\perp} + {\alpha^{\parallel}}_k M^{-1} u^k
\end{equation}
and of the definitions of $U$ and $D$ given in eq.~(\ref{eq:cont_FP_epsilon}) we can express 
the effective equation for $\eta$ (\ref{eq:effective_functional}) as
 \begin{eqnarray}\label{eq:effective_functional_alpha}
  \frac{\partial \eta}{\partial \tilde{t}} + U^i \frac{\partial \eta }{\partial x^i} 
  + \frac{1}{2} D^{ij} \frac{\partial^2 \eta }{\partial {x}^i \partial {x}^j }  
 +\left( \overline{u^i \frac{\partial}{\partial x^i} \left( - M^{-1}\right) \alpha^{\perp}  } 
  +  \frac{1}{2} \left(D^{kj}+\overline{u^j M^{-1} u^k}-\overline{u^k M^{-1} u^j} \right) \frac{\partial \alpha^{\parallel}_k}{\partial x^j}+
  + \alpha^{\parallel}_k U_k  \right)\frac{\partial \eta}{\partial A}
\\\nonumber
  + D^{jk} \alpha^{\parallel}_k \frac{\partial^2 \eta}{\partial x^j\partial A}
 +\frac{1}{2} D^{jk} \alpha^{\parallel}_j \alpha^{\parallel}_k \frac{\partial^2\eta}{\partial {A}^2} 
 + \overline{\alpha^{\perp}(-M^{-1})\alpha^{\perp}} \frac{\partial^2\eta}{\partial {A}^2} = 0
\end{eqnarray}
 This corresponds to having that the functional $A_t$ in the limit $\epsilon \to 0$ becomes then
a stochastic integral along slow trajectories plus an additional contribution
\begin{eqnarray}\label{eq:functional_Result_0_neq}
\lim_{\epsilon\to0}A_t= \int_{t'}^t  \alpha^{\parallel}_k\circ dX^k_\tau + 
\int_{t'}^t \left[\overline{u^i \frac{\partial}{\partial x^i}\left( - M^{-1}\right) \alpha^\perp}
+\frac{1}{2}\left(\overline{u^j M^{-1} u^k}-\overline{u^k M^{-1} u^j} \right) \frac{\partial \alpha^{\parallel}_k}{\partial x^j}\right] d\tau + 
\left(2 \overline{\alpha^{\perp}(-M^{-1})\alpha^{\perp}} \right)^{1/2}dW'_\tau
\end{eqnarray}
In case 
\begin{equation}\label{eq:irrotalpha}
 \frac{\partial\alpha^{\parallel}_i}{\partial x^j}=\frac{\partial\alpha^{\parallel}_j}{\partial x^i}
\end{equation}
which corresponds to an irrotational $\alpha^{\parallel}$
then \eqref{eq:functional_Result_0_neq} simplifies to
\begin{eqnarray}\label{eq:functional_Result}
 \lim_{\epsilon\to0}A_t=\int_{t'}^t  \alpha^{\parallel}_k\circ dX^k_\tau + 
\int_{t'}^t \overline{u^i \frac{\partial}{\partial x^i}\left( - M^{-1}\right) \alpha^{\perp}  } d\tau + \left(2 \overline{\alpha^{\perp}(-M^{-1})\alpha^{\perp}} \right)^{1/2}
dW'_\tau
\end{eqnarray}
where $W'_t$ is an independent Wiener process. 

We then introduce the projection operator onto the subspace spanned by $\{u_k, k=1,\ldots,n\}$, defined
as
\begin{equation}
\Pi(\bullet) = u^j \left(D^{-1}\right)_{jk} \left(\overline{\bullet (-M^{-1}) u^k}+\overline{u^k (-M^{-1})\bullet }\right)
\end{equation}
It follows that a generic $\alpha=h+r_i u^i$ with $\overline{\alpha}=0$ can be decomposed 
in its parallel an perpendicular components by means of the projector:
\begin{eqnarray}
h+r_i u^i &=& \underbrace{\left(D^{-1}\right)_{jk}\left( \overline{(h+r_i u^i) (-M^{-1}) u^k}+\overline{u^k (-M^{-1})(h+r_i u^i)}\right)}_{\alpha^{\parallel}_j} u^j +\nonumber \\
&+&\underbrace{h + \left[r_j- \left(D^{-1}\right)_{jk} \left( \overline{(h+r_i u^i) (-M^{-1}) u^k}+\overline{u^k (-M^{-1})(h+r_i u^i)}\right)\right] u^j}_{\alpha^{\parallel}}
\end{eqnarray}

 \section{Generator of a joint stochastic process. }\label{appendix:generator}
To illustrate the steps required to obtain eq.~(\ref{eq:generator}), in this appendix we show how to
 derive the generator of the joint  process $X_t,Y_t,{\cal A}_t$ following equations (\ref{eq:cont_slow}),
 (\ref{eq:cont_fast})  and (\ref{eq:def_a}) i.e.
\begin{equation}
dX^i_t = {u}^i(X_t,Y_t,t) dt + \beta^{ij}(X_t,Y_t,t) \cdot dB^j_t
\end{equation}
\begin{equation}
dY^a_t = \epsilon^{-1}{z}^a(X_t,Y_t,t) dt + \epsilon^{-1/2}\sigma^{ab}(X_t,Y_t,t) \cdot d\hat{B}_t^b
\end{equation}
\begin{equation}
d{\cal A}_t = {r}_i(X_t,Y_t,t)\cdot dX^i_t +h\,dt
\end{equation}
where as before
\begin{equation}
d^{ij}=\beta^{ik}\beta^{jk} \qquad\qquad
g^{ab}=\sigma^{ac}\sigma^{bc}\;.
\end{equation}
The case with an additional functional $B$ of the fast variables is analogous.
Similarly to the derivation of It\=o's formula we express the differential of a generic function $f(X_t,Y_t,{\cal A}_t)$ as
\begin{eqnarray}
df=&&\left[dX^i\frac{\partial }{\partial x^i}+\frac{1}{2}dX^idX^j\frac{\partial^2 }{\partial x^i\partial x^j}+
dY^a\frac{\partial }{\partial y^a}+\frac{1}{2}dY^adY^b\frac{\partial^2 }{\partial y^a \partial y^b}\right.\\\nonumber
&&\left. +  
d{\cal A} \frac{\partial }{\partial { A}}+\frac{1}{2}d{\cal A}^2 \frac{\partial^2 }{\partial { A}^2}+
dX^idY^a\frac{\partial^2 }{\partial x^i\partial y^a}
+d{\cal A}dX^i \frac{\partial^2 }{\partial { A}\partial x^i}+d{\cal A}dY^a \frac{\partial^2 }{\partial { A}\partial y^a}
\right]f
\end{eqnarray}
so that upon taking the average and retaining only the contributions of order $O(dt)$ we have that
\begin{eqnarray}
 \frac{d\langle f \rangle}{dt}= &&\left\langle\left[\underbrace{u^i\frac{\partial }{\partial x^i}+\frac{1}{2}d^{ij}\frac{\partial^2 }{\partial x^i\partial x^j}}
 _{L_0}+\underbrace{
\epsilon^{-1}z^a\frac{\partial }{\partial y^a}+\epsilon^{-1}\frac{1}{2}g^{ab}\frac{\partial^2 }{\partial y^a \partial y^b}}
_{\epsilon^{-1}M}\right.\right.\\\nonumber
&&\left.\left. +  
\left(r_iu^i+h \right) \frac{\partial }{\partial { A}}+\frac{1}{2}d^{ij}r_ir_j\frac{\partial^2 }{\partial { A}^2}+
+d^{ij}r_i\frac{\partial^2 }{\partial { A}\partial x^i}
\right]f\right\rangle
\end{eqnarray}
from which eq. (\ref{eq:generator}) follows.



\bibliographystyle{elsarticle-num}
 \bibliography{review}
  




\end{document}